\documentclass{elsarticle}

\usepackage[top=2cm,bottom=3cm,left=2cm,right=3cm]{geometry}
\usepackage{graphicx}
\usepackage{amsmath,amssymb,amsthm}
\usepackage{bm,color,url}
\usepackage{mdframed}
\usepackage{subcaption}
\usepackage{breakcites}
\usepackage{thmtools}
\usepackage{enumitem}
\usepackage{subcaption}

\newcommand{\mmu}{\bm{\mu}}
\newcommand{\ttheta}{\bm{\vartheta}}
\newcommand{\hh}{\bm{h}}
\newcommand{\pp}{\bm{p}}
\newcommand{\qq}{\bm{q}}

\newcommand{\Rset}{\mathbb{R}}

\newcommand{\qoi}{G}
\newcommand{\pparam}{\ttheta}
\newcommand{\param}{\vartheta}
\newcommand{\mybar}[1]{\bar{#1}}

\DeclareMathOperator*{\argmax}{arg\,max}
\DeclareMathOperator*{\argmin}{arg\,min}

\usepackage[linesnumbered,ruled]{algorithm2e}
\SetKwProg{Fn}{}{}{end}

\newtheorem{example}{Example}
\newenvironment{biblio}{\paragraph{\textbf{Bibliography and further reading}} \itshape}{}

\usepackage[normalem]{ulem}
\newcounter{lnote}

\newcommand{\lorenzo}[1]{\textcolor{black}{#1}}

\begin{document}
	
\begin{frontmatter}	
	
\title{A note on tools for prediction under uncertainty and identifiability of SIR-like dynamical systems for epidemiology}
\author[imati]{Chiara Piazzola}
\ead{chiara.piazzola@imati.cnr.it}
\author[imati]{Lorenzo Tamellini\corref{cor1}} 
\ead{tamellini@imati.cnr.it}
\address[imati]{Consiglio Nazionale delle Ricerche - Istituto di Matematica Applicata e Tecnologie Informatiche ``E. Magenes'' (CNR-IMATI), Via Ferrata 5/A, 27100 Pavia, Italy}
\author[aachen,kaust]{Ra\'ul Tempone}
\ead{tempone@uq.rwth-aachen.de, raul.tempone@kaust.edu.sa}
\cortext[cor1]{Corresponding author}
\address[aachen]{Alexander von Humboldt Professor in Mathematics for Uncertainty Quantification, RWTH Aachen University, Pontdriesch 14-16, 52062,
    Aachen, Germany}
\address[kaust]{King Abdullah University of Science and Technology (KAUST) - Computer, Electrical and Mathematical Sciences \& Engineering Division (CEMSE), Thuwal, 23955-6900, Saudi Arabia.}

\begin{abstract} 
	We provide an overview of the methods that can be used for prediction under uncertainty and data fitting
	of dynamical systems, and of the fundamental challenges that arise in this context.
	The focus is on SIR-like models, that are being commonly used when attempting to predict the trend of the COVID-19 pandemic. 
	In particular, we raise a warning flag about identifiability of the parameters of SIR-like models;
	often, it might be hard to infer the correct values of the parameters from data, even for very simple models,
	making it non-trivial to use these models for meaningful predictions.
	Most of the points that we touch upon are actually generally valid for inverse problems in more general setups.
\end{abstract}


\end{frontmatter}

\section{Introduction}

This work provides an overview of the methods that can be used for prediction under uncertainty
(also known as Uncertainty Quantification) and data fitting of dynamical systems, and of the fundamental challenges that arise in this context.
While this work can be easily connected with the usage of SIR-like models for the COVID-19 pandemic,
the discussion presented here is actually valid for compartmental models in epidemiology and for dynamical systems in general;
most points would actually be valid also in the context of inverse problems with spatial inhomogeneities.
We put particular emphasis on the issue of identifiability, whose possible lack might cause serious issues when attempting
long-term forecasts. To make our case clearer, in this work we use synthetic data only,
which gives us full control on the errors generated by the numerical identifiability procedure.

For the sake of compactness, we have chosen to not provide many technical details on the topics that we touch,
but rather point the reader to the relevant bibliography. For the same reason,
most of the bibliography for further reading is provided at the end of each section, rather than during the discussion.
We chose, however, to keep a rather concrete register, therefore
each section comes with one or two short examples. We use for this purpose simple
models, with the understanding that the points raised by the examples will be even more valid for more complicated models.
For a more bird's eye view on data-informed modeling and identifiability, and their ramifications in the general society,
see e.g. \cite{alahmadi,guillaume.identifiability}.
For readers' convenience, we report here the topic of each section and list the examples:
\begin{description}[itemsep=-4pt]
  \item[Section \ref{sec:SIR_models}:] SIR-like models in epidemiology
  \item[Section \ref{section:prior_forward}:] Forward Uncertainty Quantification (UQ): tools to make predictions under uncertainty
  \item[Section \ref{sect:sobol}:] Sensitivity analysis: pinpointing what parameters we need to get right, and preliminary assessment
  of feasibility of inversion
  \item[Section \ref{section:inverse}:] Inverse UQ (data fitting) as a preliminary step to tune the pdf of the parameters to the data
  \item[Section \ref{section:UQ_workflow}:] An ideal UQ workflow, from data to predictions under uncertainty
  \item[Section \ref{sect:struct_identifiability}:] Structural identifiability
  \item[Section \ref{section:pract_id}:] Practical identifiability
  \item[Section \ref{section:disc}:] Discussion and conclusions: a revisited UQ workflow
  \end{description}
\begin{description}[itemsep=-4pt]
  \item[Example \ref{ex:UQ_for_SIR}:] Forward UQ of a SIR model 
  \item[Example \ref{ex:Sobol_for_SIR}:] Computing the Sobol indices for a SIR model 
  \item[Example \ref{ex:inverseSIR}:] Inverse and posterior-based forward UQ of a SIR model 
  \item[Example \ref{ex:non-uniform-prior}:] Incorporating prior information on parameters in inverse UQ
  \item[Example \ref{ex:two-sigma}:] Inverse UQ when different data types have different noise levels
  \item[Example \ref{ex:structural_identifiability_SIR}:] Structural identifiability of a SIR model by differential algebra
  \item[Example \ref{ex:structural_identifiability_SIR_evans}:] Structural identifiability of a SIR model by mapping approach
  \item[Example \ref{ex:SIR_identifiability}:] Practical non-identifiability of a SIR model with unknown under-reporting factor 
  \item[Example \ref{ex:SEIRDz}:] Structural and practical identifiability of a SEIRD model 
\end{description}

\lorenzo{All the numerical results in the examples have been obtained with Matlab,
  and the source code is available at \url{https://sites.google.com/view/sparse-grids-kit}.
  Some of the examples rely on the functionalities of the Sparse Grids Matlab Kit,
  which is developed by some of the authors of this manuscript and can be downloaded from the same website.}

\section{SIR-like models in epidemiology}\label{sec:SIR_models}

The recent COVID-19 pandemic has triggered an unprecedented effort among researchers worldwide\footnote{On July 14th: 1600+ preprints on \url{arxiv.org}, 5100+ preprints on \url{medrxiv.org}, 1400+ preprints on \url{biorxiv.org}}.
In the field of applied mathematics, a large share of this effort has been focusing on devising tools to forecast the trends of the epidemics.

The most widely used tools to this end are compartmental models, where individuals of a population are categorized
in compartments (Infected, Recovered, Dead, etc.) and can transition from one compartment to another according to
some ``transition rates''. The origin of these models can be traced back to the work of Kermack and McKendrick \cite{Kermack}.
The actual model in that paper was a system of integro-differential equations. Some simplifications allow to rewrite
those equations as a non-linear system of ordinary differential equations (ODEs), whose simplest form is the SIR model:
\begin{equation}\label{eq:SIR}
  \begin{cases}
    \displaystyle \dot{S} = -\frac{\beta}{N_{pop}}IS\\[6pt]
    \displaystyle \dot{I} = \frac{\beta }{N_{pop}}IS - r I\\[6pt]
    \displaystyle \dot{R} = r I,    
  \end{cases}
\end{equation}
which describes the time-evolution of three compartments: individuals (S)usceptible to the disease, individuals (I)nfected with the disease,
and finally individuals (R)emoved from the disease dynamics (either because they recovered, assuming immunity after having contracted the disease, or died).
The total number of individuals in the population $N_{pop}=S+I+R$ is supposed constant,
and individuals transition from one compartment to the next one with certain transition rates
$\beta,r$. Besides the ODE, the Kermack and McKendrick integro-differential equations can also be rewritten as a stochastic differential equation
whose limit is the ODE equation; see \cite{Capistran}.

Of course, a simple SIR model is insufficient to capture the dynamics of the COVID-19 disease, due to its biological peculiarities,
such as the incubation time and the presence of asymptomatic carriers of the disease,
as well as human interventions such as individuals
in quarantine (hence with limited transmissivity) and hospitalized. Therefore, many works in the COVID-19 literature
consider more complex variations of the simple SIR model (\ref{eq:SIR}) with, for example, more compartments, time-dependent coefficients, or by introducing
network models, in an attempt to better describe the dynamics of the pandemic and provide reliable forecasts of its evolution.
Of course, one should always keep in mind that while more complex models have potentially a greater predictive power,
they are also more complex to analyze and tune, so that one should ideally look for the model with the optimal trade-off between these two aspects.

\begin{biblio}
\begin{itemize}
\item For a survey of SIR-like models ``pre-COVID-19'' for diseases such as Zika, Dengue, Ebola, H1N1, see e.g.
  \cite{Capaldi,Capistran,Tuncer,Roosa.Chowell,Chowell,Eisenberg,Roberts,Tonsing}.
\item For some examples of SIR-like models for COVID-19, see e.g. \cite{Giordano,Russo,Peng,Wang,DellaRossa,Gatto,Crisanti}.
\item A somewhat different approach is proposed in \cite{Pugliese}, where the underlying model is a simple SIR,
  with a more complex model for the probability distribution of the delays between infection and the observed events
  (hospitalization, recovery, death).
\item Control strategies for SIR-like systems are also an important topic, see e.g. \cite{Zanella}.
\end{itemize}  
\end{biblio}


\section{Forward Uncertainty Quantification (UQ): tools to make predictions under uncertainty}\label{section:prior_forward}

In general, SIR-like models can be written as ODE systems for a state vector $X$ with $N_{states}$ components.
The evolution of the system depends on $N_{coef}$ coefficients $\pp = [p_1, \ldots, p_{N_{coef}}]$ and on the $N_{states}$ initial conditions $\qq = [q_1, \ldots,q_{N_{states}}]$.
Moreover, we might be interested in monitoring not only the states of the system but also some
related quantities $Y$ (\emph{Quantities of Interest}, say we have $N_{qoi}$ of them), which can be derived from $X$ by an observation operator $\qoi$,
that in turn might depend on $N_{hyp}$ hyper-parameters $\hh = [h_1, \ldots, h_{N_{hyp}}] $:
\begin{equation} \label{eq:ode_system}
  \begin{cases}
    \dot{X} = f(X,\pp) \\
    X(t_0) = \qq\\
    Y(t) = \qoi(X(t),\hh),
  \end{cases}
\end{equation}
where $\forall t \in [0,T]$ we have $X \in \Rset^{N_{states}}$,  $Y \in \Rset^{N_{qoi}}$, and 
$f(\cdot,\pp):\Rset^{N_{states}} \rightarrow \Rset^{N_{states}}$,
$\qoi(\cdot,\hh):\Rset^{N_{states}} \rightarrow \Rset^{N_{qoi}}$, $\pp \in \Rset^{N_{coef}}$,
$\qq \in \Rset^{N_{states}}$, $\hh \in \Rset^{N_{hyp}}$.

We collect coefficients and initial conditions in a vector $\pparam = [\pp,\qq]$ with $N_{\pparam} = N_{coef} +N_{states}$ components.
Throughout the manuscript, we refer to $\pparam$ as \emph{parameters}, and we will write $X(\pparam), Y(\pparam,\hh)$ to emphasize the
dependence of the states and quantities of interest on parameters and hyper-parameters.
For SIR, $\pp=[\beta,r]$, and $Y$ might be for instance:
\begin{itemize}
  \item \lorenzo{the \emph{prevalence data}, i.e., the number of infected individuals at a specific time}, $\qoi(X(t)) = I(t)$;
  \item \lorenzo{the \emph{incidence data}, i.e., the number of new cases over the reference time period, $\qoi(X(t)) = \frac{\beta}{N_{pop}} I(t) S(t)$;}
  \item the peak-time of number of infected persons: $\qoi(X) = \argmax_{t \in [0,T]}I(t)$;
  \item the peak-time of the \lorenzo{cumulative incidence data i.e., the} new infected persons in a time-window of length $\Delta$:
    $\qoi(X,\Delta) = \argmax_{t} \int_{t}^{t + \Delta} \frac{\beta}{N_{pop}} S(s)I(s)ds$.
\end{itemize}
Another important scenario is under-reporting, where we assume that due to insufficient measurements,
we observe only a fraction $K$ of the total number of infected, $\qoi(X(t),K) = \frac{1}{K} I(t)$
($K$ being possibly unknown);
\lorenzo{similar under-reporting scenarios could be of course conceived also for other quantities.
  Observe that at this point of the manuscript we are not concerned whether these quantities are reasonably easy to measure and obtain:
  for instance, for most infections data available are typically incidence rather than prevalence. At this level, we are just giving examples
  of the mathematical setup of the problem, in an ideal scenario where we have access to all sort of measurements.}

Typically, most of the parameters (and possibly the hyper-parameters as well) are not known exactly and they are either taken
from literature or calibrated from data.
We can then assume that these parameters are random variables with a certain probability density function (pdf):
for instance, uniform random variables over a variability range, or Gaussian random variables centered around a most likely value.%
\footnote{Technically, most parameters of compartmental models must be positive, so a Gaussian random variable is not suitable and
  one should consider other random variables, e.g. Beta or log-normal. However, here we are keeping things on a simple/introductory level,
  and in any case a Gaussian random variable can be truncated to ensure positivity if needed.}
Then, a natural question is: how does the variability of the parameters impact the quantities of interest $Y$ of the SIR-like model at hand?
Or otherwise, what is the variability range of $Y$ as the parameters range over their values?

This kind of analysis is known as \emph{forward UQ} in computational science and engineering.
The most straightforward way to accomplish this task is by sampling methods, i.e., by generating $M$ samples
of the parameters $\pparam_1,\pparam_2,\ldots \pparam_M $ according to their probability distribution,
solving the SIR-like system for each $\pparam_i$, and estimating statistics such as mean, standard deviation, confidence bands,
and the probability density function of $Y$ from the corresponding values $Y(\pparam_1,\hh), Y(\pparam_2,\hh), \ldots Y(\pparam_M,\hh)$.
The easiest sampling scheme is Monte Carlo, but more advanced sampling techniques can be used (Latin Hypercube Sampling,
Stratified Sampling, Quasi-Monte-Carlo, Sparse grids, among others; see the bibliography at the end of this section).
Within sampling schemes, quantities such as mean, standard deviations and higher moments
with respect to the parameters can be computed by averaging over the $M$ samples of the model results:
\begin{equation}\label{eq:stoc.coll}
  \mathbb{E}_{\pparam}[Y(\pparam,\hh,t)] \approx \sum_{i=1}^M \omega_iY(\pparam_i,\hh,t),
\end{equation}
where the parameter samples $\pparam_i$ and the weights $\omega_i$ depend on the specific sampling method used.
For instance, Monte Carlo employs random $\pparam_i$ and $\omega_i = \frac{1}{M}$.
The probability density function of the quantity of interest can be approximated by, for example, histograms
or kernel density estimates, see e.g. \cite{rosenblatt:kde,parzen:kde}.


\begin{example}[Forward UQ of a SIR model]\label{ex:UQ_for_SIR}
Consider a SIR model with initial conditions $S(0)=0.95, I(0)=0.05, R(0)=0$.
The survey on the literature performed by \cite{DellaRossa} suggests these ranges for the parameters:
$\beta \in [0.25,0.35]$, $r \in [0.06,0.18]$. We assume that a-priori we have no knowledge that
any value of $\beta,r$ is more plausible than others; therefore, we assume that $\beta,r$
are uniform independent random variables.
We solve the SIR system with Matlab's \texttt{ode45} up to final time $T=150$.

Figure \ref{fig:SIR_UQ}-left shows the SIR dynamics obtained by 100 Monte Carlo samples.
The black lines are the average trajectories of SIR obtained by sampling values of $\beta$ and $r$
(65 samples with sparse grids sampling). The remaining panels show pdfs of quantities of interest of SIR:
SIR states at $T=30$ (after the average peak position) and $T=100$ (when the dynamics is over),
again computed both with sparse grids (solid line) and Monte Carlo (circle markers);
\emph{peak time} and \emph{peak intensity} (we only show the pdf obtained by sparse grids). 
Figure \ref{fig:SIR_resp} shows the so-called \emph{response surface}, i.e., a plot showing how the quantity of interest
changes as $\beta$ and $r$ vary in their range (we report only those obtained by sparse grids).
Response surfaces are useful to derive information on the general trends of the system, and to quickly
approximate the value of a quantity of interest without evaluating the full model,
i.e., they act as a surrogate model for the dynamical system; the pdfs of the quantities of interest obtained
by sparse grids have actually been obtained by querying these response surfaces rather than the full model.

\begin{figure}[tbp]
  \centering
  \includegraphics[width=0.19\linewidth]{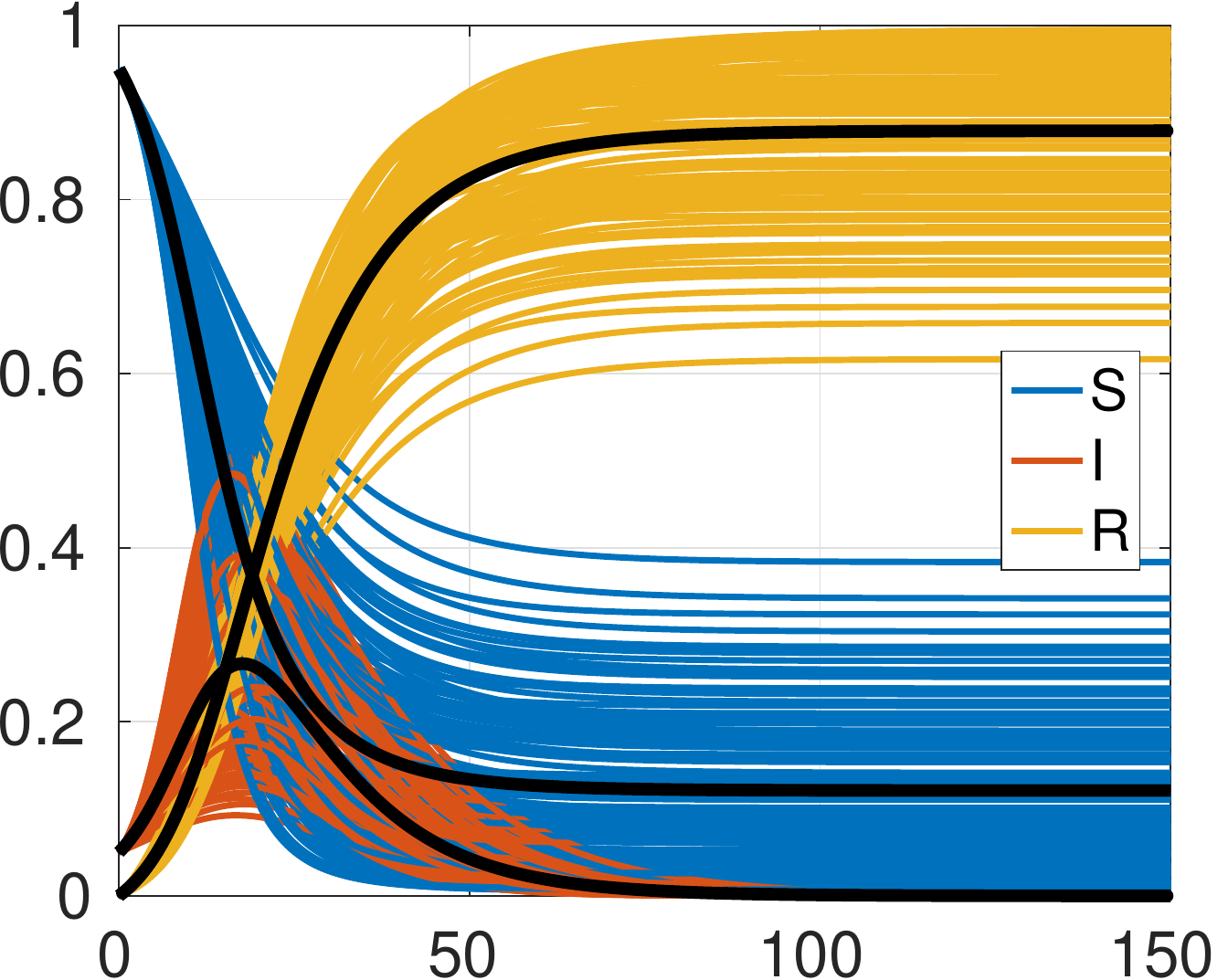}
  \includegraphics[width=0.19\linewidth]{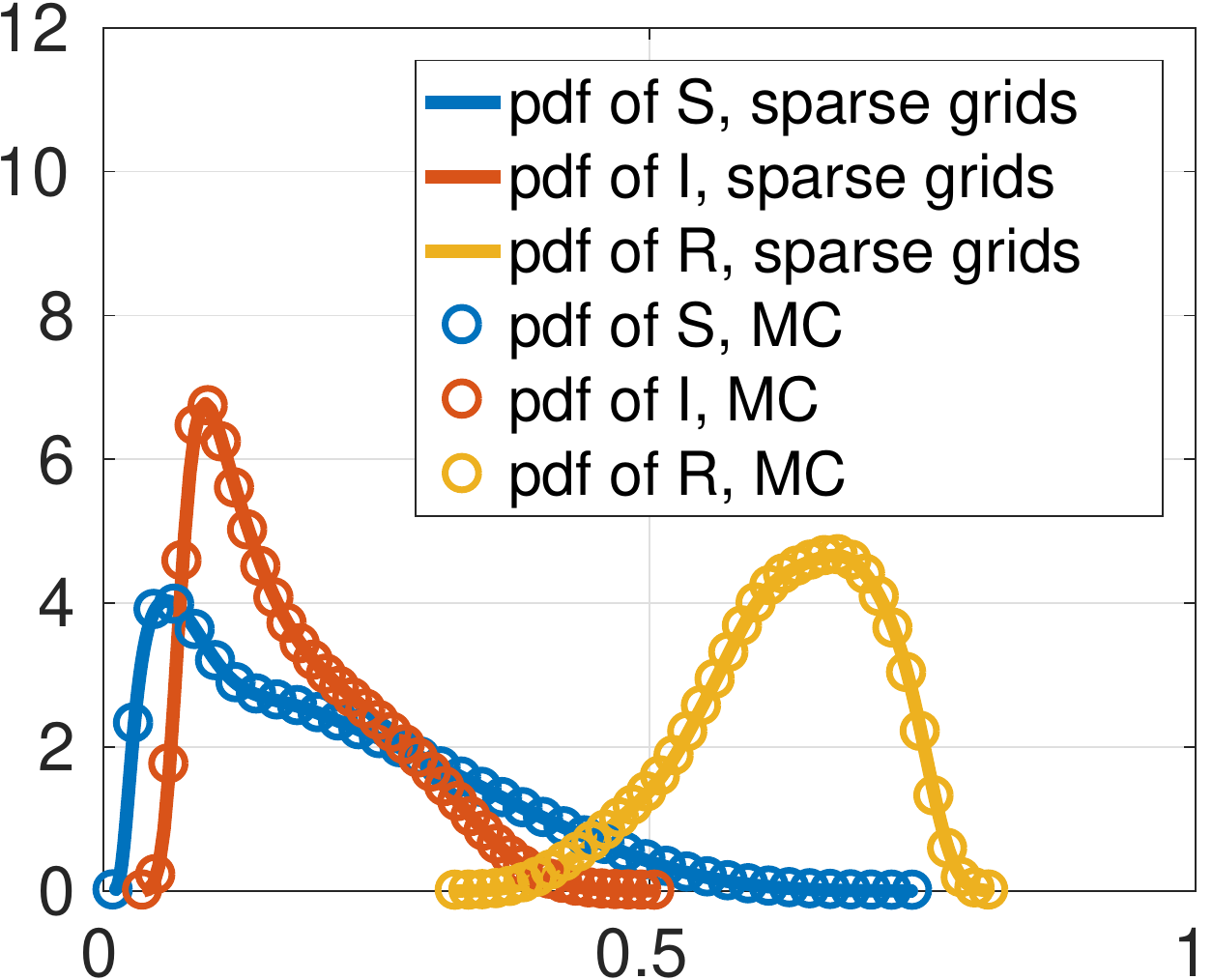}
  \includegraphics[width=0.19\linewidth]{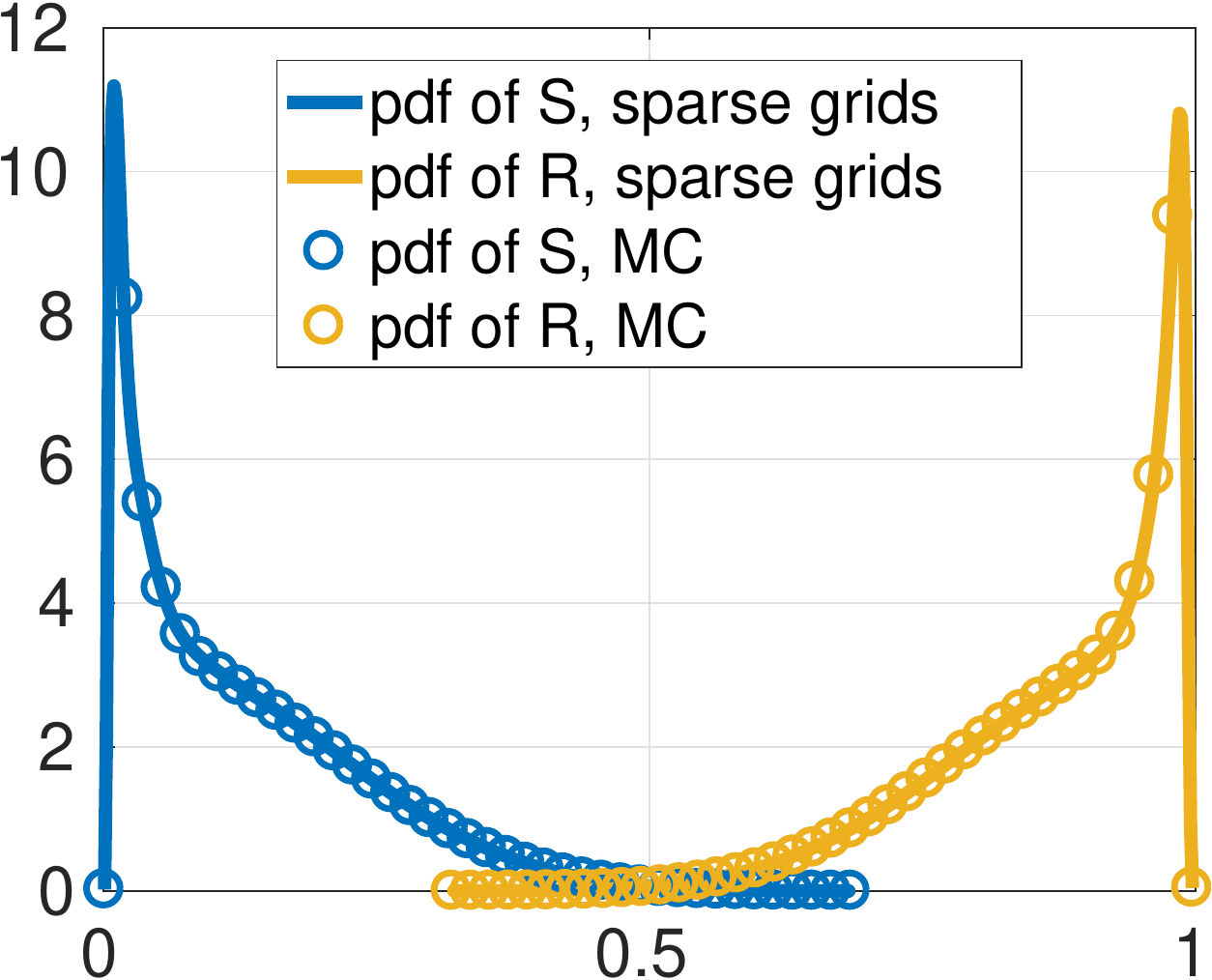}
  \includegraphics[width=0.20\linewidth]{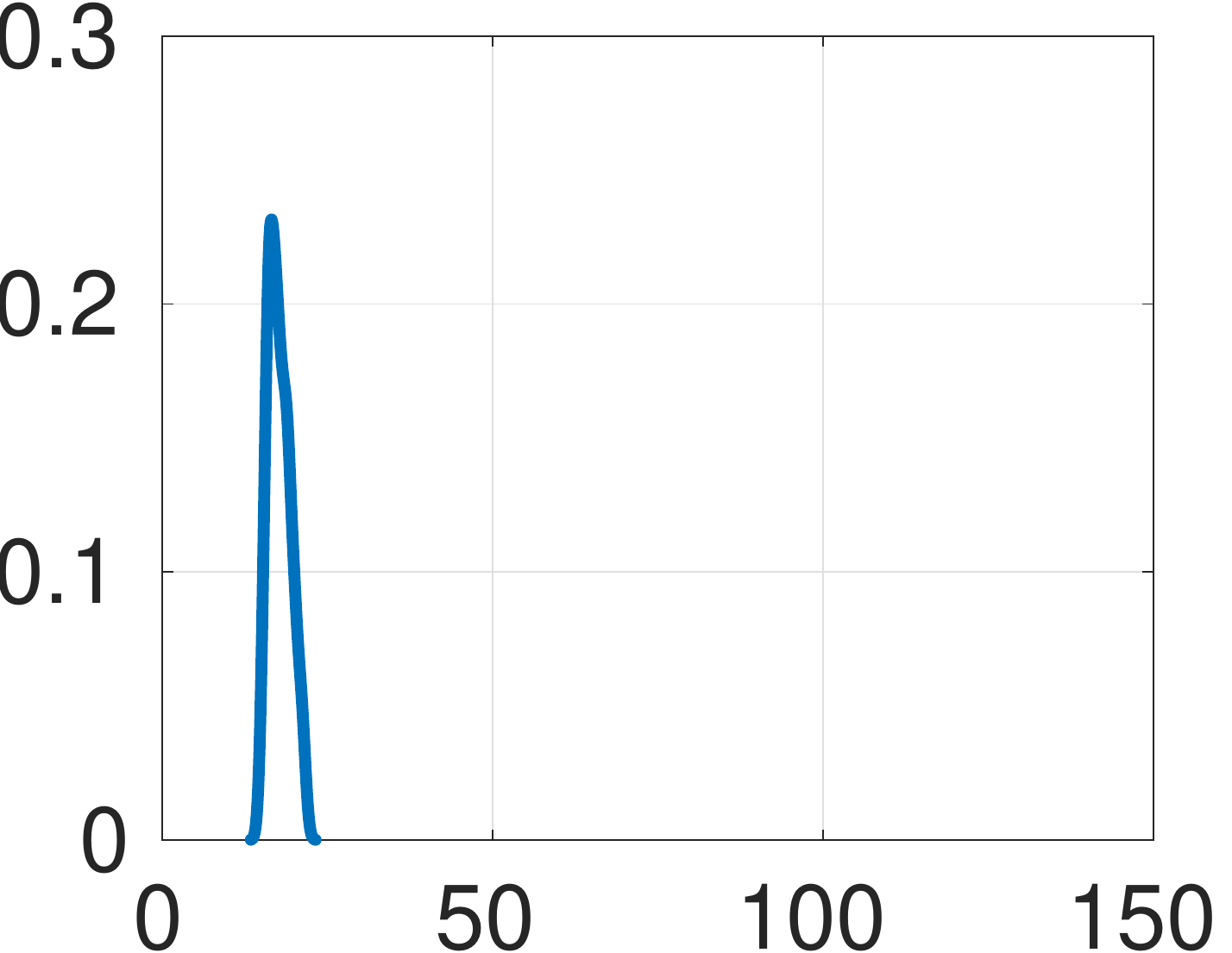}
  \includegraphics[width=0.175\linewidth]{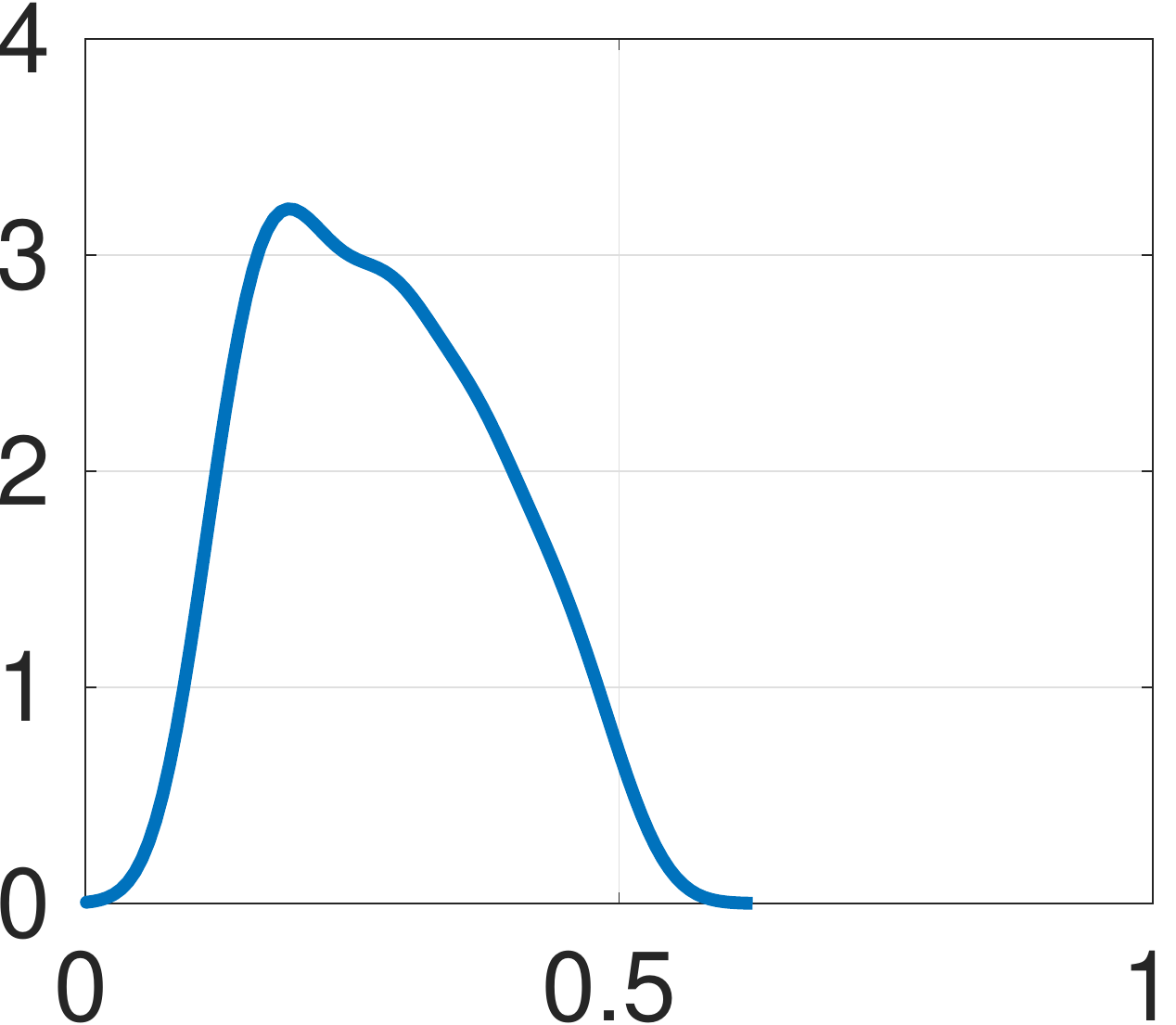}
  \caption{Left: SIR dynamics. The colored lines represent the SIR dynamics obtained by Monte Carlo samples of $\beta,r$, and 
    the thick black lines represent the average computed by sparse grids. The other panels represent pdfs of quantities
    of interest: SIR states at $T=30,100$ (by Monte Carlo and sparse grids), peak-time for $I$ and peak-value for $I$
    (sparse grids only).}
  \label{fig:SIR_UQ}
\end{figure}

\begin{figure}[tbp]
  \centering
  \includegraphics[width=0.19\linewidth]{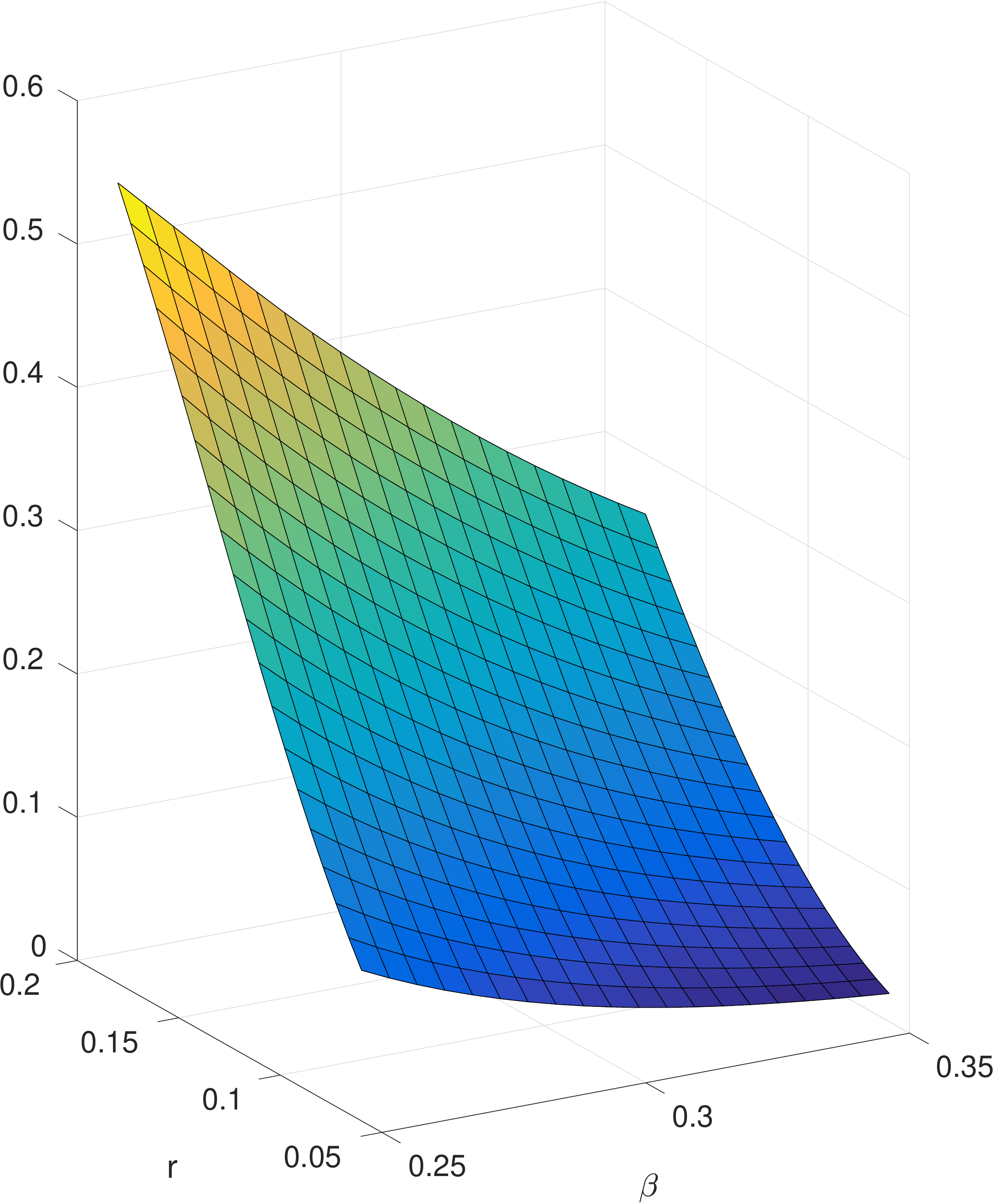}
  \includegraphics[width=0.19\linewidth]{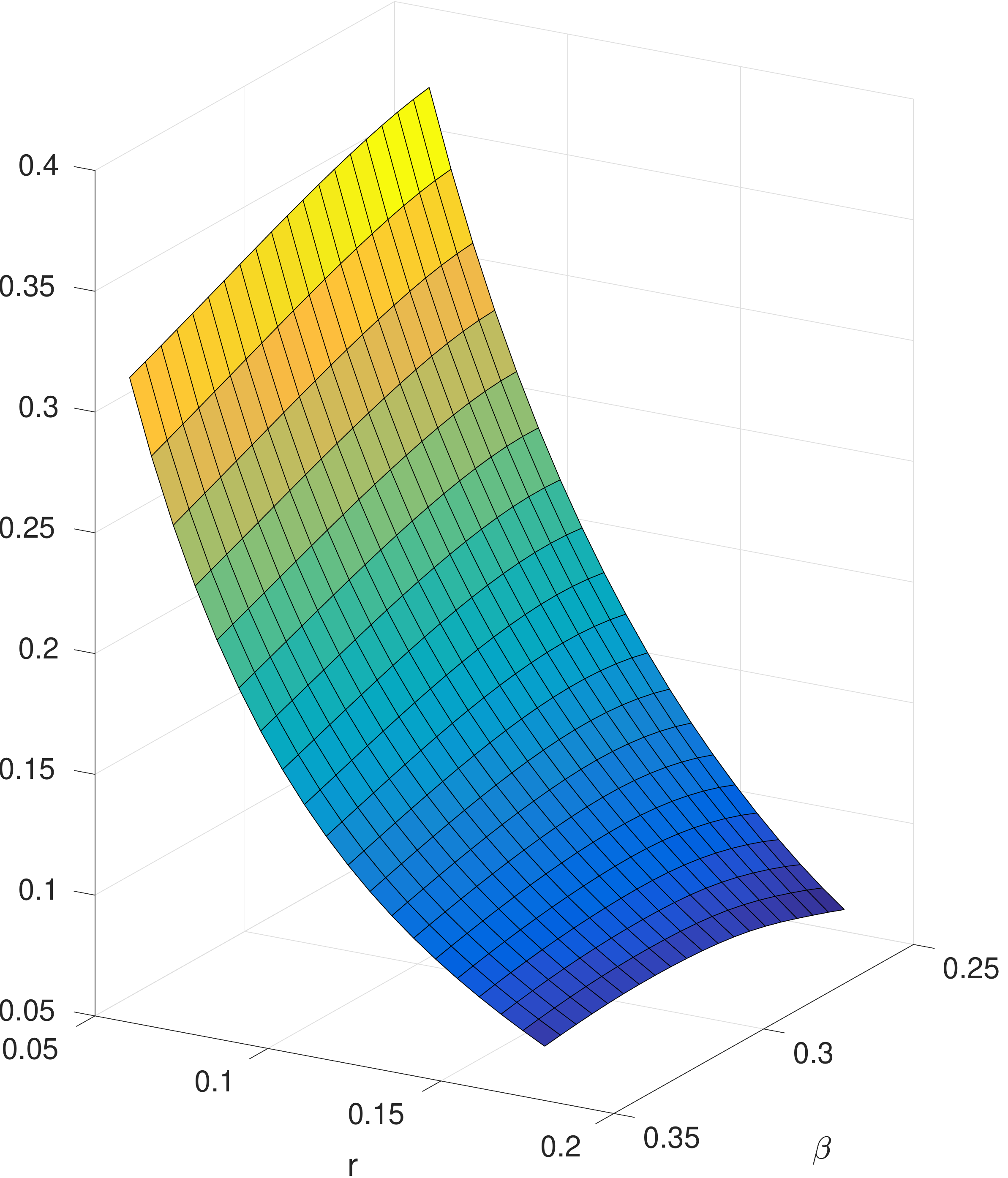}
  \includegraphics[width=0.19\linewidth]{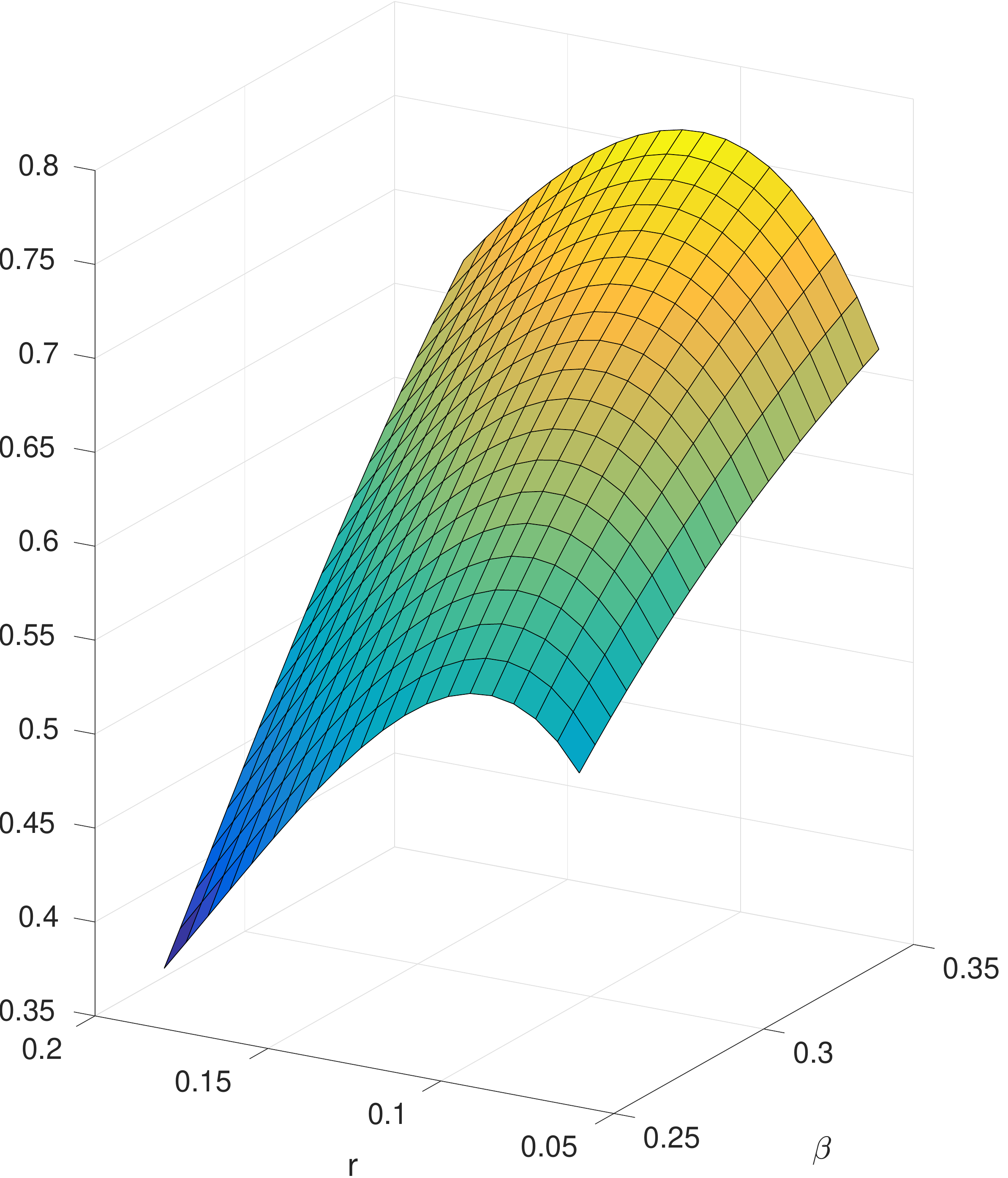}
  \includegraphics[width=0.19\linewidth]{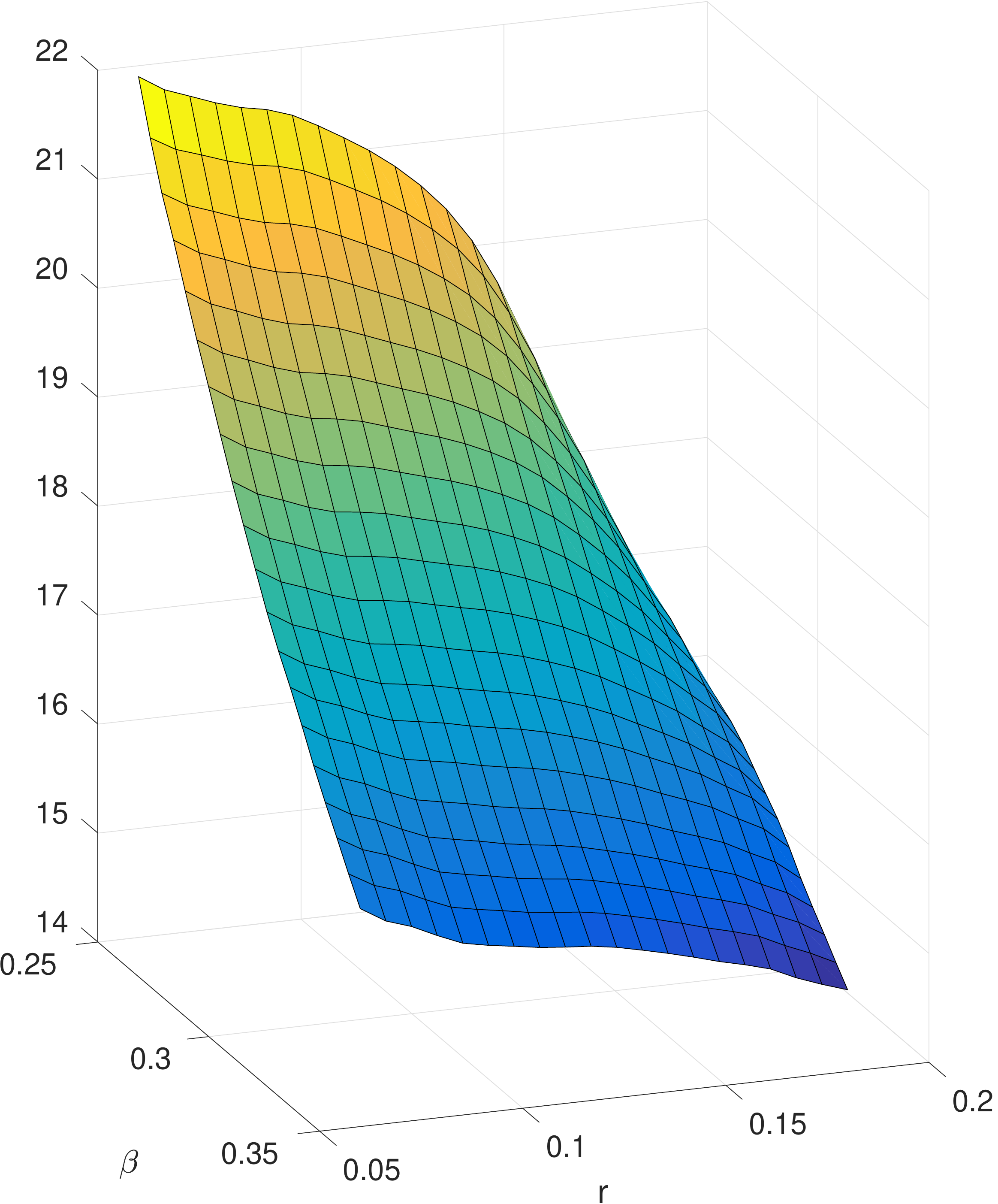}
  \includegraphics[width=0.19\linewidth]{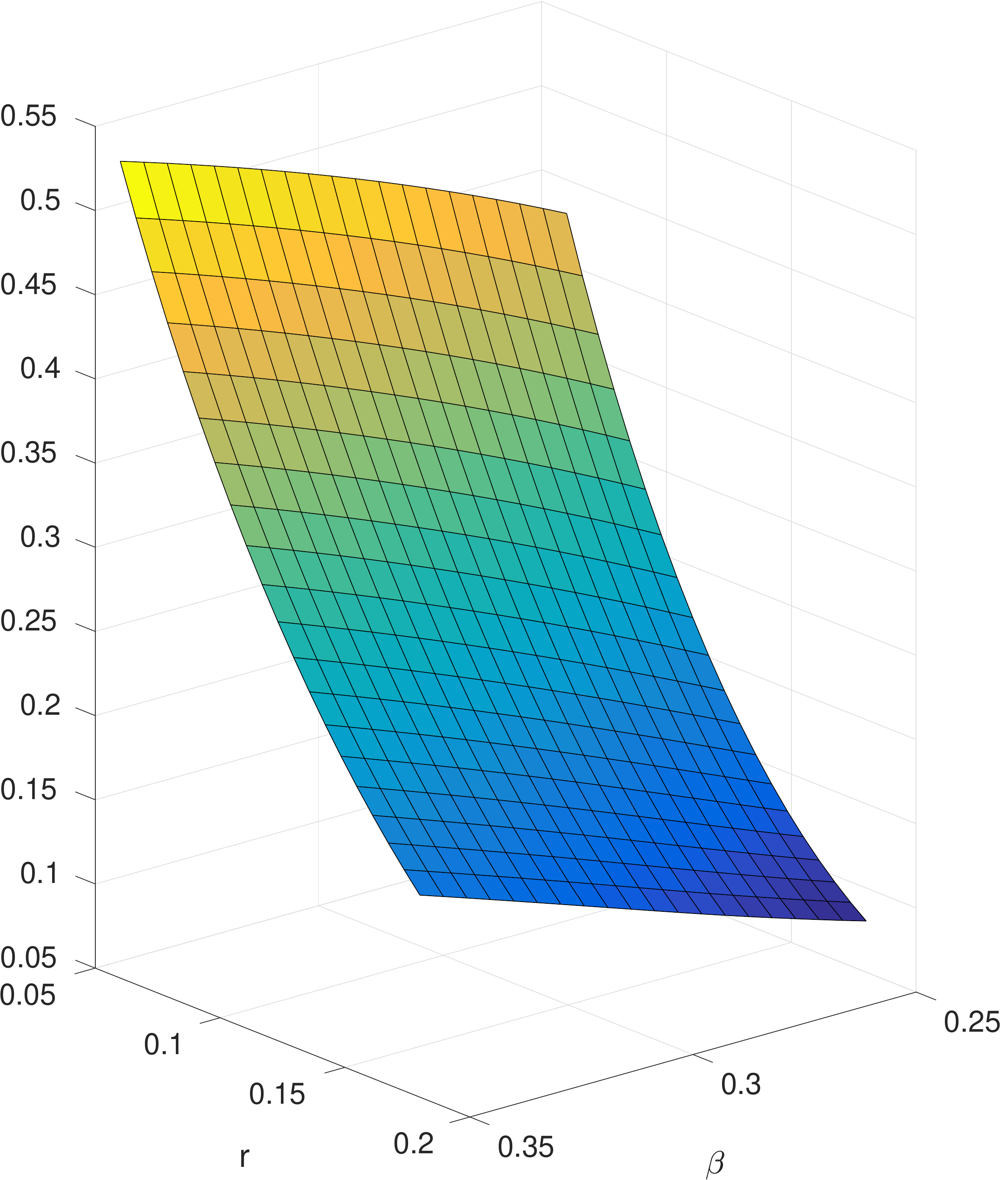}
  \caption{Sparse grids response surfaces for the SIR quantities of interest.
    From left to right: S at $T=30$, I at $T=30$, R at $T=30$, peak time, and peak value.}\label{fig:SIR_resp}
\end{figure}
\end{example}

\begin{biblio}
\begin{itemize}
\item For random sampling methods (Monte Carlo, Latin Hypercube Sampling, Stratified Sampling), see \cite{cafl:mcreview,nied:qmcbook,McKay:sampling}.
  Random sampling methods are robust and easy to implement, but have a poor accuracy (typically proportional to $M^{-1/2}$, with $M$ denoting the number of samples).
\item For sparse grids sampling methods, see \cite{babuska.nobile.eal:stochastic2,xiu.hesthaven:high,cohen_devore_2015,gunzburger_webster_zhang_2014}.
  These are deterministic (i.e., non-random) sampling schemes that generalize tensor (cartesian) grid sampling when the parameter space
  is high-dimensional, in which case a cartesian grid sampling scheme would be too expensive.
  They are less straightforward than random sampling methods, but guarantee greater accuracy, at least for
  problems up to a few tens of parameters. These tools have been developed in the context of UQ
  for models that are expensive to evaluate, whereas evaluating a SIR-like model is typically
  very fast. Therefore, their use is not as crucial in the context of COVID-19,
  and random sampling methods might be favored for their straightforwardness.
  Sparse grids have still an advantage over random sampling for sensitivity analysis, see the next section.
\item A somewhat intermediate possibility are Quasi Monte Carlo sampling methods, such as Sobol or Halton sequences.
  These are also deterministic sampling schemes as well, that aim at covering the space of parameters in the ``most uniform way''
  (space filling) \cite{cafl:mcreview,nied:qmcbook,sloan.woz:sobol}. They typically have accuracy proportional to $M^{-1}$, with $M$ denoting the number of samples.
  
\end{itemize}
\end{biblio}

\section{Sensitivity analysis: pinpointing what parameters we need to get right, and preliminary assessment
of feasibility of inversion} \label{sect:sobol}

Sensitivity analysis aims at assessing which parameters have the largest impact on the quantities of interest.
This information is crucial to determine which parameters should be subjected to further investigations to reduce their variability.
The sensitivity analysis can be \emph{local} or \emph{global}:
\begin{itemize}
  \item local sensitivity analysis is usually based on the derivatives of the quantities of interest with respect to $\pparam$,
    upon fixing each $\param_i$ at some representative value (average, median, mode).%
    \footnote{Of course here we are assuming that the quantity of interest depends smoothly on $\pparam$, so that
      it is possible to compute the derivatives. This is not always obvious and should be checked.}
  \item global sensitivity analysis considers the total variability of a quantity of interest
    and decomposes such variability into elementary components, each due to $\param_i$ individually
    or to mixed effects such as $\param_i\param_j$, $\param_i\param_j\param_k,\ldots$: 
    the larger the component, the more sensitive the quantity of interest to $\param_i$ is. 
\end{itemize}

In these short notes we focus on the Sobol indices for global sensitivity analysis, which are variance-based indices,
i.e. the variability of the quantity of interest is measured as its variance \cite{sobol93}.
The total variance is decomposed as follows: one term due to each $\param_i$; one term due to
each mixed effect composed of two parameters, $\param_i \param_j$; one term for each mixed effect composed of three parameters  $\param_i\param_j\param_k$, and so forth.
The sum is then normalized to one, and each quantity thus obtained is called Sobol index:
\begin{equation}
  \label{eq:sobol.dec}
  1 = \sum_{i=1}^{N_{\pparam}} s_i + \sum_{i,j=1, i \neq j}^{N_{\pparam}} s_{ij} + \sum_{i,j,k=1, i \neq j \neq k}^{N_{\pparam}} s_{ijk} + \ldots 
\end{equation}
The Sobol index of each parameter $\param_i$ per se, $s_i$, is usually reported as an indicator of the importance of each parameter,
and is called the \emph{principal Sobol index}. Another relevant quantity is the \emph{total Sobol index} of a parameter, $s_{i}^T$,
which is obtained by adding to the principal Sobol index of $\param_i$ all the Sobol indices of mixed effects of which $\param_i$ is part,
e.g. for $\param_1$:
\[
s_1^T = s_1 + \sum_{j=1,j\neq 1}^{N_{\pparam}} s_{1j}  + \sum_{j,k=1,j\neq 1,k \neq 1}^{N_{\pparam}} s_{1jk} + \ldots
\]
Note that in general $\sum_{i=1}^{N_{\pparam}} s_i < 1$ and $\sum_{i=1}^{N_{\pparam}} s^T_i > 1$. This approach bears many similarities
with the ANOVA decomposition in statistics.
An important observation is that, with an eye to parameter identification, we can expect that if the Sobol index of a parameter is small,
it will be hard to recover its value from measurements of the quantity of interest. 
We also remark that as a general rule of thumb, the larger the range of values of a parameter, the larger its corresponding Sobol indices.

\begin{example}[Computing the Sobol indices for a SIR model]\label{ex:Sobol_for_SIR}
Consider again the SIR example of the previous section. The Sobol decomposition of any quantity of interest $Y$ reads
$
  1 = s_\beta + s_r + s_{\beta r}
$
and the total Sobol indices can be computed as
$
  s_\beta^T = s_{\beta} + s_{\beta r}, \   s_r^T = s_{r} + s_{\beta r}.
$
Figure \ref{fig:sobolSIR} shows the time-evolution of the Sobol indices (principal and total) for the SIR states:
the principal indices are represented by the solid line, while the total indices are represented by the dashed line.
The principal and total indices behave very similarly, indicating that the interaction between the two parameters is quite limited.
Note that the Sobol indices are not constant in time and behave differently for the different compartments.
More specifically, the asymptotic regime is mostly dictated by $r$ for all the compartments,
while $\beta$ impacts more in the transient regime, especially in the case of the compartment $R$.
This has an impact on the inversion procedure. In particular,
severe difficulties in the estimation of $\beta$ can be encountered if the data of $R$ are missing or too noisy.
Moreover, note that the influence of $r$ is larger, in general.
Further evidence of this is shown in the right-most panel, where we show the variability of the trajectories if the range of $r$ is reduced
to $[0.06,0.1]$. The overall variability is greatly reduced, as expected. 
\begin{figure}[tbp]
  \centering
  \includegraphics[width=0.72\linewidth]{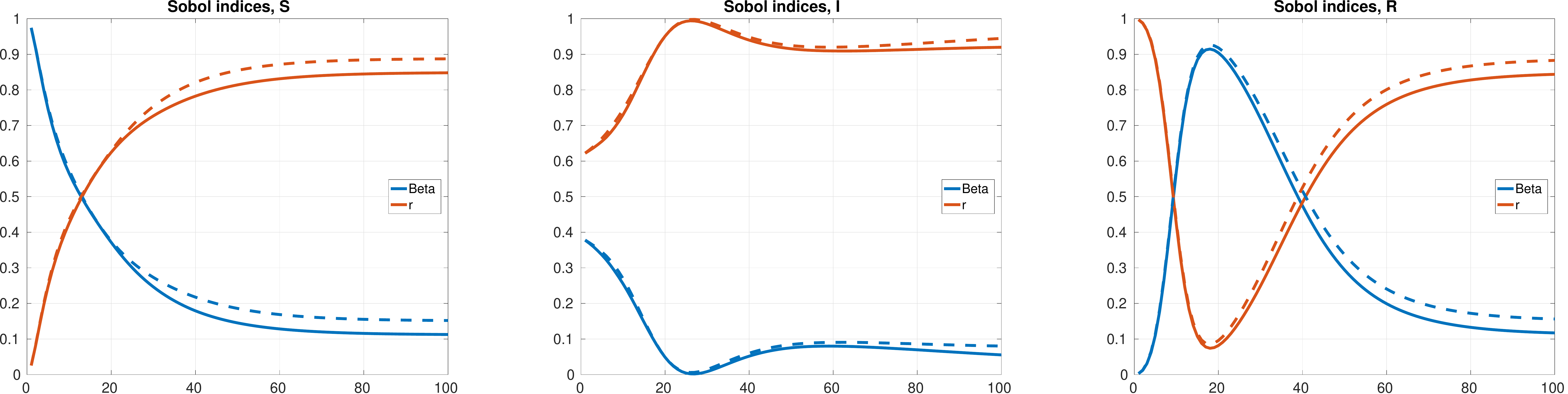}
  \includegraphics[width=0.27\linewidth]{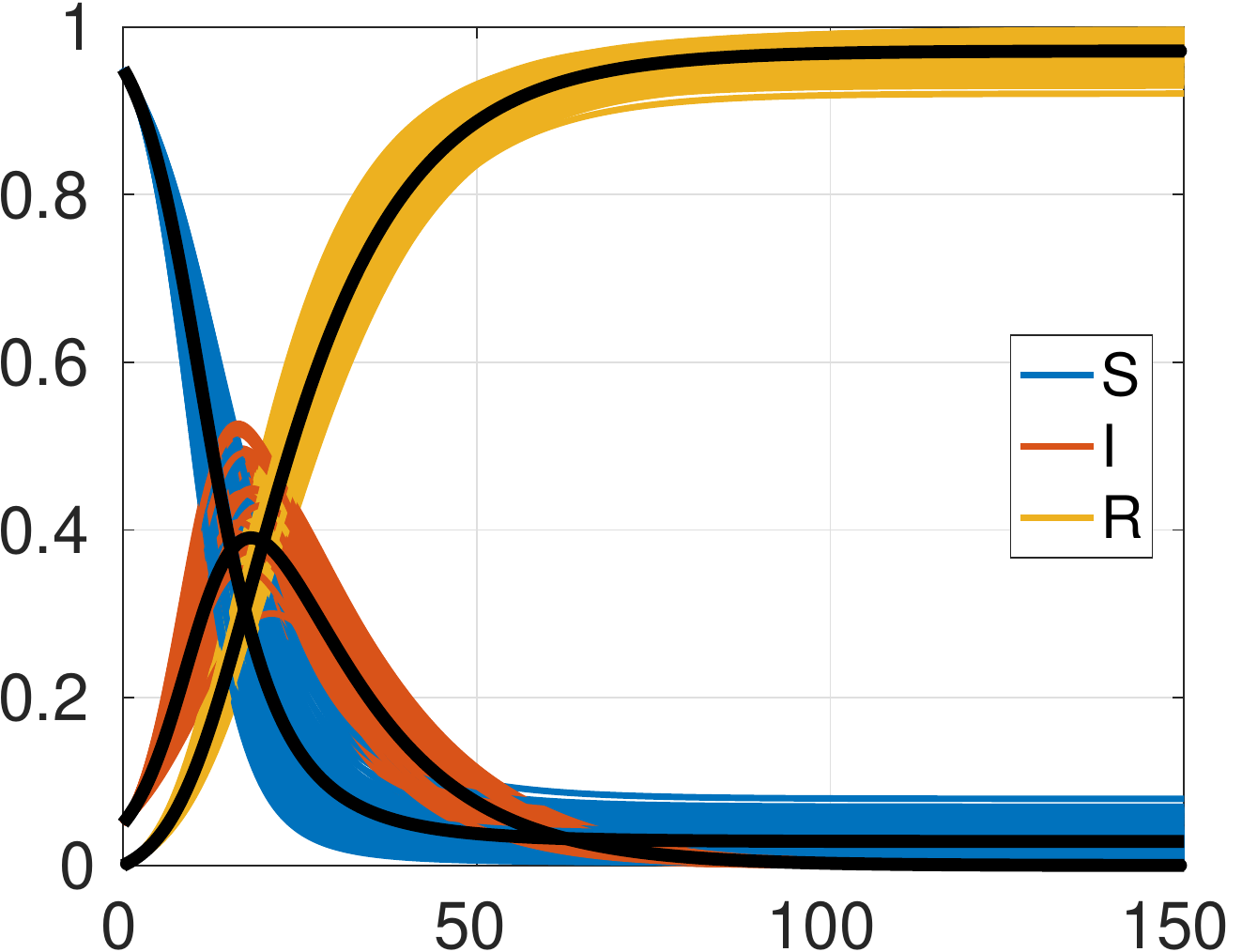}
  \caption{Time-evolution of Sobol indices for the SIR compartments,
    and SIR trajectories obtained by reducing the range of variability of $r$.
    The principal indices are represented by the solid line, while the total indices are represented by the dashed line.}
  \label{fig:sobolSIR}
\end{figure}
\end{example}

\begin{biblio}
\begin{itemize}
	\item Classical books on sensitivity are e.g. \cite{cacuci2003sensitivity,saltelli:book}.
	\item Sobol indices can be computed either by Monte Carlo sampling, see e.g. \cite{saltelli:book},
	or perhaps more conveniently by polynomial expansions or sparse grids sampling, see e.g. \cite{feal:compgeo}.
	\item Sobol indices analyses for SIR-like problems can be found in \cite{Borgonovo,Capaldi}.
	\item An alternative to Sobol indices are the Morris indices \cite{morris:sensitivity}. 
	\item Sobol indices can also be computed with respect to measures of variability other than variance,  
	see e.g. \cite{delloca:sobol}.
\end{itemize}
\end{biblio}

\section{Inverse UQ (data fitting) as a preliminary step to tune the pdf of the parameters to the data}\label{section:inverse}

The sections above discussed some general elementary tools to perform predictions under uncertainty once the pdfs for the parameters
have been chosen. This section discusses the preliminary step to the UQ process,
i.e. how to construct pdfs for the parameters, and in particular how to do so by merging
prior information on the parameters and the available data. Upon deriving these data-informed pdfs,
we will use them to carry out the UQ analysis. In literature, data-informed pdfs are often
called \emph{posterior pdfs}, $\rho_{post}$, as opposed to \emph{prior pdfs}, $\rho_{prior}$, before the data are available.

Computing the posterior pdfs of the parameters can be done by means of the Bayes theorem on conditional probabilities.
This procedure is quite general and includes as a special case the least-squares approach for data fitting
(this connection will be made clearer later).
For ease of exposition, we exemplify the procedure over a specific example - extension to other problems is
relatively straightforward (see e.g. Examples \ref{ex:non-uniform-prior}, \ref{ex:two-sigma} later on).
For now we assume that: 
\begin{itemize}
\item We have at our disposal $N_{meas}$ measurements of the $I$ state and $N_{meas}$
  measurements of the $R$ state, at equispaced times $t_i = i\Delta t$, $i=1,2,3,\ldots N_{meas}$.
  In total we have $2N_{meas}$ data, $\mathcal{D} = \{\hat{I}_1,\hat{I}_2,\ldots,\hat{R}_1,\hat{R}_2,\ldots \}$.
  \lorenzo{As stated previously, please note that we are not hinting in any way that having measurements of both $I(t)$
    (i.e., prevalence data) and $R(t)$ is what happens in real scenarios: we make this (quite restrictive in practice) assumption
    to put ourselves in the easiest scenario to illustrate the mathematical procedure. 
    Using more realistic data would add mathematical technicalities without giving any further insight;}
\item These data correspond to some values $\pparam_{true}$ of coefficients and initial conditions of the system (\ref{eq:ode_system});
\item The prior pdfs for $\pparam$ are uniform (see Example \ref{ex:non-uniform-prior} for an example with Gaussian priors);
\item Our measurements are under-reported by a factor $K$, i.e., we are able to measure only a fraction
  of the actual compartments;
\item Data are noisy, i.e. affected by some random errors $\epsilon_{I,i}$, $\epsilon_{R,i}$,
  that are modeled by independent random variables with zero mean and standard deviation $\sigma$;
\item the standard deviation $\sigma$ is identical for $I$ and $R$
  (see Example \ref{ex:two-sigma} for the generalization to the case where the two compartments have different $\sigma$);
\item $\epsilon_{I,i}$, $\epsilon_{R,i}$ are Gaussian random variables $\mathcal{N}(0,\sigma^2)$. See discussion at the
  end of the section for bibliography on more general models;
\item $K,\sigma$ are hyper-parameters constant in time. We assume for the moment that $K$ is known and $\sigma$ is unknown
  (see Examples \ref{ex:structural_identifiability_SIR}, \ref{ex:structural_identifiability_SIR_evans},  \ref{ex:SIR_identifiability}
  for a discussion on how to determine $K$ in case it is assumed unknown as well).
\end{itemize}
In formulas our data model is the following: 
\begin{equation}\label{eq:data-model}
  \begin{cases}
    \hat{I}_i = \qoi(I(\pparam_{true},t_i),K) + \epsilon_{I,i} = \frac{1}{K}I(\pparam_{true},t_i) + \epsilon_{I,i}, \quad i=1,2,\ldots N_{meas} \\[6pt] 
    \hat{R}_i = \qoi(R(\pparam_{true},t_i),K) + \epsilon_{R,i} = \frac{1}{K}R(\pparam_{true},t_i) + \epsilon_{R,i},  \quad i=1,2,\ldots N_{meas}.
  \end{cases}
\end{equation}
We also introduce the $2N_{meas}$ misfits $\mathcal{M} = \{M_{I,1},M_{I,2},\ldots,M_{R,1},M_{R,2},\ldots\}$
between the data and the model predictions, obtained upon fixing the parameters at some estimate
$\pparam_{guess}$ of $\pparam_{true}$: 
\begin{equation}\label{eq:misfits}
  \begin{cases}
    M_{I,i}(\pparam_{guess}) = \hat{I}_i -\frac{1}{K}I(\pparam_{guess},t_i), \quad i=1,2,\ldots N_{meas}  \\
    M_{R,i}(\pparam_{guess}) = \hat{R}_i -\frac{1}{K}R(\pparam_{guess},t_i), \quad i=1,2,\ldots N_{meas}.  
  \end{cases}
\end{equation}

\subsection{Bayes Theorem and posterior distributions} \label{sect:likelihood}

The Bayes theorem provides us with a practical formula to compute the posterior pdf of the parameters $\pparam$,
i.e., with a means of adjusting the prior pdf to the data at hand. An informal writing of the Bayes formula is
\begin{equation}
  \label{eq:posterior-pdf-informal}
  \textrm{pdf($\pparam$ given $\mathcal{M}$)} = \textrm{pdf($\mathcal{M}$ given $\pparam$)} \times \textrm{pdf($\pparam$)}  \times \frac{1}{\textrm{ pdf($\mathcal{M}$)}}
\end{equation}
where ``pdf($\pparam$ given $\mathcal{M}$)'' is the pdf of the parameters when given the misfits, hence given the data
(i.e., the posterior pdf that we aim at computing), while ``pdf($\pparam$)'' is the pdf of the parameters based only on a-priori information.
The ``pdf($\mathcal{M}$)'' can be simply considered to be the normalization constant
such that the posterior pdf is actually a pdf (i.e., its integral is equal to 1).
Therefore, to make the computation of the posterior pdf practical we only need to know the expression of
the ``pdf($\mathcal{M}$ given $\pparam$)'', which is the so-called \emph{likelihood function}; we will denote this
quantity as $\mathcal{L}(\pparam)$.

Deriving an expression for $\mathcal{L}(\pparam)$ is quite straightforward. If $\pparam_{guess}$ were the true values, then the probability that the misfits $\mathcal{M}$ have certain values is the probability
that the measurement errors $\epsilon_{I,i}, \epsilon_{R,i}$ have those values (cf. Equations (\ref{eq:data-model}) and (\ref{eq:misfits})).
By assumption, we know that $\epsilon_{I,i}, \epsilon_{R,i}$ are independent Gaussian random variables
with zero mean and standard deviation $\sigma$, therefore,
\begin{equation}
  \label{eq:likelihood}
  \mathcal{L}(\pparam) =
    \prod_{i=1}^{N_{meas}} \frac{1}{\sqrt{2\pi\sigma^2}}e^{\frac{-1}{2\sigma^2}(\frac{1}{K}I(\pparam,t_i) - \hat{I}_i)^2}
    \prod_{i=1}^{N_{meas}} \frac{1}{\sqrt{2\pi\sigma^2}}e^{\frac{-1}{2\sigma^2}(\frac{1}{K}R(\pparam,t_i) - \hat{R}_i)^2}
\end{equation}
so that the posterior pdf of the parameters reads
\begin{equation}\label{eq:posterior}
  \rho_{post}(\pparam | \mathcal{D}) \propto
  \mathcal{L}(\pparam) \rho_{prior} (\pparam) =
  \prod_{i=1}^{N_{meas}} \frac{1}{\sqrt{2\pi\sigma^2}}e^{\frac{-1}{2\sigma^2}(\frac{1}{K}I(\pparam,t_i) - \hat{I}_i)^2}
  \prod_{i=1}^{N_{meas}} \frac{1}{\sqrt{2\pi\sigma^2}}e^{\frac{-1}{2\sigma^2}(\frac{1}{K}R(\pparam,t_i) - \hat{R}_i)^2}
  \rho_{prior} (\pparam)
\end{equation}
where the $\propto$ symbol is used to signify that we have omitted the normalization constant.

\subsection{Computational challenges of working with the posterior pdf}

Equipped with \eqref{eq:posterior}, we would then only need to proceed as in Section \ref{section:prior_forward}
and perform the UQ analysis.
Although conceptually straightforward, this approach can be practically challenging, because it is not
easy to obtain samples of the random parameters $\pparam$ distributed according to the posterior pdf (\ref{eq:posterior}).
The classical computational tool to this end is the so-called Markov-Chain Monte Carlo - MCMC \cite{stuart:acta.bayesian},
which generates a sequence of proposed values of $\pparam$ that are \emph{asymptotically}
distributed according to $\rho_{post}$.
A nice feature of MCMC algorithms is that they do not require knowledge of the normalization constant.
The use of MCMC for forward UQ has however some drawbacks:
\begin{enumerate}
\item the likelihood function has to be evaluated at every proposed $\pparam$, which requires evaluating the SIR-like model.
  Even if evaluating SIR-like models for a single choice of $\pparam$ is quite cheap,
  this procedure can be overall expensive, bearing in mind that until the sequence of generated $\pparam$
  enters in the asymptotic regime, the values generated have to be discarded because they are not distributed according to $\rho_{post}$.%
\item most MCMC algorithms proceed by acceptance-rejection criteria, where a new value of $\pparam$
  is generated and then rejected if doesn't agree with certain criteria; this leads to a further increase in the number of model evaluations;
\item the forward UQ analysis based on the MCMC samples is a Monte Carlo analysis, which needs many samples of $\pparam$ to provide an accurate estimate
  (the accuracy being proportional to the inverse of the square root of the number of samples as already discussed
  -- or more precisely, the inverse of the square root of the number of accepted samples upon having entered the asymptotic regime).
\item the design of an efficient MCMC algorithm (effective proposal strategies with low rejection rate, quick to enter the asymptotic regime) might be non-trivial.
\end{enumerate}

\subsection{Gaussian approximation of the posterior: Maximum Likelihood Estimate (MLE) and Fisher approximation} \label{sect:posterior} 

Instead of using an MCMC approach, the strategy we employ here is to approximate $\rho_{post}$ with a multi-variate Gaussian distribution
with mean $\mmu_G$ and covariance matrix $\Sigma_G$; this is also called \emph{Fisher approximation}.
This approximation, provided that the available data are sufficient to determine the parameters,
is in general more and more accurate as more data become available, i.e., as $N_{meas} \rightarrow \infty$, and it has
the advantage that upon doing so, it is much easier to perform the UQ
analysis, because obtaining samples from Gaussian random variables is a standard task.
It has, however, some disadvantages that will be made clearer in the later sections, when discussing identifiability of the system:
in a nutshell, we can already reveal that the problem is that the Fisher approximation \emph{assumes identifiability of the system},
but this is not always true in practice and whether the system is identifiable or not should be checked beforehand.
MCMC instead does not assume identifiability, and can in principle be used even when the system is not identifiable:
dealing with a non-identifiable system is, however, intrinsically difficult and care needs to be taken also when tackling it using MCMC methods, as will be made clearer later on.

The Gaussian approximation is centered at the point of maximum of the posterior pdf (maximum a-posteriori
estimate, MAP). Since we have further made the assumption that the prior pdf for $\pparam$ is uniform
(see Example \ref{ex:non-uniform-prior} for the extension to the case of non-uniform prior),
this is equivalent to computing the value $\pparam$ where the likelihood function is maximized; this point is generally
known as the \emph{Maximum Likelihood Estimate (MLE)} for $\pparam$: 
\[
\mmu_G = \ttheta_{MLE} = \argmax_{\pparam} \mathcal{L}_{\mathcal{D}}(\pparam). 
\]
In practice, it is numerically more convenient to work with the logarithm of the likelihood,
and to recast the problem as a minimization problem,  i.e., to compute $\mmu_G$ as
\begin{equation}\label{eq:NLL}
  \ttheta_{MLE} = \argmin_{\pparam} NLL(\pparam), \qquad NLL(\pparam) = - 2 \log ( \mathcal{L}_{\mathcal{D}}(\pparam) ).  
\end{equation}
The function $NLL(\pparam)$ is called \emph{negative log-likelihood}, and in the particular case where the noise
affecting the data is assumed to be Gaussian random variables (such as in our case), this
problem is equivalent the least-squares estimate of $\pparam$. Indeed, it is straightforward to
combine (\ref{eq:likelihood}) and (\ref{eq:NLL}) to obtain%
\footnote{the full NLL includes additional terms in $\log(\sigma)$ and $\log(2\pi)$ that we can however drop in the minimization
  process, since they do not depend on $\ttheta$.}
\begin{equation}
  \label{eq:MLE=LS}
  \ttheta_{MLE} = \argmin_{\pparam} \frac{1}{\sigma^2}\left[ \sum_{i=1}^{N_{meas}} \left( \frac{1}{K} I(\pparam,t_i) - \hat{I}_i \right)^2
                                   + \sum_{i=1}^{N_{meas}} \left( \frac{1}{K} R(\pparam,t_i) - \hat{R}_i \right)^2 \right]. 
\end{equation}                              
Observe that this formulation does not require prior information on the value of the noise variance $\sigma^2$:
if $\sigma^2$ is unknown, it can be recovered as the sample variance of the misfits at $\ttheta_{MLE}$
\begin{equation}
  \label{eq:MLE-sigma}
  \sigma^2 \approx \sigma_{MLE}^2 = \frac{1}{2N_{meas}} \left[ \sum_{i=1}^{N_{meas}} \left( \frac{1}{K} I(\ttheta_{MLE},t_i) - \hat{I}_i \right)^2
                                      + \sum_{i=1}^{N_{meas}} \left( \frac{1}{K} I(\ttheta_{MLE},t_i) - \hat{I}_i \right)^2 \right].
\end{equation}
The covariance matrix $\Sigma_G$ of the Gaussian approximation can be chosen as the inverse of the Hessian of the NLL at the MLE of the parameters $\ttheta_{MLE}$,
see e.g. \cite{carrera:MLE}:
\begin{equation}
  \label{eq:Sigma_MLE}
  \Sigma_G = H^{-1}, \quad H_{i,j} = \frac{\partial^2}{\partial_{\param_i}\partial_{\param_j}} NLL(\pparam) \Big\rvert_{\pparam = \ttheta_{MLE}}.
\end{equation}
The matrix $H$ is also called the \emph{Fisher Information Matrix}.
The diagonal entries of $\Sigma_G$ are the variances of the posterior pdfs of the Gaussian approximations
of the parameters, i.e.,
\begin{equation}
  \label{eq:posteriors}
  \param_i \sim \mathcal{N}([\pparam_{MLE}]_{i}, [\Sigma_{G}]_{i,i}).
\end{equation}
This formula quantifies the intuitive fact that the precision of the MLE is related to how narrow the minimum of the NLL at $\ttheta_{MLE}$ is. A deep, narrow minimum means that moving even slightly from $\ttheta_{MLE}$ will change consistently the value of NLL; therefore, we have a significant evidence that the estimate is precise. Conversely, a shallow minimum means that the MLE is
not very reliable. At $\ttheta_{MLE}$ the Hessian is positive definite, with large eigenvalues if the minimum is narrow;
therefore, its inverse has small eigenvalues, and in general small diagonal entries, that can be used as variances of the parameters
(the opposite is true for a shallow minimum: the Hessian has small eigenvalues, which means that the diagonal entries of its inverse will
be large, and consequently the variances of the parameters will be large).
As already mentioned, approximating the true posterior with equation \eqref{eq:posteriors} is in general more and more valid as more data become available,
provided that the system is identifiable, as we will make clear below.%
\footnote{The statistical interpretation of this fact is that
the MLE is asymptotically Gaussian distributed with covariance matrix equal to the inverse of the
Fisher Information Matrix. This means that it is an \emph{efficient} estimator,
because it reaches the Cramer-Rao lower bound on the variance of estimators \cite{kay:signal.processing}.}
Given the expression of the likelihood in equation (\ref{eq:likelihood}), we can derive an expression for $H$ as follows:
\begin{align}
  H_{i,j} = \sum_{m=1}^{N_{meas}} \frac{1}{K\sigma^2}
    \left[
      \frac{1}{K} \frac{\partial}{\partial_{y_i}} I_m(\ttheta_{MLE}) \frac{\partial}{\partial_{y_j}} I_m(\ttheta_{MLE}) +
      \left(\frac{1}{K} I_m(\ttheta_{MLE}) - \hat{I}_m \right) \frac{\partial^2}{\partial_{\param_i,\param_j}} I_m(\ttheta_{MLE}) \right] + \nonumber \\
      \sum_{m=1}^{N_{meas}} \frac{1}{K\sigma^2}
    \left[
      \frac{1}{K} \frac{\partial}{\partial_{y_i}} R_m(\ttheta_{MLE}) \frac{\partial}{\partial_{y_j}} R_m(\ttheta_{MLE}) +
      \left(\frac{1}{K} R_m(\ttheta_{MLE}) - \hat{R}_m \right) \frac{\partial^2}{\partial_{\param_i,\param_j}} R_m(\ttheta_{MLE})
    \right],\label{eq:hessian}
\end{align}
with $I_m(\ttheta_{MLE}) := I(\ttheta_{MLE},t_m)$ and $R_m(\ttheta_{MLE}) := R(\ttheta_{MLE},t_m)$.  
Usually, the terms involving the second derivatives of $I$ and $R$ are dropped because they are smaller than the other terms:
this is because either the misfits at $\pparam = \pparam_{MLE}$ are small or because of near-linearity
of the models $I_m(\pparam), R_m(\pparam)$ close to the solution, i.e., $\partial_{\param_i,\param_j} I_m(\ttheta_{MLE})$
and $\partial_{\param_i,\param_j} R_m(\ttheta_{MLE})$ are small \cite[Chap.~10]{nocedal.wright:opt}.
Collecting all the derivatives of the model predictions with respect to the parameters in the Jacobian matrix $J_{IR}$,
we can write in compact form\footnote{Sometimes the term \emph{Fisher Information Matrix} is used to indicate this approximation rather than the full Hessian.}
\begin{align}
  & H_{i,j} \approx \frac{1}{K^2\sigma^2} J_{IR}(\ttheta_{MLE})^T J_{IR}(\ttheta_{MLE}),   \label{eq:hessian.dropped} \\  
  &[J_{IR}]_{m,i} = \frac{\partial}{\partial_{\param_i}} I_m(\ttheta_{MLE}), \quad m = 1,2,\dots,N_{meas}, \nonumber \\
  & [J_{IR}]_{m+N_{meas},i} = \frac{\partial}{\partial_{\param_i}} R_m(\ttheta_{MLE}), \quad m = 1,2,\dots,N_{meas}. \nonumber 
\end{align}

Finally, we make an important remark: minimizing the NLL to compute $\ttheta_{MLE}$ requires repeatedly evaluating the
SIR-like model for the various parameters $\pparam$ proposed by the optimizer. The minimization procedure should be
repeated several times with different starting guesses, to avoid local mimima.

\begin{example}[Inverse and posterior-based forward UQ of a SIR model]\label{ex:inverseSIR} 
In this example we show the results of the inversion procedure using \emph{artificial/synthetic data},
by fixing the values of the parameters to $\pparam_{true}=[0.29, 0.09]$, adding numerical Gaussian noise 
with $\sigma=0.025$, discount factor to $K=3$, considering data collected at $t=1,2,\ldots,30$,
and verifying the results of the inversion procedure. We then perform the forward UQ
based on the posterior pdf.

Regarding the results of the inversion procedure, we expect to see that
$\rho_{post}$ is centered close to the true value of the parameters with a reasonably small variance,
i.e. $\pparam_{MLE} \approx \pparam_{true}$ and $[\Sigma_G]_{i,i}$ such that the support of $\rho_{post}(\param_i)$
is smaller than the support of  $\rho_{prior}(\param_i)$. We also expect $\sigma_{MLE}^2$
to be a reasonable approximation of the true $\sigma^2$.
Regarding the subsequent forward UQ, we expect to see that
the uncertainty in the prediction is smaller than what would be obtained by
using the prior information only, and the expected values of the quantities of interest
is closer to the true values when using the posterior pdf than when using the prior.
The resulting estimates for the parameters  obtained from the inverse UQ are
\[
  \beta=0.2848, \quad r=0.0861, \quad \sigma=0.02791.
\]
The estimated covariances computed using the full Hessian (that we can compute directly by centered
finite differences in this simple test) and with Equation (\ref{eq:hessian.dropped}) (where the Jacobian
entries are also computed by centered finite differences) are respectively
\[
\Sigma_{G,Jac} = 10^{-4} \times
  \left[
  \begin{array}{cc}
    0.7995   &  0.1064 \\
    0.1064   &  0.2609 
  \end{array}
\right],
\quad
\Sigma_{G,Hessian} = 10^{-4} \times
\left[
  \begin{array}{cc}
    0.8073   &  0.1206 \\
    0.1206   &  0.2616 
  \end{array}
\right].
\]

\begin{figure}[tpb]
  \centering
  \includegraphics[width=0.22\linewidth]{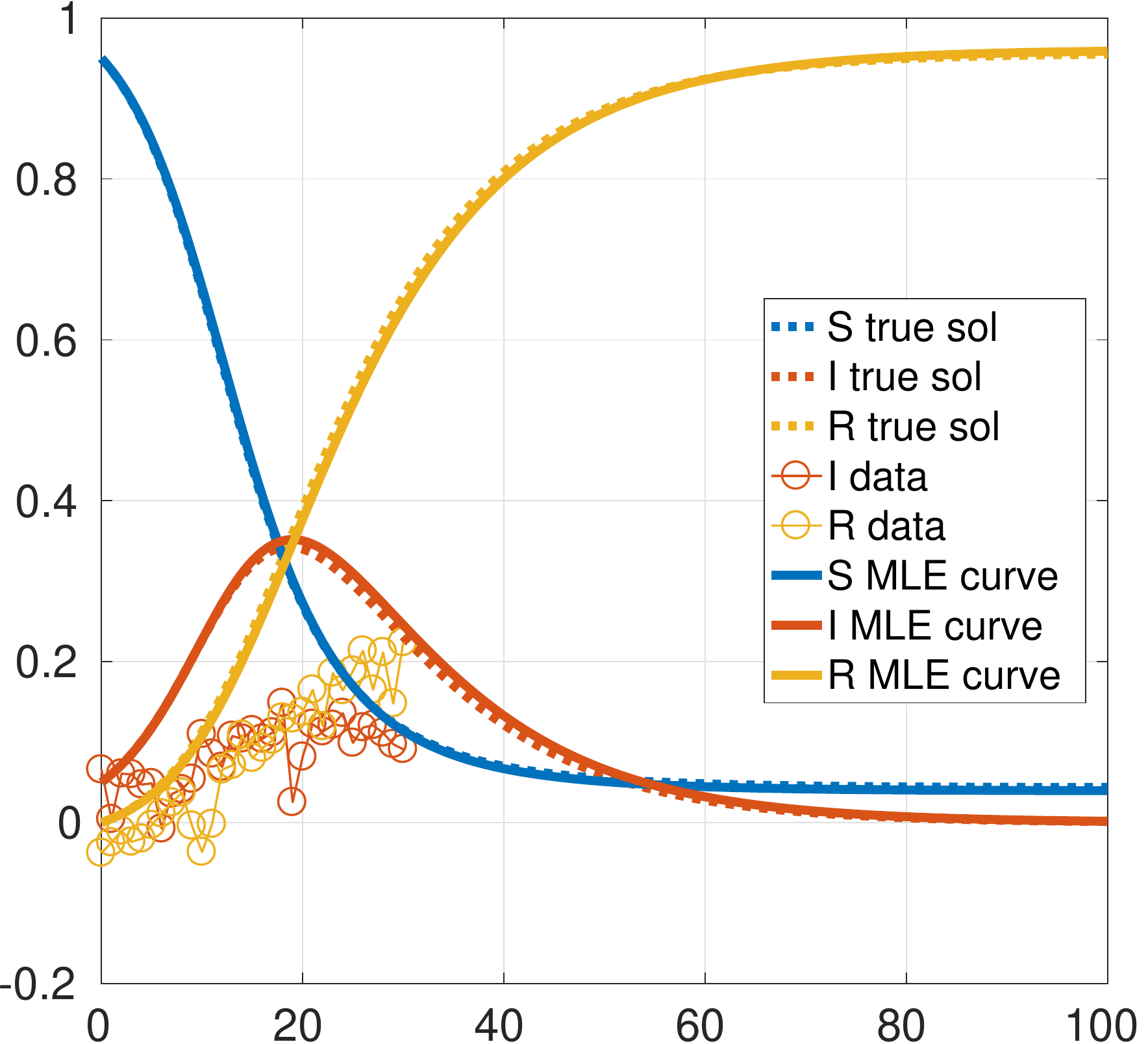}
  \includegraphics[width=0.22\linewidth]{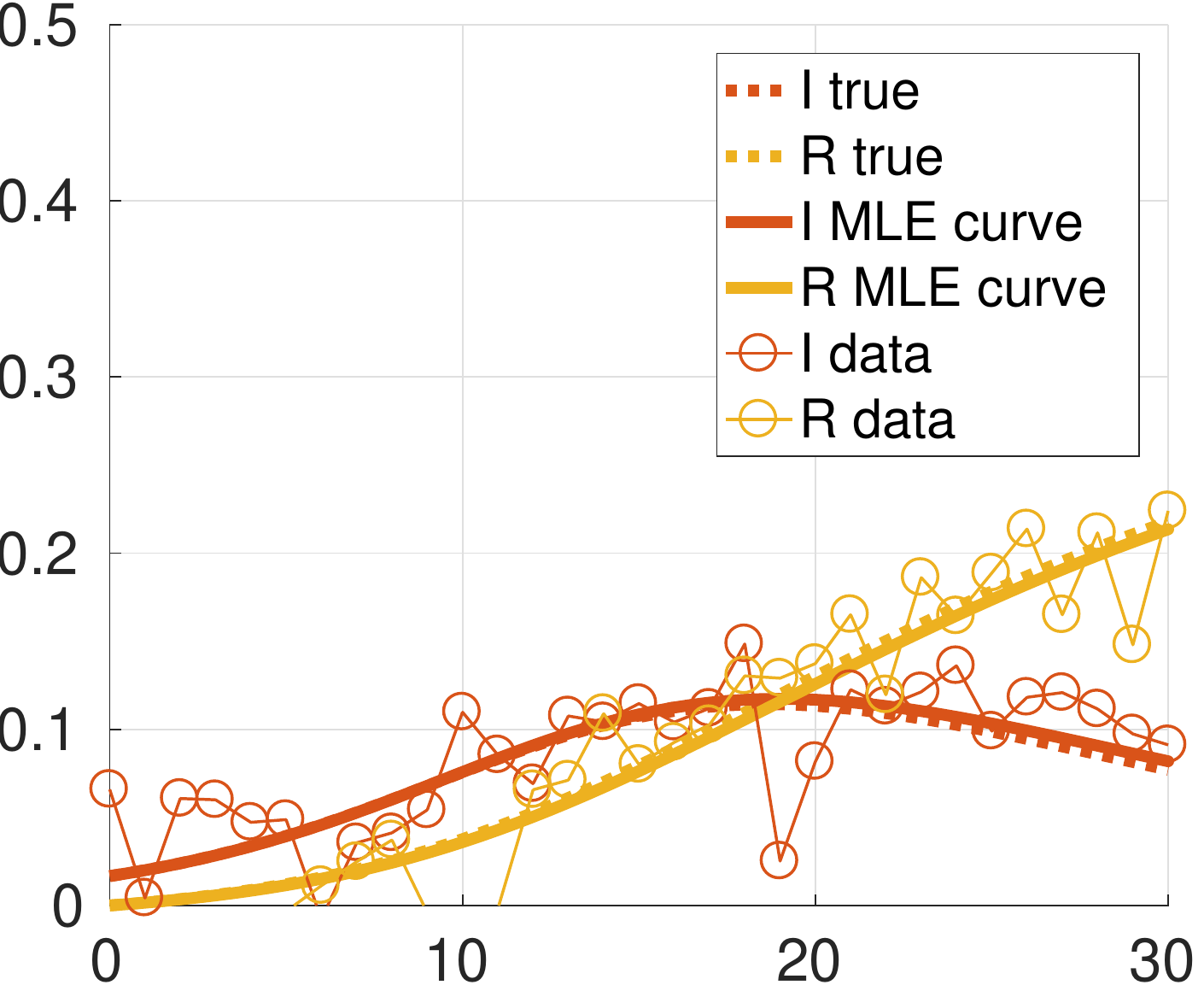}
  \includegraphics[width=0.22\linewidth]{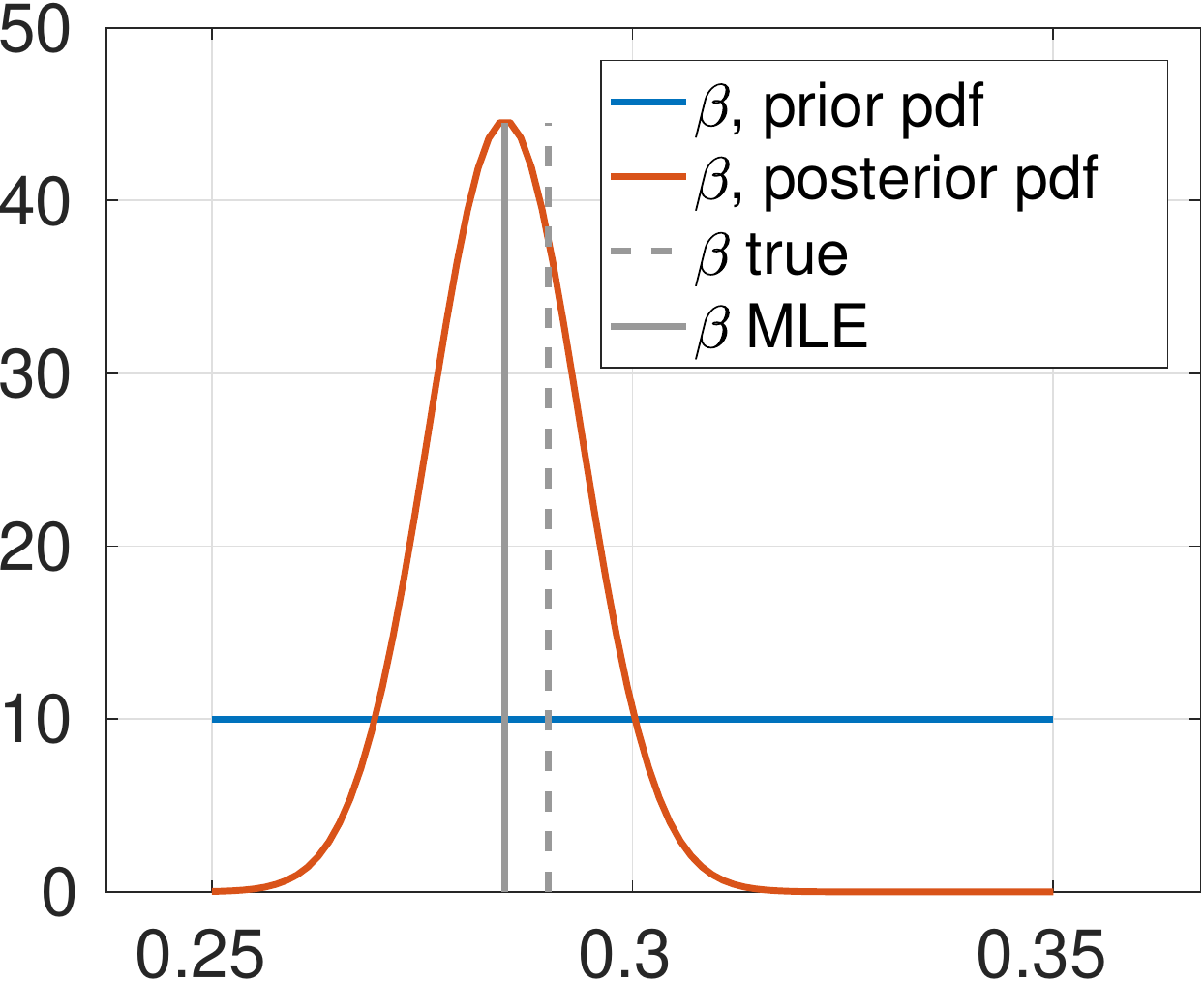}
  \includegraphics[width=0.22\linewidth]{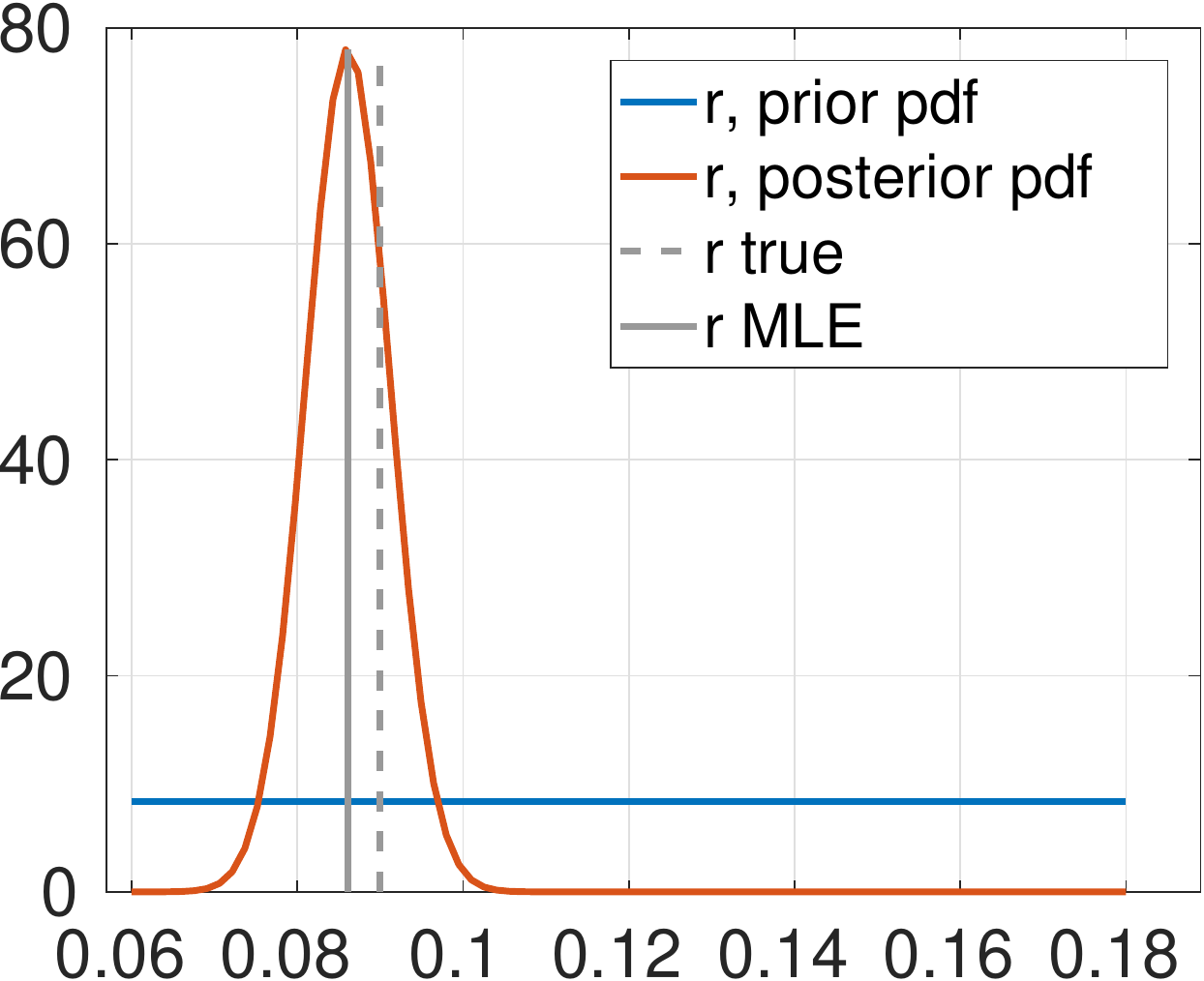} \\
  \includegraphics[width=0.22\linewidth]{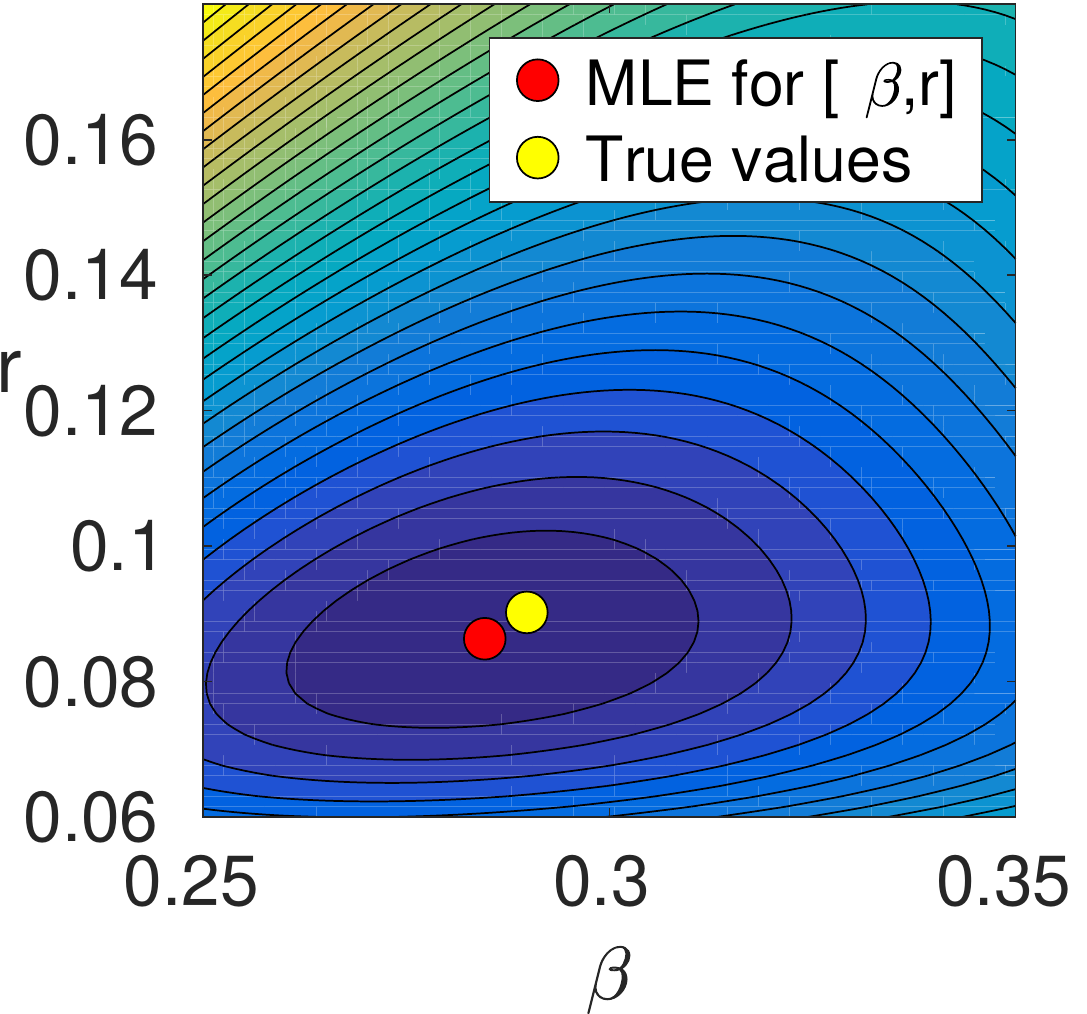} 
  \includegraphics[width=0.22\linewidth]{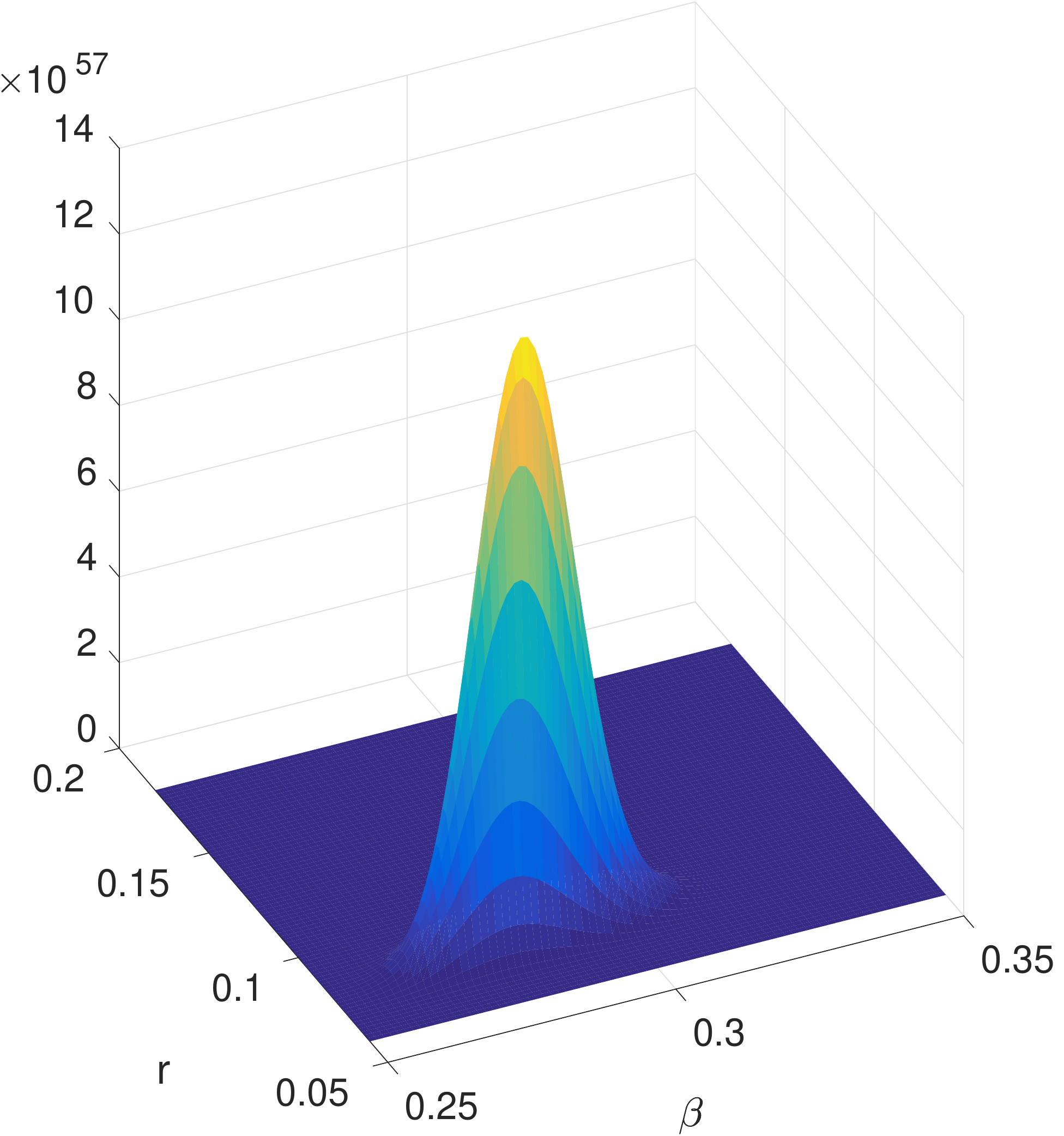} 
  \includegraphics[width=0.22\linewidth]{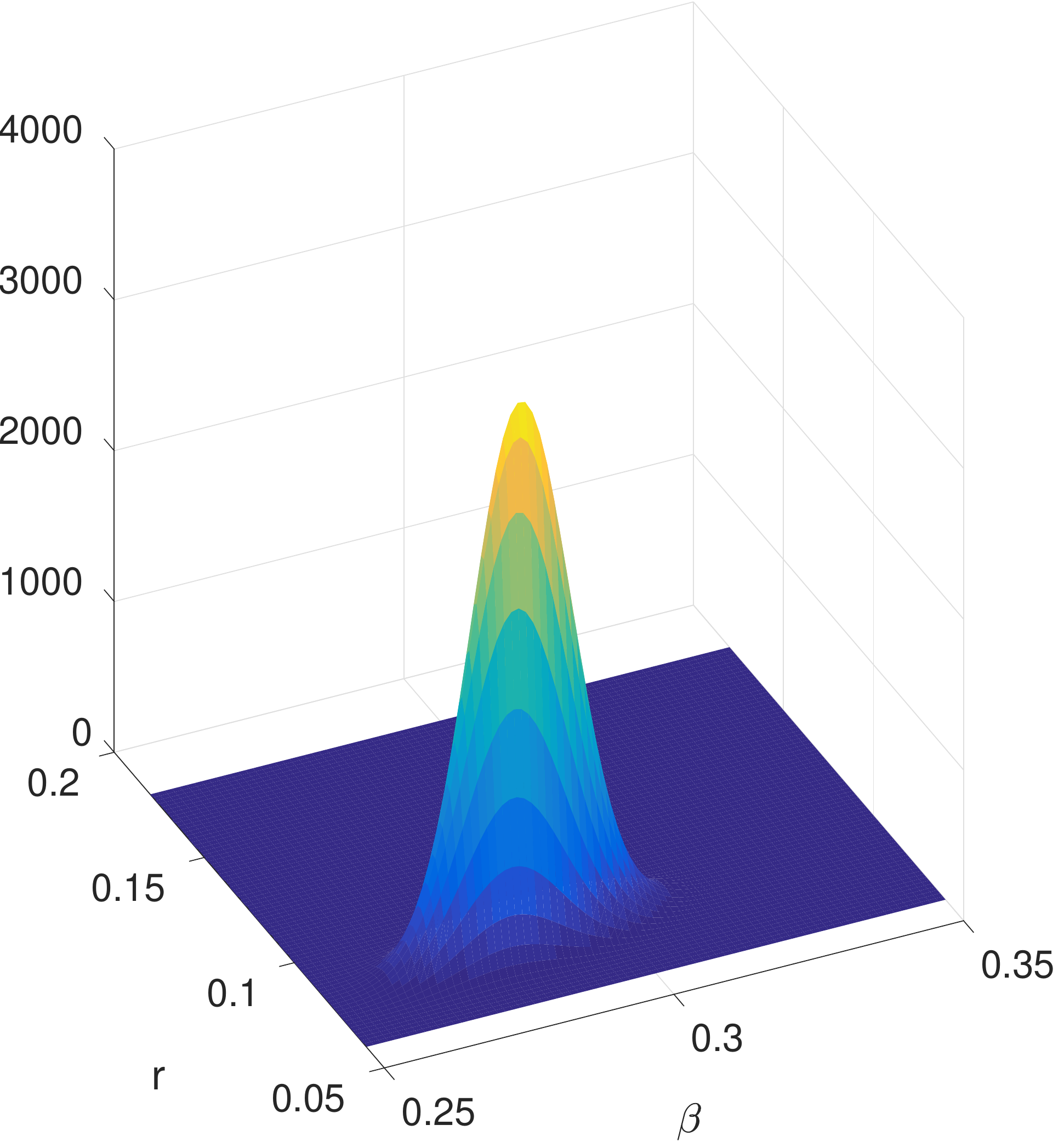} 
  \includegraphics[width=0.22\linewidth]{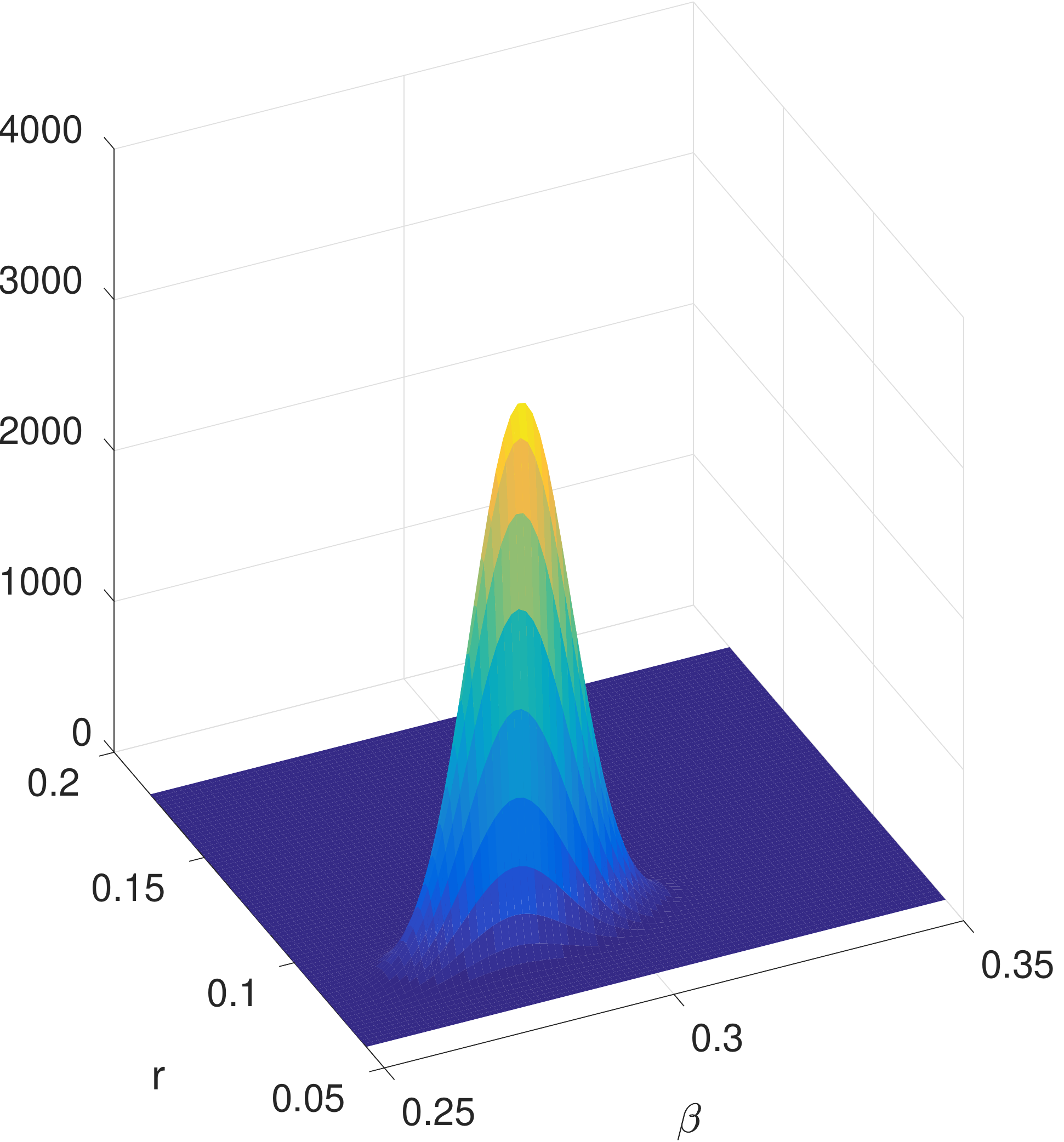}
  \caption{Result of the inverse UQ analysis for SIR. Top row, from left to right:
    trajectories corresponding to $\pparam = \pparam_{true}$ and $\pparam = \ttheta_{MLE}$,
    and synthetic data obtained dividing the trajectories for $\pparam = \pparam_{true}$ by the under-reporting
    factor $K$ and adding the Gaussian noise; zoom on the data, and trajectories for $\pparam = \pparam_{true}$  and $\pparam = \ttheta_{MLE}$
    rescaled by $K$; prior and posterior pdfs for $\beta$ and $r$, as well as the true and MLE values of the parameters.
    Bottom row, from left to right: isolines of $NLL$; surface-plot of the full likelihood; surface-plot of the likelihood after the
    Fisher approximation, cf. equation (\ref{eq:hessian}); surface-plot of the likelihood after
    having further dropped the second derivatives of $I,R$ in the definition of $H$, cf. equation (\ref{eq:hessian.dropped}). 
  }
  \label{fig:inverse-SIR-K=3}
\end{figure}

Figure \ref{fig:inverse-SIR-K=3} provides more details on the results.
The top row shows on the left the true trajectories from which the data were generated (dotted thick lines),
the noisy data (circles with thin line) 
and the trajectory obtained by fixing the parameters as $\pparam=\ttheta_{MLE}$ (solid line). The next panel provides a zoom
on the data. We have also rescaled both the true trajectory and the MLE trajectory by $K$, to emphasize
the match with the data.
The match between true and MLE trajectories is very good, although not perfect
(the distance between the trajectories would further reduce for smaller standard deviations $\sigma$ of the noise).
The last two panels of the row compare the prior and (Gaussian approximation of) the posterior pdfs of the parameters.
It can be seen that the posterior are centered close to the true value, and the Gaussian pdf is quite
concentrated in comparison to the prior interval.
The bottom row provides details about the minimization procedure. More specifically,
the leftmost panel shows the contour of the NLL function,
the true values of the parameters (yellow dot) and the MLE (red dot).
The presence of noise prevents a perfect match between the true values and the MLE,
but the match is nonetheless good and the isolines are nicely rounded, which suggest a unique, narrow (hence trustworthy) minimum.
The next panel shows the corresponding likelihood function, taken by exponentiating the NLL
(remember that since we have assumed uniform prior, the posterior is proportional to the likelihood),
which shows a clear Gaussian profile. The next panel shows the Gaussian approximation
where the covariance approximation has been computed by inverting the true Hessian of the NLL,
while the approximation obtained by using the Jacobian matrix only is shown in the right-most panel,
cf. equation (\ref{eq:hessian}) and (\ref{eq:hessian.dropped}).
The three latter plots match well (other than by the rescaling factor), indicating that the approximation steps
are not introducing significant errors. The MLE has been computed with the \texttt{fminsearch} algorithm in Matlab, which implements the derivative-free Nelder--Mead (simplex) algorithm.

Finally, upon calibrating the pdfs of the parameters to the data, the forward UQ analysis can be performed,
using the sampling methods discussed in Section \ref{section:prior_forward} to
compute the mean of the quantities of interest (S,I,R compartments, location and intensity of the peak),
and their pdfs. 
Results are shown in Figure \ref{fig:posterior-based-UQ}, where we compare the results obtained with
prior and posterior pdfs, to appreciate the improvement in the quality of the predictions if data are provided.
The top row compares the true and the expected SIR trajectories after the forward UQ analysis based on prior and posterior pdfs. The thick solid lines are the expected values (we compute these with sparse grids sampling), the thick dotted lines are the true trajectories
and the colored lines are Monte Carlo trajectories based on the prior/posterior
distribution. The results clearly show that the posterior forward UQ is more centered around the
true trajectories, and the uncertainty in the prediction is smaller.
Similar conclusions can be obtained by looking at the pdfs of the quantities of interest computed based on either the prior or the posterior
pdfs, where we have marked with vertical dotted lines the true values (mid-row: prior-based, bottom row: posterior based).

\begin{figure}[tpb]
  \centering
  \includegraphics[width=0.45\linewidth]{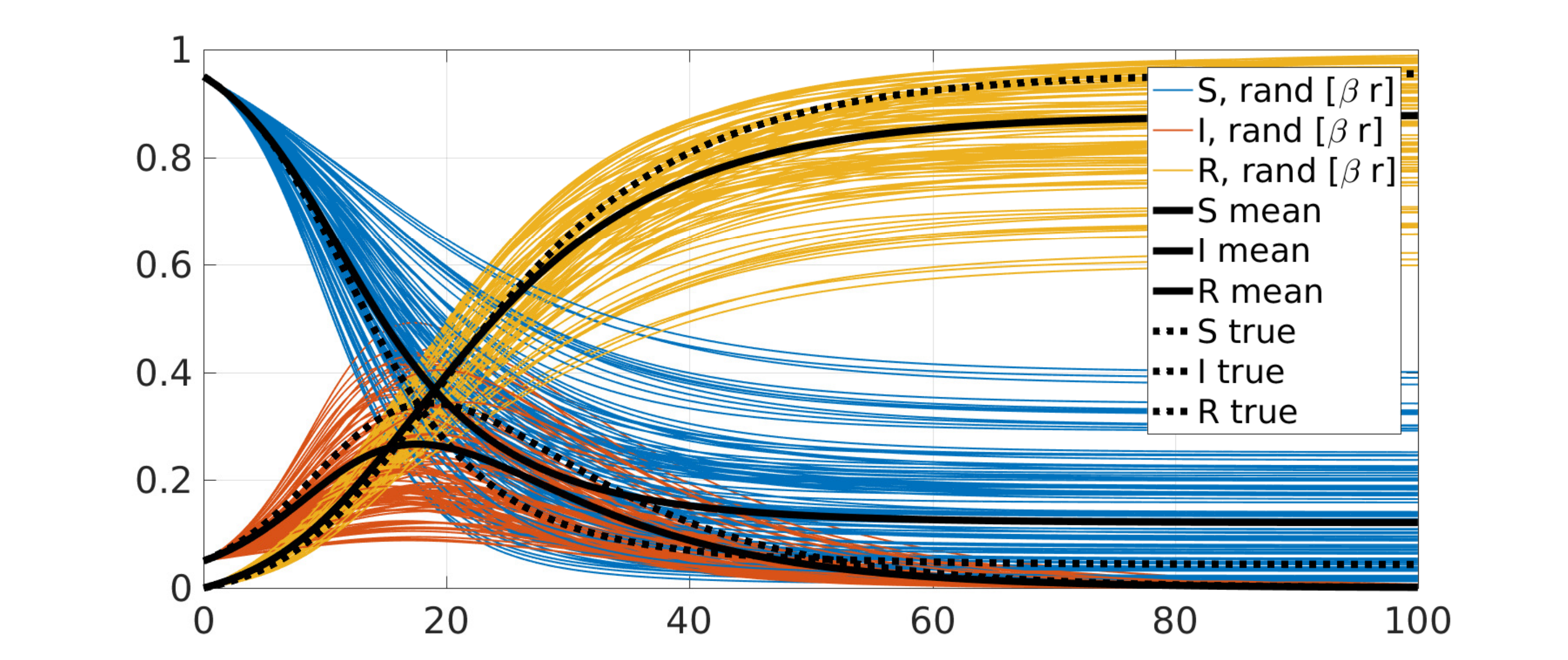}
  \includegraphics[width=0.45\linewidth]{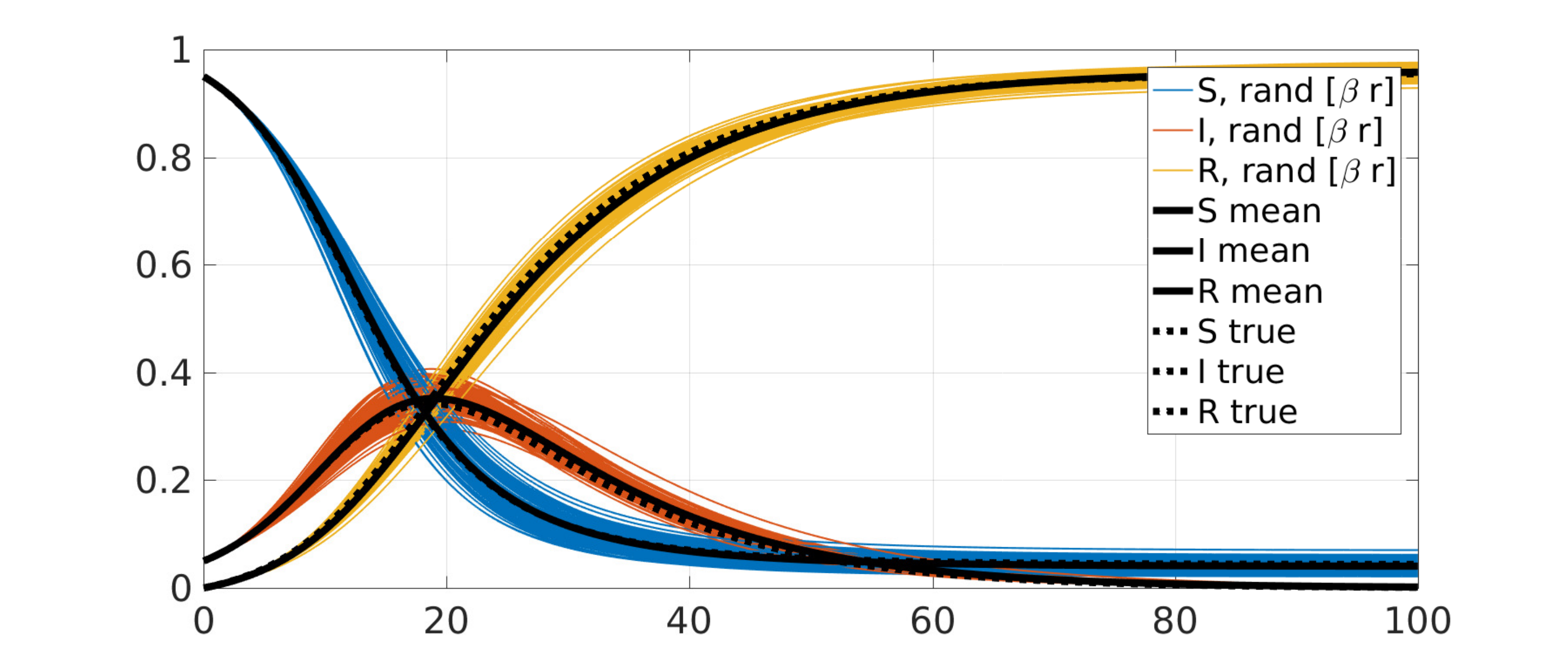}\\
  \includegraphics[width=0.95\linewidth]{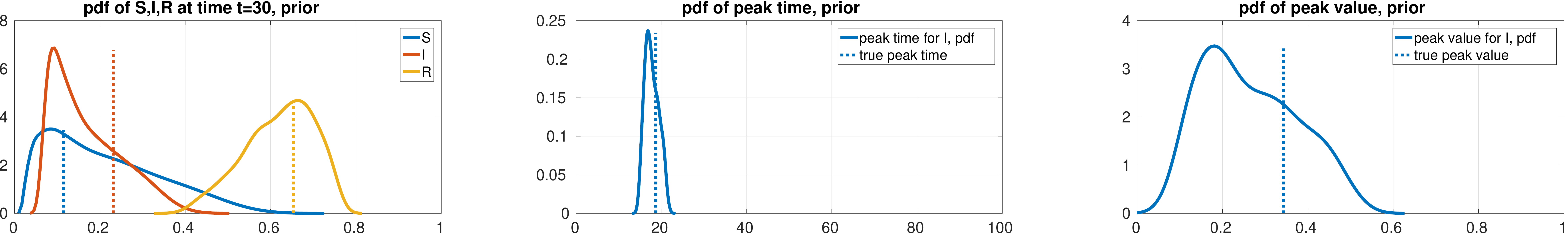} \\
  \includegraphics[width=0.95\linewidth]{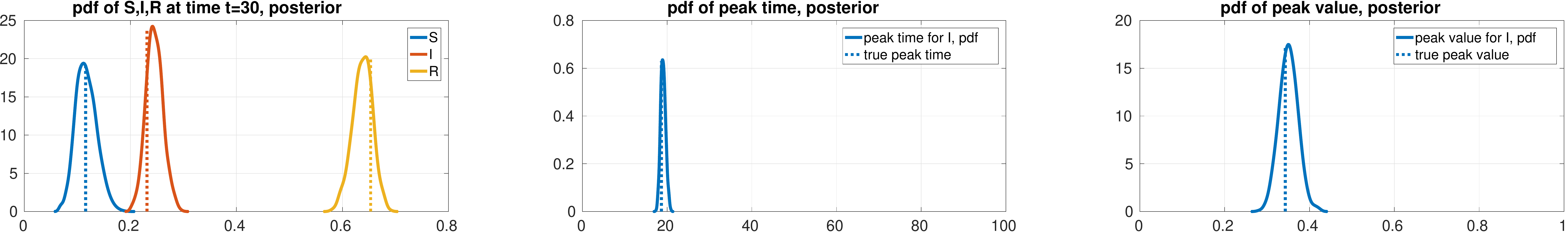}\\
  \caption{Forward UQ analysis for SIR based on either the prior or the posterior pdfs of the parameters.}
  \label{fig:posterior-based-UQ}
\end{figure}

\end{example}

\begin{example}[Incorporating prior information on parameters in the inverse UQ]\label{ex:non-uniform-prior}
  Suppose that we provide prior information about the parameters $\pparam_{true}$ different from the uniform distribution.
  For instance, we could for simplicity assume that each $\param_i$ is a Gaussian random variable
  with mean $\bar{\param}_i$ and standard deviation $s_i$, and these variables are all independent
  (in this way we are allowing $\param_i$ to assume negative values with a non-zero probability; we will fix this issue later in the example).
  Then, the posterior distribution (\ref{eq:posterior}) becomes 
  \begin{equation}
    \label{eq:posterior_with_gaussian_prior}
      \rho_{post}(\pparam | \mathcal{D}) \propto
      \mathcal{L}(\pparam) \rho_{prior} (\pparam) =
      \prod_{i=1}^{N_{meas}} \frac{1}{\sqrt{2\pi\sigma^2}}e^{\frac{-1}{2\sigma^2}(\frac{1}{K}I(\pparam,t_i) - \hat{I}_i)^2}
      \prod_{i=1}^{N_{meas}} \frac{1}{\sqrt{2\pi\sigma^2}}e^{\frac{-1}{2\sigma^2}(\frac{1}{K}R(\pparam,t_i) - \hat{R}_i)^2}
      \prod_{i=1}^{N_{\pparam}} \frac{1}{\sqrt{2\pi s_i^2}}e^{\frac{-1}{2s_i^2}(\param_i - \bar{\param}_i)^2}    
  \end{equation}
  and the Fisher approximation should now be centered at the maximum of the posterior pdf (we referred to this value as MAP)
  \[
    \mmu_{MAP} = \argmin_{\pparam} \left[ -2 \log ( \mathcal{L}_{\mathcal{D}}(\pparam)\rho_{prior}) \right], 
  \]
  which in practice recovers a form of Tikhonov regularization of the least-squares problem, with penalization parameters
  $\frac{\sigma^2}{s_i^2}$:
  \[
    \mmu_{MAP} = \argmin_{\pparam} \left[ \frac{1}{\sigma^2}\sum_{i=1}^{N_{meas}} \left( \frac{1}{K} I(\pparam,t_i) - \hat{I}_i \right)^2
      + \frac{1}{\sigma^2} \sum_{i=1}^{N_{meas}} \left( \frac{1}{K} R(\pparam,t_i) - \hat{R}_i \right)^2
      + \sum_{i=1}^{N_{\pparam}}  \frac{1}{s_i^2}(\param_i - \bar{\param}_i)^2
    \right]. 
  \]
  The computation of the sample variance estimator $\sigma_{MLE}^2$ and of the Hessian $H$ should be of course updated accordingly.
  Coming back to the non-positivity issue of the Gaussian prior pdf, a possible workaround is e.g. to assume a log-normal
  prior for the parameters and then work with the log of the parameters in the model.
  The expression derived in equation (\ref{eq:posterior_with_gaussian_prior}) for 
  the posterior pdf would still be valid, the change being ``hidden'' in the mappings $\pparam \rightarrow R(\pparam,t)$,
  $\pparam \rightarrow I(\pparam,t)$.
\end{example}

\begin{example}[Inverse UQ when different data types have different noise levels]\label{ex:two-sigma}
  Let us consider the data model \eqref{eq:data-model} and assume that the noises $\epsilon_{I,i}$ and $\epsilon_{R,i}$
  are independent Gaussian random variables distributed according to $\mathcal{N}(0,\sigma_I^2)$ and $\mathcal{N}(0,\sigma_R^2)$, respectively.
  The posterior distribution of the parameters can be computed with a slight modification of the procedure explained above
  to account for the different variances. We detail the new procedure here below, following closely \cite{carrera:MLE,lever.eal:inversion}.
   	
  We begin by assuming that the ratio $\lambda$  between the variances of the two sets of data $\lambda := \frac{\sigma_I^2}{\sigma_R^2}$ is known.
  In this case minimizing the NLL \eqref{eq:NLL} is equivalent to minimizing the following quantity 
    \begin{equation*}
    T(\pparam) = \sum_{i=1}^{N_{meas}} \left( \frac{1}{K} I(\pparam,t_i) - \hat{I}_i \right)^2 + \lambda
    \sum_{i=1}^{N_{meas}} \left( \frac{1}{K} R(\pparam,t_i) - \hat{R}_i \right)^2,
    \end{equation*}
    which is obtained combining \eqref{eq:likelihood}, opportunely modified to incorporate the different standard deviations, and \eqref{eq:NLL}.
    The parameter $\lambda$ weighs the sum of squared residuals of $I$ and $R$, highlighting a different level of trust in the first or second set of data.
    If $\lambda$ is small, the residuals of $I$ condition more the quantity $T$, whereas the data for $R$ are more influential  if $\lambda$ is large.
    Finally, if $\lambda=1$ the data sets are equally weighted and we recover the least-squares formula \eqref{eq:MLE=LS}.
    Hence, minimizing the quantity $T$ can be seen as a weighted least-squares criterion.  
        
    As $\lambda$ is however unknown in general, we vary iteratively $\lambda$ within an appropriate range of values,
    and minimize $T$ for each value of $\lambda$ to find the corresponding $\pparam_{MLE}$. Among the considered $\lambda$,
    we then select the one that realizes the minimum of $NLL$. The corresponding value of $T$ is also needed, and we denote it by $T_{min}$.
    Indeed, the values of the empirical variances of the two data sets are then recovered in the following way
    \[ 
    \sigma_I = \sqrt{T_{min}/2N_{data}}, \quad \sigma_R = \sqrt{\sigma_I^2/\lambda}.
    \]
    
    Note that value of $\lambda$ can be selected also by means of criteria other than the minimum of the NLL.
    In particular, in \cite{lever.eal:inversion} the Kayshap Information Criteria has been employed, that penalizes values of $\lambda$
    that result in shallow NLL.  
    
    Finally, we illustrate this minimization procedure with the help of an example. We consider the SIR model, and generate 41
    equispaced synthetic data in the time interval $[0,20]$ with $\pparam_{true}=[0.29, 0.09]$,
    adding numerical Gaussian noises with $\sigma_I = 0.2$ and $\sigma_R = 0.05$, where $\lambda_{\text{true}}= 16$.
    We consider integer values of $\lambda$ in the range $[1,90]$, and for each of them we minimize $T$.
    In Figure \ref{fig:sigma_I_and_sigma_R} on the left we plot the minimum value of the NLL for each considered $\lambda$.
    We observe that the value of $\lambda$ resulting in the minimum NLL can be approximately correctly identified ($\lambda_{min}\approx 21$),
    resulting in the following estimates of the parameters 
    \[
    \pparam_{MLE} = [0.2917, 0.0950], \quad \sigma_I = 0.2330 \quad \sigma_R = 0.0508. 
    \]
    The figure in the middle displays the posterior distribution of the parameters corresponding to $\lambda_{min}$
    and to the two extremes of the considered range of $\lambda$.
    The figures on the right show 200 Monte Carlo trajectories based on the posterior distribution of the parameters.
    The value $\lambda=\lambda_{min}$ (left-most panel) gives a narrow bundle of trajectories matching the true ones. The other bundles ($\lambda=1,90$ i.e., the smallest and largest values of $\lambda$ tested) are more spread and
    in the case of $\lambda = 90$ they are also far from being centered around the true trajectories.   
    \begin{figure}
    	\centering
    	\begin{subfigure}{0.3\textwidth}
    		\includegraphics[width=0.9\linewidth]{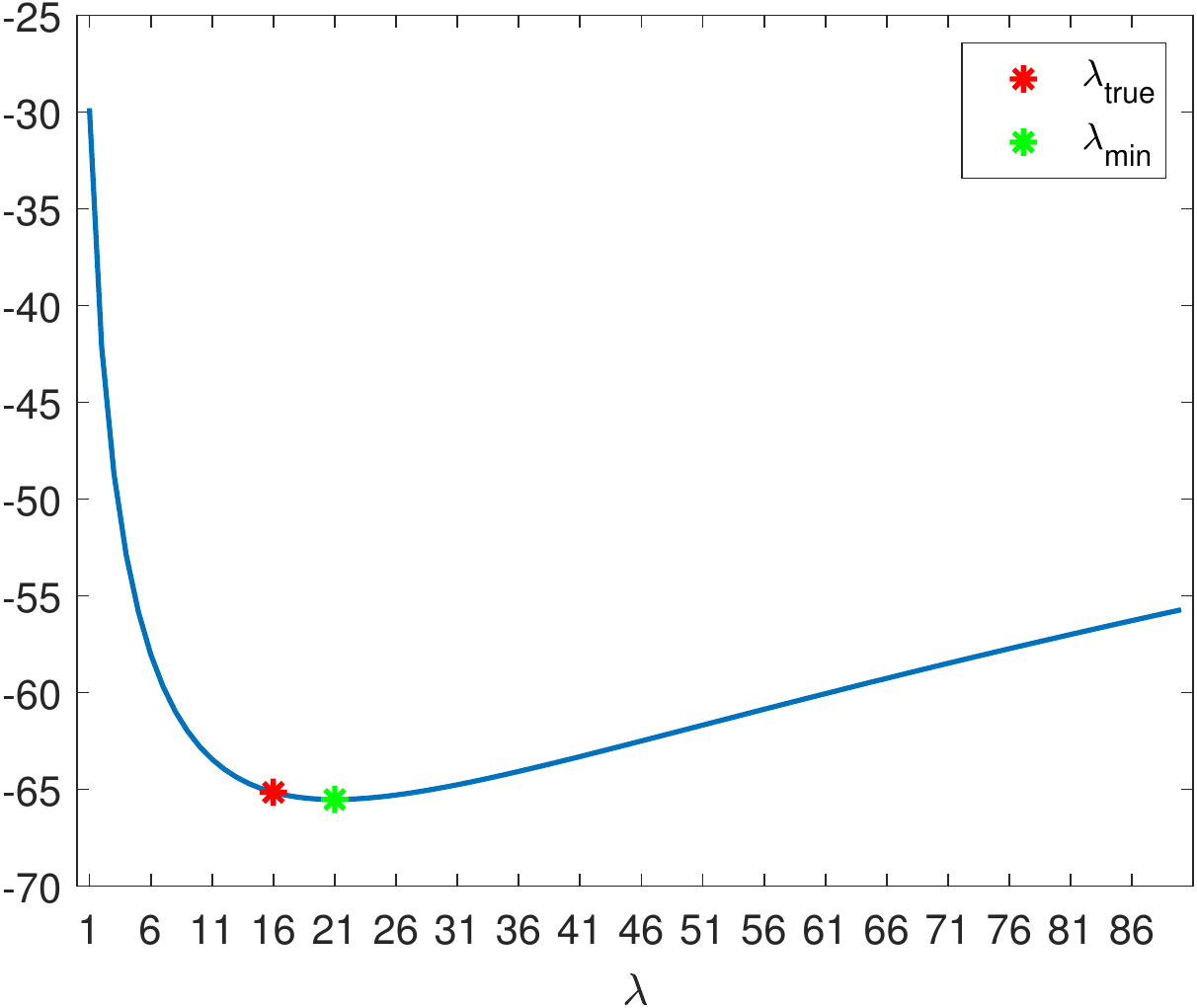} 
    	\end{subfigure} \hfill
   		\begin{subfigure}{0.32\textwidth}
   			\includegraphics[width=0.92\linewidth]{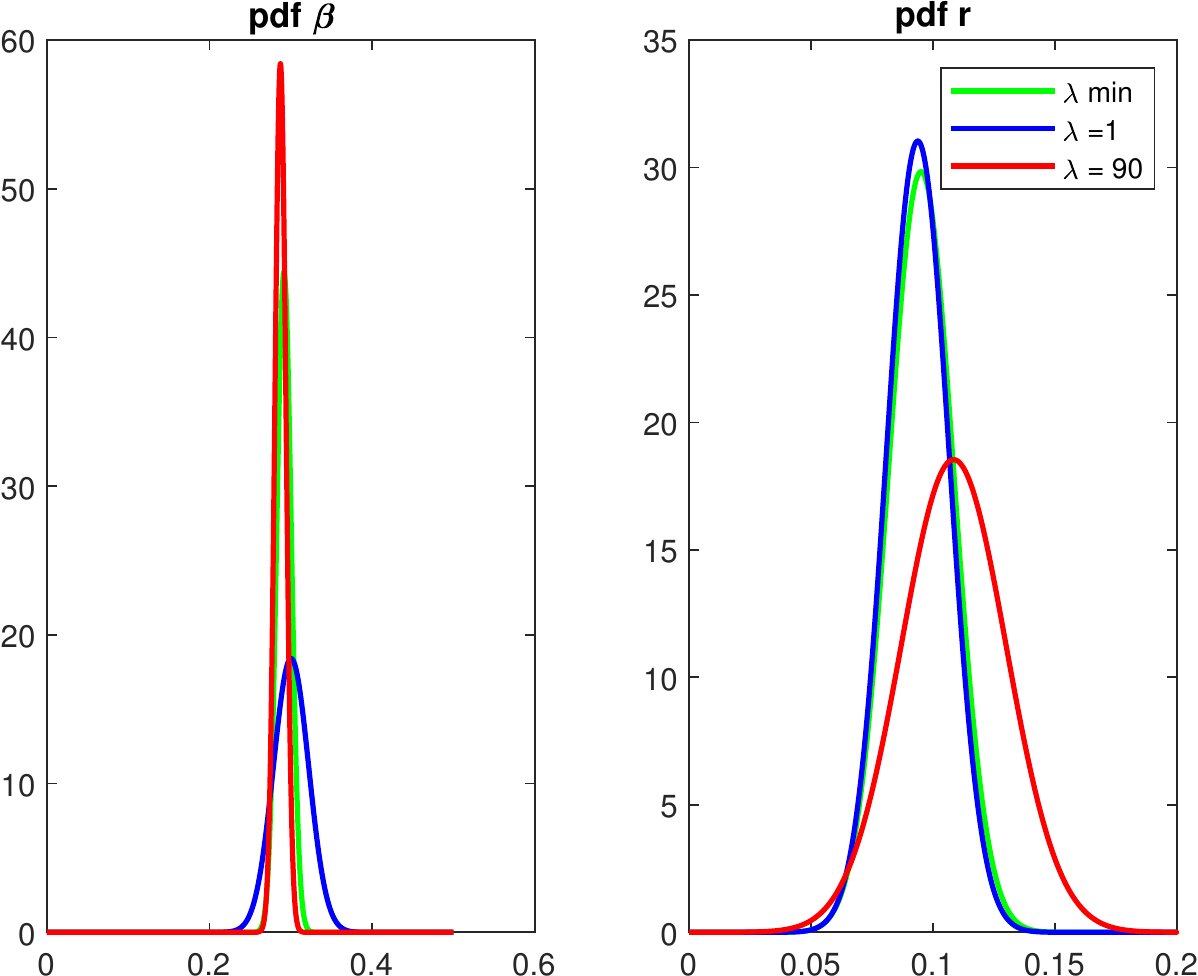} 
   		\end{subfigure}
    	\begin{subfigure}{0.35\textwidth}
    		\includegraphics[width=\linewidth]{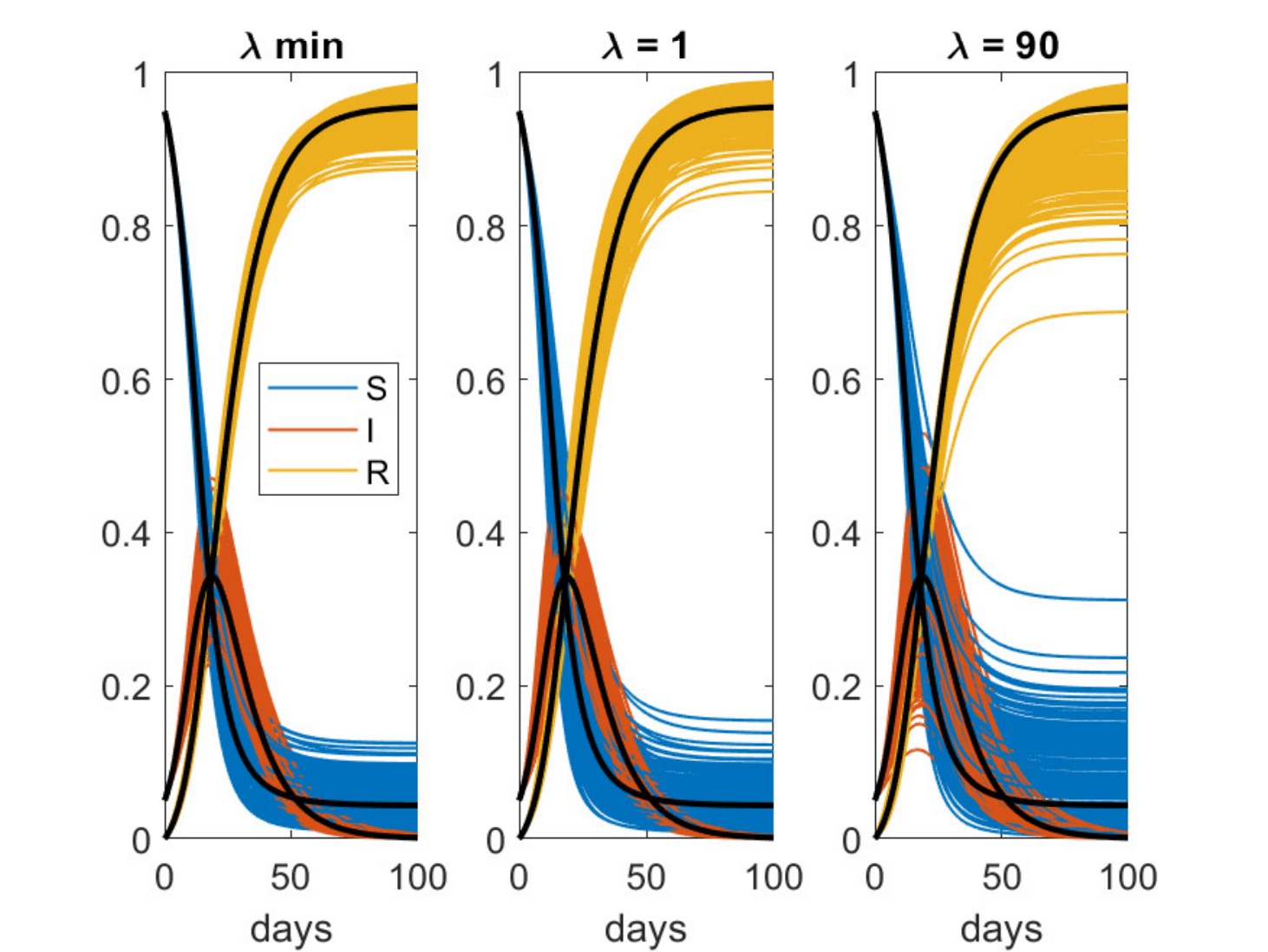} 
    	\end{subfigure}
       	\caption{Left: values of $NLL$ across the range of tested values of $\lambda$ for the test in which the data for $I$ and $R$ have different standard deviations. The smallest NLL is reached for $\lambda_{min}$ close to the exact one $\lambda_{true}$; middle: posterior distribution of $\beta$ and $r$ for $\lambda_{min}$, $\lambda = 1$, and $\lambda = 90$; right: Monte Carlo trajectories based on the posterior distribution of the parameters, for $\lambda_{min}$, $\lambda = 1$, and $\lambda = 90$, respectively. The black lines are the true trajectories. }
    	\label{fig:sigma_I_and_sigma_R}
    \end{figure} 
 
\end{example}

\begin{biblio}
\begin{itemize}

\item MCMC methods for deriving the posterior pdfs of parameters of SIR-like models for COVID-19 have
  been used in e.g. \cite{Crisanti,Gatto}.

\item Several variations of Gaussian approximation of the posterior pdfs can be conceived, depending on the choice
    of the mean (e.g., MLE, MAP, expected value of the posterior) and of the covariance matrix, see e.g. \cite[Result 8, p.224]{berger}.
  
\item For further approaches to the Bayesian inversion problem, see
  Approximate Bayesian Computation (ABC, see e.g. \cite{Drovandi:abc}) and the Integrated Nested Laplace Approximation (INLA, \cite{rue:INLA}).
  ABC is used in cases when it is difficult to evaluate (or even define) the likelihood function.
  INLA is instead useful when certain conditional posteriors can be reasonably approximated by Gaussian distributions.
  
\item In this section we have assumed a simple error model with additive Gaussian errors. See e.g. \cite{Tuncer,Capaldi}
  for more general error models. 
  \lorenzo{
    Furthermore, one can model under-reporting and additional features of the measurements,
    such as the fact that these kind of data are intrinsically integer numbers, by choosing more
    suitable likelihood functions than in equation (\ref{eq:likelihood}), such as negative binomial or
    quasi-Poisson likelihooods}, see e.g. \cite{verhoef,Linden,Pugliese}.
  
\item In this section we have assumed that $\rho_{prior}$ are either uniform or Gaussian random variables; other choices with
  good properties are possible, e.g. the Jeffreys non-informative priors \cite{Firth:Jeffreys}.

\item Using sparse grids for UQ with exact posterior pdf in equation (\ref{eq:posterior})
  is not straightforward, because the functional shape of $\rho_{post}$ does not fall into
  classical families of pdfs (e.g. uniform, Gaussian, gamma) for which these methods are developed.
  A possible remedy would be to compute ad-hoc polynomials, in the spirit of \cite{witteveen:arbitrary,Oladyshkin:arbitrary}.
  Another possibility is to keep sampling according to the prior pdf, see e.g. \cite{schillings.schwab:inverse},
  but this is possibly suboptimal if the prior and the posterior are significantly different (e.g., a uniform prior and a very peaked
  posterior).

\item Conversely, sparse grids sampling for Gaussian random variables has been discussed multiple times in literature,
  see e.g. \cite{ernst.eal:Leja+Levy}; therefore, sparse grids computation are easy to use upon having performed the Fisher approximation.

\item Replacing the evaluation of the full-model with a response surface either in the MCMC sampling of posterior pdf
  or in the minimization of NLL is possible (this operation is routine in computationally-heavy inverse problems,
  see e.g. \cite{marzouk.xiu:stoc-coll-bayesian}), but in this context this operation does not dramatically
  speed-up the computational time because SIR-like models are rather cheap to evaluate.

\item We could extend the error model for $\epsilon_{I}, \epsilon_R$ to account for the various
  sources of error (model error, response surface error if used, numerical discretization errors), see e.g. \cite{pagani:errorRB}.

\item Minimizing the NLL function is an example of non-linear least-squares optimization problems,
  for which ad-hoc algorithms exist, like Gauss--Newton, Levenberg-Marquardt, VarPro, Trust-region reflective,
  see e.g. \cite{nocedal.wright:opt,Golub2003varpro}.

\item \lorenzo{Inverse UQ / data fitting methodologies can also be applied to stochastic models, where the dynamics
    of the system are influenced by a ``process noise'', and measurements are further corrupted by
    ``observation noise'' (i.e., the errors $\epsilon_{I,i}$, $\epsilon_{R,i}$ in equation \eqref{eq:data-model}).
    The methodology for data fitting described above of course needs to be adjusted accordingly, see e.g. \cite{deValpine,He,MartinFernandez}.}

\end{itemize}
\end{biblio}

\section{Summary: an ideal UQ workflow, from data to prediction}\label{section:UQ_workflow}

Summarizing the discussion so far, the \textit{ideal UQ workflow} for predictions under uncertainty would consist of the steps
reported in Algorithm \ref{workflow:UQ}.
This ideal workflow however is missing one step, i.e., the identifiability analysis, which is a crucial preliminary analysis
to perform. We discuss it in details in the next sections. The adjusted ideal UQ workflow will then be presented in the final Section
\ref{section:disc} (Discussion and conclusion), see Algorithm \ref{workflow:UQ2}.

    
\begin{algorithm}
  \caption{Ideal UQ workflow}  \label{workflow:UQ}
  Choose a model and the prior distributions for its parameter (literature, expert opinion)\;
  
  Compute Sobol indices to assess which parameters are more influential and can be inferred from data. Fix the remaining parameters to some reasonable value\;
  
  Perform the inverse UQ analysis to adjust the prior distribution to the data evidence\; 

  Perform the forward UQ analysis based on the posterior distribution to obtain statistical information about the quantities of interest of the model
      (e.g. expected value, variance, full pdf of the outputs)\;
\end{algorithm}

\section{Structural identifiability} \label{sect:struct_identifiability}

The fundamental assumption underlying the inversion approach proposed in Section \ref{section:inverse} is that
there exists a certain set of parameters and hyper-parameters $[\pparam_{true},\hh_{true}]$ that generated
the observed data from the system (\ref{eq:ode_system}), as expressed in Equation (\ref{eq:data-model}).
We then embrace the fact that the data are noisy, and that this noise might prevent us from correctly
determining the values $[\pparam_{true},\hh_{true}]$; we therefore give up on giving a ``one-shot'' estimate
of  $[\pparam_{true},\hh_{true}]$, and rather content ourselves with computing a posterior pdf,
which quantifies our degree of belief on each possible value of  $[\pparam_{true},\hh_{true}]$.
The Fisher approximation then further assumes that the NLL has a unique, well-shaped minimum,
which means that the posterior pdf of the parameters
is sufficiently well-approximated by a Gaussian pdf centered at $[\pparam_{MLE},\hh_{MLE}]$,
whose variance 
gets smaller as we acquire more data.
If instead we believe that there exists a certain set $[\pparam_{true},\hh_{true}]$ but
for some reason we think that our data do not support the assumption that the posterior is Gaussian
(for instance, because we have only limited data), we could
consider the ``full'' posterior given by equation (\ref{eq:posterior}) instead
(using e.g. MCMC as computational tool), 
and ideally three scenarios might then occur:
\begin{enumerate}
\item the posterior pdf is actually close to Gaussian;
\item the posterior pdf is unimodal but it departs from Gaussian in that it might show ``heavy tails'' and/or some degree of skewness.
  This would indicate that we might be introducing a bias that leads to over/underestimates;
\item the posterior pdf is multi-modal. This would mean that the inversion procedure is suggesting a few
  ``likely'' combinations of parameters $\pparam$, each corresponding to one peak of the posterior pdf:
  in this case the heights of the peaks represent our belief on the plausibility that such $\pparam$ is  the ``true one''.
\end{enumerate}
In any case, the crucial point that one has to address is: can we guarantee that \emph{there is} a unique set $[\pparam_{true},\hh_{true}]$
that generates the observed outputs? Or, equivalently, is the inverse problem well-posed?
If not, the Fisher approximation is bound to fail (for instance, item 3 in the list above)
and the MCMC approach also needs to be handled with care.
In the field of mathematical epidemiology (and more generally of dynamical systems/systems control),
this question falls into the study of the so-called \emph{system identifiability}, which can be divided in two consecutive steps:
\begin{description}
\item[Structural identifiability:] studying from a theoretical point of view the well-posedness of the identifiability (inverse) problem, assuming that
  perfect information is available, i.e., that infinitely many noise-free observations of the outputs are available. It is an intrinsic
  property of the system and is the topic of this section. This topic is well-studied in the epidemiological literature:
  a list of references is available at the end of the section. 

\item[Practical identifiability:] addressing the identifiability of the system given limited and noisy observations of the outputs.
  It depends not only on the properties of the system but also on the quality of the data. In other words, the structural identifiability
  is a necessary but not sufficient condition for the practical identifiability of the system. The practical identifiability is
  the topic of the next Section \ref{section:pract_id}.
\end{description}

Mathematically, a system is structurally identifiable if the model map, i.e. the function $[\pparam,\hh] \mapsto Y(\pparam,\hh)$ 
mapping each realization of $\pparam,\hh$ to the corresponding values of the outputs / quantities of interest is injective.
Of course, numerical estimates of parameters obtained by UQ techniques for structurally non-identifiable systems
are not reliable and might lead to very wrong predictions. We will show some results on this in Example \ref{ex:SIR_identifiability}.
In the following we discuss two approaches to structural identifiability, one based on differential algebra and a mapping approach. Other methods are available, see the bibliography at the end of the section.

\subsection{Differential algebra}

The differential algebra approach to the structural identifiability problem is based on deriving a set of differential equations for the
model outputs $Y$ of the form $\mathcal{P}(Y,\dot{Y},\ddot{Y},\ldots,\pparam,\hh) = 0$, 
where $\mathcal{P}$ is a monic differential polynomial including only $Y$, their derivatives and the model parameters/hyperparameters;
see e.g. Example \ref{ex:structural_identifiability_SIR}.
These equations are known as input-output equations and are an implicit form of the model map,
as they generate the same output as the original model.
The coefficients of the input-output equations give indication on the identifiability of the system:
if the map from $[\pparam,\hh]$ to the coefficients of the input-output equation is injective, the system is structurally identifiable;
conversely, if there are multiple values of $[\pparam,\hh]$ that generate the same input-output equations, the system is structurally non-identifiable.

One approach to obtain such equations is by ad-hoc substitution and differentiation to eliminate the unwanted quantities,
starting from the original system (\ref{eq:ode_system}) -- of course, one must be careful not to remove/introduce additional solutions
e.g. by canceling/multiplying every term by $Y$.
For systems with many compartments, the manual ad-hoc substitution method might be impractical, and a more algorithmic approach is needed:
one possibility consists in generating the input-output equations as part of the so-called characteristic set of the
algebraic ideal generated by the polynomials defining the model, see \cite{ritt}.
Finally, note that the calculations required to derive the input-output equations can be done also using symbolic calculus software,
e.g. Mathematica and Maple.

We refer to \cite{eisenberg-diff-alg} for the theoretical background underlying this procedure:
in particular, in that work it is shown that the identifiability result does not depend on the particular method
employed to derive the input-output equations, as long as a certain property, called \emph{mutual reduction}, holds true
 -- see Example \ref{ex:structural_identifiability_SIR}.
This property is always true in the case of a single output quantity,
whereas it needs to be enforced in the case of multiple output quantities.
The fact that the identifiability result does not depend on the specific form of the input-output equation stems
from the fact that by definition all the input-output equations generate the same output trajectory as the original model.
From this, it follows that all forms of such equations contain the same identifiability information of the original system.

\lorenzo{Another (quite technical) preliminary condition for the differential algebra approach to be valid
  is the so-called \emph{solvability condition} \cite{saccomani},
  which can be safely be assumed to hold for a wide class of practical problems, and we do this
  in this work as well. We refer again to \cite{eisenberg-diff-alg} for a discussion on why this condition can be ``safely assumed'',
  and to \cite{ovchinnikov1} for an ad-hoc example where this condition is not valid and the differential algebra approach fails.}


  \begin{example}[Structural identifiability of a SIR model by differential algebra]\label{ex:structural_identifiability_SIR}
    In this example, we focus on the case of the SIR model \eqref{eq:SIR} with output $Y = \frac{1}{K} I$.
    Our argument here is similar to those in \cite{Tuncer,Eisenberg};
    \lorenzo{as already mentioned, having prevalence data is somehow unrealistic, yet it is a scenario already rich enough for our
      ``didactical'' purposes.}
    We consider the following system  
    \[
     \begin{cases}
    \displaystyle \dot{S} = -\frac{\beta}{N_{pop}}IS\\[6pt]
    \displaystyle \dot{I} = \frac{\beta }{N_{pop}}IS - r I\\[6pt]
    \displaystyle Y = \frac{1}{K} I.    
    \end{cases}
    \]
    Note that we have neglected the equation for $R$ in \eqref{eq:SIR} as it does not influence the dynamics of the system.
    Combining the differential equation for $I$ and $Y$ yields 
    \[
      \dot{Y} = \frac{\beta }{N_{pop}}YS - r Y,
    \]
    which can be solved for $S$; from the latter, an expression for $\dot{S}$ can then be derived.
    Then, replacing these expressions for $S$ and $\dot{S}$ in the differential equation for $S$ of the SIR model, we obtain the following equation
    \[
      N_{pop} \ddot{Y}Y- N_{pop}\dot{Y}^2 + K\beta\dot{Y}Y^2 + K r \beta Y^3=0.
    \]
    We then divide all the terms by $N_{pop}$ to obtain the following monic polynomial, i.e. a polynomial with the coefficient of the highest order term equal to 1: 
    \[
      \ddot{Y}Y-\dot{Y}^2 + K\frac{\beta}{N_{pop}}\dot{Y}Y^2 + K r\frac{\beta}{N_{pop}} Y^3=0.
    \]
    Assuming $N_{pop}$ and $K$ known, the map $ [\beta,r] \mapsto [K\frac{\beta}{N_{pop}},K r\frac{\beta}{N_{pop}}]$ is injective, i.e., the system
    \[
      \begin{cases}
        \displaystyle C_1 = K\frac{\beta}{N_{pop}}, \\[10pt]
        \displaystyle C_2 = K r\frac{\beta}{N_{pop}},
      \end{cases}
    \]
    can be solved for $\beta,r$. This means that it is possible to uniquely identify $\beta$ and $r$, and the model is structurally identifiable.
    Conversely, if $K$ is unknown, only $r$ and the combination $K \beta$ can be uniquely estimated, i.e., the model is structurally non-identifiable. 
    
    If instead we consider the case of having also a second quantity of interest $Z = \frac{1}{K}R$, all the parameters of the SIR model result
    to be structurally identifiable, as we show in the following.
    Since we have data of $I$ and $R$, we drop the differential equation for the compartment $S$ of the SIR model and consider the following equivalent system 
    \[
    \begin{cases}
    \displaystyle \dot{I} = \frac{\beta }{N_{pop}}I(N_{pop}-I-R) - r I\\[6pt]
    \displaystyle \dot{R} = r I   \\[6pt] 
    \displaystyle Y = \frac{1}{K} I \\[6pt]
    \displaystyle Z = \frac{1}{K} R.    
    \end{cases}
    \]
    The practice of eliminating a compartment is possible as we assume that $N_{pop} = S+I+R$ for all times, which means that the dynamics
    of the compartment that we eliminate is completely determined by the remaining ones.
    In particular, if we let $S=N_{pop}-I-R$, we can immediately see that
    \[
      \dot{S} = -\dot{I} - \dot{R} = - \frac{\beta }{N_{pop}}I(N_{pop}-I-R) + r I - r I = - \frac{\beta }{N_{pop}}I S,
    \]
    i.e., we recover the initial equation for $S$.
    In this second example we have two outputs, therefore we have to derive two input-output equations.
    We rewrite the differential equations for $I$ and $R$ in terms of $Y$ and $Z$ and obtain 
    \[ 
    K \dot{Y} - \frac{\beta}{N_{pop}}(N_{pop}-KY-KZ)KY+rKY = 0 \qquad \text{and} \qquad K\dot{Z} - KrY = 0. 
    \] 
    To make the polynomials monic we have to introduce a ranking among the variables.%
    \footnote{i.e., a total ordering of the variables and of their derivatives. For a formal definition, see \cite{Miao}.} 
    A common choice for the ranking is $Y <Z<\dot{Y}<\dot{Z}$ (any other ranking would lead to the same results as already mentioned, of course after different computations),
    from which it follows that the leading monomials of the input-output equations are $\dot{Y}$ and $\dot{Z}$, respectively.
    We then divide all the terms of the equations above by the coefficient of the corresponding leading term and obtain the following monic input-output equations 
    \[
    \dot{Y} - \beta Y + \frac{\beta K}{N_{pop}}Y^2 +  \frac{\beta K}{N_{pop}} YZ + rY = 0 \qquad \text{and} \qquad  \dot{Z} - rY = 0. 
    \]
    Since we have more than one output, before concluding on the identifiability of the system we have to check
    that the two equations are mutually reduced; if not, the analysis could lead to spurious results, as pointed out in \cite{eisenberg-diff-alg}.
    The concept of reduction is again based on the chosen ranking of the variables:
    a polynomial $\mathcal{P}_i$ is reduced with respect to the polynomial $\mathcal{P}_j$
    if it does not contain neither the leading monomial of $\mathcal{P}_j$ with equal or greater degree nor its derivatives.
    In our case it can be easily seen that the input-output equations are mutually reduced,
    and by looking at their coefficients we conclude that $\beta$, $r$ and $K$ can be simultaneously identified.
    Hence, the SIR model is structurally identifiable from prevalence data of $I$ and $R$.

    \lorenzo{We close this example mentioning that \cite{Tuncer} discusses structural
      identifiability of SIR (with no under-reporting factor) in the more realistic scenario
      where cumulative incidence data rather than prevalence are available.
      The result obtained is that SIR is structurally identifiable even from such kind of data
      (but other problems that we will discuss in Example \ref{ex:SIR_identifiability} occur, so that one should
      restrain from attempting parameter identification of SIR from cumulative incidence data).}
  \end{example}

\subsection{Mapping approach}

  This approach is discussed in \cite{Evans,Evans2}; we refer the reader interested to the theoretical background
  to these two references and only sketch the main idea and the ``practical recipe'' here.
  Example \ref{ex:structural_identifiability_SIR_evans} gives an example of the application of this method to the SIR model.
  For ease of notation, in this discussion the vector $\pp \in \Rset^p$ collects all the uncertain elements of (\ref{eq:ode_system}),
  i.e., not just the coefficients, but also the initial conditions and the hyper-parameters, i.e. $p = N_{coef} + N_{states} + N_{hyp}$.
  Moreover, we change the notation in (\ref{eq:ode_system}) to a more compact form and write
  $f^{\pp}(X)$ instead of $f(X,\pp)$ and similarly for $\qoi$, and let $N_{states}=n, N_{qoi}=m$. Summarizing the new notation,
  (\ref{eq:ode_system}) becomes
  \begin{equation} \label{eq:ode_system_evans}
    \begin{cases}
      \dot{X}(t,\pp) = f^{\pp}(X(t,\pp)) \\
      X(0,\pp) = X_0(\pp)\\
      Y(t,\pp) = \qoi^{\pp}(X(t,\pp)), 
    \end{cases}
  \end{equation}
  with $X \in \Rset^n$,  $Y \in \Rset^m$, $f^{\pp}:\Rset^n \rightarrow \Rset^n$,
  $\qoi^{\pp}:\Rset^n \rightarrow \Rset^m$, $\forall t \in [0,t_{max}]$, and $X_0:\Rset^p \rightarrow \Rset^n$.

\begin{figure}[tbp]
    \centering
    \includegraphics[width=0.8\linewidth]{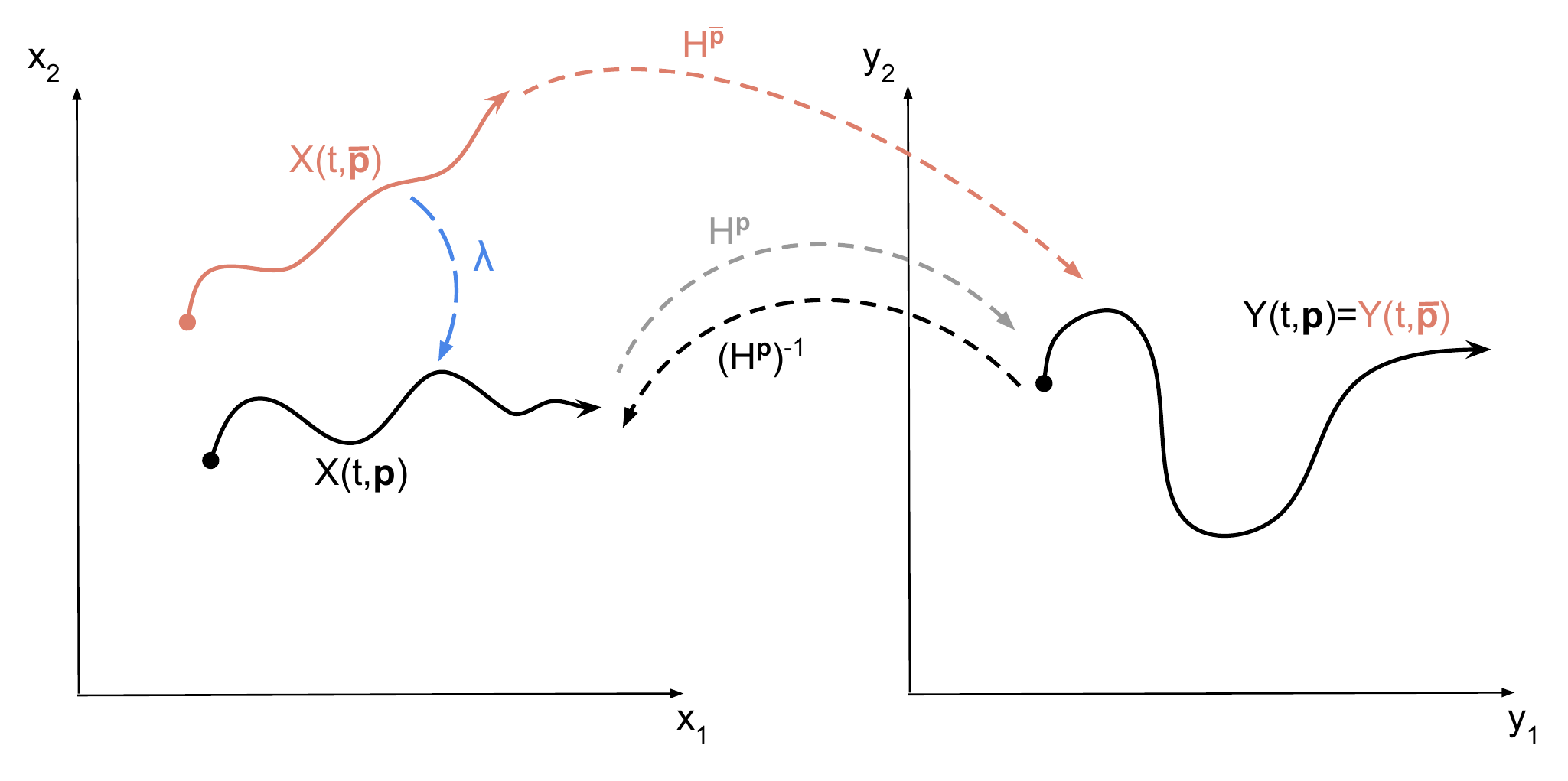}
    \caption{The mapping approach for structural identifiability.}\label{fig:evans}
  \end{figure}

  We explain the method with the support of Figure \ref{fig:evans}.
  If the system (\ref{eq:ode_system_evans}) is not identifiable, 
  then there are two different sets of parameters, say $\pp$ and $\mybar{\pp}$, such that
  the trajectories $X(t,\pp)$ and $X(t,\mybar{\pp})$ are different but the corresponding
  outputs are identical, $Y(t,\pp) = Y(t,\mybar{\pp})$ $\forall t \in [0,t_{max}]$.
  If these conditions hold true, then it is possible to rework the equality $Y(t,\pp) = Y(t,\mybar{\pp})$
  to explicitly construct a map $\lambda: \Rset^n \rightarrow \Rset^n$
  that maps the trajectories $X(t,\pp)$ and $X(t,\mybar{\pp})$ to one another: $\lambda(X(\mybar{\pp}))=X(\pp)$ (we will come back to this point with more details
  later on). 
  
  Since the map $\lambda$ has been derived by enforcing equality between the outputs $Y(t,\pp)$ and $Y(t,\mybar{\pp})$, without taking into account
  the dynamics of the system, one has to further check that $\lambda(X(\mybar{\pp}))$ still solves the dynamical system. 
  Taking time derivatives of both sides of $\lambda(X(\mybar{\pp}))=X(\pp)$ and using (\ref{eq:ode_system_evans})
  one gets (removing the dependence on $t$ for ease of notation):
  \begin{align}
    \lambda(X(\mybar{\pp}))& =X(\pp) \nonumber \\
    \nabla \lambda \big\vert_{X(\mybar{\pp})} \dot{X}(\mybar{\pp}) & = \dot{X}(\pp) \nonumber \\
    \nabla \lambda \big\vert_{X(\mybar{\pp})} f^{\mybar{\pp}}(X(\mybar{\pp})) & = f^{\pp}(X(\pp)) \nonumber \\
    \nabla \lambda \big\vert_{X(\mybar{\pp})} f^{\mybar{\pp}}(X(\mybar{\pp})) & = f^{\pp}(\lambda(X(\mybar{\pp}))). \label{eq:lambda_cond_evans}
  \end{align}
  Therefore, one has to check that the last equation, i.e. equation \eqref{eq:lambda_cond_evans} is valid for the proposed $\lambda$. This will result in a set of conditions for the components of $\pp$ and $\mybar{\pp}$: if the resulting conditions are $\pp = \mybar{\pp}$ the system is identifiable;
  otherwise, there will be non-trivial conditions between some of the components of $\pp$ and some of the components of $\mybar{\pp}$,
  (see e.g. Example \ref{ex:structural_identifiability_SIR_evans}), which means that the system is not identifiable.

  We now come back to the issue of constructing the map $\lambda$.
  As already mentioned, the idea is to construct $\lambda$ from  the condition $Y(\mybar{\pp}) = Y(\pp)$, i.e. from
  \[
    \qoi^{\mybar{\pp}}(X(\mybar{\pp})) = \qoi^{\pp}(X(\pp)) = \qoi^{\pp}(\lambda(X(\mybar{\pp})).
  \]
  In principle, it would be enough to solve for $\lambda$ in the latter, i.e. $\lambda(X)=  (\qoi^{\pp})^{-1} [\qoi^{\mybar{\pp}}(X)]$,
  but the system might be underdetermined if we have $m<n$ observables.
  Then, the idea is to complement the observables with additional equations and to create an ``augmented'' observables vector
  $H^{\pp}: \Rset^n \rightarrow \Rset^n$, as follows, by taking into account the 
  directional derivatives of the outputs $Y_j$ along the direction $f^{\pp}(X)$:
  \begin{align*}
    H^{\pp}_1 & = Y_1 = \qoi_1^{\pp}(X(\pp)) \\
    H^{\pp}_2 & = Y_2 = \qoi_2^{\pp}(X(\pp)) \\[-4pt]
              & \ldots \\
    H^{\pp}_m & = Y_m = \qoi_m^{\pp}(X(\pp)) \\[6pt]
    H^{\pp}_{m+1} & = \nabla_X Y_1 \cdot f^{\pp}(X) \\
    H^{\pp}_{m+2} & = \nabla_X Y_2 \cdot f^{\pp}(X) \\
    \ldots \\
    H^{\pp}_{n} & = \nabla_X Y_{n-m} \cdot f^{\pp}(X).
  \end{align*}
  The next step is to verify that $H^{\pp}$ is a bijective map (this condition is named Observability Rank Condition, ORC),
  i.e. that the Jacobian of $H^{\pp}$ with respect to $X$ is non-singular
  in the domain of definition of the trajectories $X$ for every admissible $\pp$.
  Finally, the map $\lambda$ can be computed by solving for $\lambda$ the system of equations:
  \[
    H^{\mybar{\pp}}(X(\mybar{\pp})),\mybar{\pp}) = H^{\pp}(\lambda(X(\mybar{\pp})),\pp),
  \]
  leading to (cf. Figure \ref{fig:evans})
  \[
   \lambda(X) = (H^{\pp})^{-1} H^{\mybar{\pp}}(X),
  \]
  where the well-posedness of $(H^{\pp})^{-1}$ is guaranteed by the ORC.
  We close this discussion with a couple of remarks on the construction of $H^{\pp}$:
  \begin{itemize}
  \item In case the number of observables $m<n/2$, adding the directional derivatives $\nabla_X Y_j \cdot f^{\pp}(X),  j=1,\ldots m$
    will not be enough to reach $n$ observables. In this case, one should also add derivatives of higher order.
  \item In the opposite case, in which $2m>n$, one can choose among multiple directional derivatives $\nabla_X Y_j \cdot f^{\pp}(X)$:
    after having dropped the choices that make the mapping $H^{\pp}$ singular in the domain of definition of the trajectories $X$,
    one should check all remaining combinations (the fact that one choice results in identifiability in this case
    does not rule out the possibility that another choice might result in non-identifiability). 
    If $m>n$, then there are more observables than states, and the same principle applies.
  \end{itemize}    
  
  \begin{example}[Structural identifiability of a SIR model by mapping approach]\label{ex:structural_identifiability_SIR_evans}
    Let us consider the SIR model with output $y = \frac{1}{K} I$ as in Example \ref{ex:structural_identifiability_SIR},
    and rewrite it replacing $S,I$ as $X=[x_1,x_2]$:
    \[
      \begin{cases}
        \displaystyle \dot{x_1} = -\frac{\beta}{N_{pop}} x_1 x_2\\[6pt]
        \displaystyle \dot{x_2} = \frac{\beta }{N_{pop}}x_1 x_2 - r x_2\\[6pt]
        \displaystyle y = \frac{1}{K} x_2.   
      \end{cases}
      \quad
      \Rightarrow
      \quad
      f^{\pp}(x_1,x_2) =
      \left(
        \begin{array}{c}
          \displaystyle -\frac{\beta}{N_{pop}} x_1x_2 \\
          \displaystyle  \frac{\beta}{N_{pop}} x_1x_2 - r x_2
        \end{array}
      \right).
    \]
    Our goal is to recover the structural identifiability results already obtained in Example \ref{ex:structural_identifiability_SIR}
    with the differential algebra approach.
    The first step is to build the two maps $H^{\pp}$ and $\lambda$,  where $\pp = [\beta,r,K,N_{pop}]$.
    As for $H^{\pp} = (H^{\pp}_1(x_1,x_2),\, H^{\pp}_2(x_1,x_2))^T$, we need to augment the observable $y$ with a directional derivative:
    \[
    \begin{cases}
      \displaystyle H^{\pp}_1 = y = \frac{1}{K}x_2\,, \\[4pt]
      \displaystyle H^{\pp}_2 = \nabla_X y \cdot f^{\pp}(X) = \frac{1}{K} \frac{\beta}{N_{pop}} x_1 x_2 - \frac{1}{K}r x_2 \,.
    \end{cases}
    \]
    The Jacobian of this mapping has non-zero determinant whenever $x_2 \neq 0$, which is a value never attained by the trajectories $X$
    unless the initial condition is $x_2(0)=0$ (in which case the trajectory is the uninteresting case $X=[0,0]^T$). Therefore, we can apply the methodology.
    The mapping $\lambda = (\lambda_1(x_1,x_2),\,\lambda_2(x_1,x_2))^T$ is obtained by solving the equation $H^{\pp}(\lambda(X)) = H^{\mybar{\pp}}(X)$, i.e.
    \[
      \begin{cases}
        \displaystyle \frac{1}{K} \lambda_2 = \frac{1}{\mybar{K}} x_2\,, \\[6pt]
        \displaystyle \frac{1}{K} \frac{\beta}{N_{pop}} \lambda_1 \lambda_2-\frac{1}{K}r \lambda_2 = \frac{1}{\mybar{K}} \frac{\mybar{\beta}}{\mybar{N}_{pop}} x_1x_2 - \frac{1}{\mybar{K}}\mybar{r} x_2\,, 
      \end{cases}
      \Rightarrow
      \begin{cases}
        \displaystyle \lambda_1 = \frac{N_{pop}}{\beta} \left( \frac{\mybar{\beta}}{\mybar{N}_{pop}} x_1 - \mybar{r} + r  \right)\,, \\[8pt]
        \displaystyle \lambda_2 = \frac{K}{\mybar{K}} x_2\,, 
      \end{cases}      
    \]
    whose Jacobian is
    \[
      \nabla_x \lambda =
      \left[
        \begin{array}{cc}
           \displaystyle \frac{N_{pop} \mybar{\beta}}{\mybar{N}_{pop} \beta} & 0 \\[8pt]
          0 &  \displaystyle\frac{K}{\mybar{K}}
        \end{array}
      \right].
    \]
    Then, enforcing condition \eqref{eq:lambda_cond_evans} results in the following equations:
    \[
      \begin{cases}
         \displaystyle - \frac{N_{pop} \mybar{\beta}}{\mybar{N}_{pop} \beta} \frac{\mybar{\beta}}{\mybar{N}_{pop}} x_1(\mybar{\pp})x_2(\mybar{\pp}) =
        -\frac{\beta}{N_{pop}} \frac{N_{pop}}{\beta} \left( \frac{\mybar{\beta}}{\mybar{N}_{pop}} x_1(\mybar{\pp}) - \mybar{r} + r \right) \frac{K}{\mybar{K}} x_2(\mybar{\pp})\, , \\[10pt]
         \displaystyle \frac{K}{\mybar{K}} \left(\frac{\beta}{\mybar{N}_{pop}} x_1(\mybar{\pp}) x_2(\mybar{\pp}) - \mybar{r} x_2(\mybar{\pp})\right) =
        \frac{\beta}{N_{pop}} \frac{N_{pop}}{\beta} \left( \frac{\mybar{\beta}}{\mybar{N}_{pop}} x_1(\pp) - \mybar{r} + r \right) \frac{K}{\mybar{K}} x_2(\mybar{\pp}) - r \frac{K}{\mybar{K}}x_2({\mybar{\pp}})\,.
      \end{cases}
    \]
    The second equation is identically verified. Conversely, the first one holds true for every $x_1,x_2$ if
    \[
      r = \mybar{r}, \quad \frac{K \beta}{N_{pop}} =  \frac{\mybar{K} \mybar{\beta}}{\mybar{N}_{pop}},
    \]
    i.e. we obtain the same non-trivial condition previously obtained in Example \ref{ex:structural_identifiability_SIR}: 
    at most one parameter out of $K, \beta, N_{pop}$ can be identified.
    If both observations of $I$ and $R$ are available instead, then we have two observables
    \[
      \begin{cases}
        \displaystyle y_1 = \frac{1}{K} x_2\,, \\[8pt]
        \displaystyle y_2 = \frac{1}{K} \left(N_{pop}-x_1 -x_2 \right),
      \end{cases}
    \]
    and the analysis must be repeated. In particular,  the map $H^{\pp}$ can now be constructed without using directional derivatives, as
    \[
    \begin{cases}
      \displaystyle H^{\pp}_1 = \frac{1}{K}x_2 \,, \\[8pt]
      \displaystyle H^{\pp}_2 = \frac{1}{K}\left(N_{pop} - x_1 - x_2\right).
    \end{cases}
    \]
    This map is linear, therefore it is bijective for every $X$ and we can carry on with the analysis.
    The mapping $\lambda$ is obtained by solving the equations $H^{\pp}(\lambda(X)) = H^{\mybar{\pp}}(X)$, resulting in
    \[
      \begin{cases}
        \displaystyle \lambda_1 = \frac{K}{\mybar{K}} x_1 + N_{pop} \left( 1 - \frac{K}{\mybar{K}} \right), \\[8pt]
        \displaystyle \lambda_2 = \frac{K}{\mybar{K}} x_2,
      \end{cases}
    \]
    whose Jacobian is $\frac{K}{\mybar{K}}$ times the identity matrix.
    Then, enforcing condition \eqref{eq:lambda_cond_evans} results in the following equations:
     \[
       \begin{cases}
         \displaystyle
        -\frac{K}{\mybar{K}} \frac{\mybar{\beta}}{\mybar{N}_{pop}} x_1(\mybar{\pp}) x_2(\mybar{\pp})
        = -\frac{\beta}{N_{pop}} \left[ \frac{K}{\mybar{K}} x_1(\pp) + N_{pop}\left(1 - \frac{K}{\mybar{K}} \right) \right] \frac{K}{\mybar{K}} x_2(\mybar{\pp}),\\[8pt]
         \displaystyle
          -\frac{K}{\mybar{K}} \left( \frac{\mybar{\beta}}{\mybar{N}_{pop}} x_1(\mybar{\pp}) x_2(\mybar{\pp}) - \mybar{r} x_2(\mybar{\pp}) \right)
        = \frac{\beta}{N_{pop}} \left[ \frac{K}{\mybar{K}} x_1(\pp) + N_{pop}\left(1 - \frac{K}{\mybar{K}} \right) \right] \frac{K}{\mybar{K}} x_2(\mybar{\pp}) - r \frac{K}{\mybar{K}}x_2(\pp), 
      \end{cases}
    \]
    which result in the conditions $r = \mybar{r}$, $K = \mybar{K}$, $\displaystyle \frac{\beta}{N_{pop}} = \frac{\mybar{\beta}}{\mybar{N}_{pop}}$,
    i.e., the system is now structurally identifiable if we assume $N_{pop}$ to be known, as already discussed in Example \ref{sect:struct_identifiability}.
  \end{example}

  \subsection{Sensitivity analysis}

  As already mentioned in Section \ref{sect:sobol}, if an output /  quantity of interest $Y$ is only weakly influenced by a parameter,
  such parameter might be non-identifiable from $Y$: therefore, computing the sensitivity of a quantity to a parameter
  gives an indication about the structural identifiability of the parameter.
  As explained in Section \ref{sect:sobol}, sensitivity analysis can be local or global.
  In the first case, a small gradient of the quantity of interest value indicates
  that the quantity of interest is not very sensible to small variations in the parameter, which might then be non-identifiable.
  The global sensitivity analysis can be performed by means of Sobol indices: parameters
  with low Sobol index do not strongly influence the quantity of interest $Y$, and might be non-identifiable from observations of $Y$.

\begin{biblio}
  \begin{itemize}
  \item The concept of structural identifiability was first introduced in \cite{Bellman}.
  \item Some theoretical background on the differential algebra approach is provided in \cite{Miao,eisenberg-diff-alg}. 
  \item Many methods we have not directly mentioned here (e.g. Taylor series approach, generating series approach, and methods based on the implicit function theorem) are explained in \cite{Miao}.
  \item See \cite{Tuncer} for the detailed discussion of structural identifiability of a SEIR model by differential algebra.
    \lorenzo{As already mentioned,} \cite{Tuncer} also shows that the SIR model is structurally identifiable also
    in the case of {cumulative incidence data}.
  \item \lorenzo{Further discussion on the solvability condition for the differential algebra approach can be found in \cite{ovchinnikov2,ovchinnikov3}.}
  \item Sensitivity analysis is discussed in, e.g., \cite{Capaldi} and \cite{Miao}, where several methods based on the Jacobian matrix $J$ and on the Fisher matrix $H$
are discussed, see equations (\ref{eq:Sigma_MLE}) and (\ref{eq:hessian.dropped}). 
  \end{itemize}	
\end{biblio}

\section{Practical identifiability}\label{section:pract_id}

Upon having assessed the structural identifiability of a system, it is still not obvious that the system can be identified
from limited, noisy data, that possibly cover only a fraction of the time-span of the dynamics
(e.g., when one measures the initial part of a trajectory and wants to assess the parameters for long-term forecast).
With reference to Figure \ref{fig:evans}, if the output trajectories $Y(\pp),Y(\mybar{\pp})$
are distinct but close, they might become indistinguishable from one another if we only have at our disposal a noisy cloud
of points around them rather than the entire exact trajectories.
The study of this setting is called \emph{practical identifiability analysis}, and is typically
performed on synthetic data 
to see under which conditions the inversion procedure obtains results ``close enough''
to the true values of the parameters.
In the following, we give a short outlook on the main tools; see the bibliography at the end of the section for further readings on each method.

\begin{description}
\item[Monte Carlo simulations/bootstrap:] generate $M$ sets of synthetic data and 
  for each data set, compute the MLE $\ttheta^{(k)}$, $k=1,\ldots,M$.
  Then, compute dispersion indices for the $M$ MLE of each parameter, such as their sample variance or their average relative error, which is defined as 
  \[
    \text{ARE}(\vartheta_i) = \frac{1}{M} \sum_{k=1}^{M} \frac{\lvert \vartheta_i^{(k)} -\vartheta_{true,i} \rvert}{\vartheta_{true,i}}.
  \]
  Repeat the procedure for synthetic data with increasing levels of noise and observe the trend of the dispersion indices
  as the noise increases. If the ARE of the estimates is e.g. higher than the noise level, the parameters are not-identifiable
  (other criteria in the same spirit might be used as well).

  The same procedure should be repeated by considering each time a different number of parameters
  to be jointly estimated, until all the parameters are included.
  In this way, it is possible to detect the influence of each parameter on the quality of the joint estimate.

  Finally, note that the amount and the time-span of the available data might be relevant for practical identifiability,
  and the analysis should ideally be repeated while varying these settings. 
  For instance, \cite{Tuncer} discusses the bootstraps analysis for the SIR model computing
  the ARE of parameter estimates using data sets of increasing time-span: the results indicate 
  that the model with unknown $\beta,r$ is practically identifiable from data on the $I$ compartment only after that the epidemic peak is reached
  (despite the fact the model is structurally identifiable regardless of the time-span of the data,  see Example \ref{ex:structural_identifiability_SIR}).

\item[Fisher Information Matrix:] as already explained in Section \ref{sect:posterior}, the Fisher Information Matrix
  is the Hessian of the NLL at the maximum log-likelihood estimate (or its approximation in Equation (\ref{eq:hessian.dropped})).
  If the Fisher Information Matrix is positive definite with large eigenvalues, the NLL has a narrow minimum 
  and we can conclude \emph{local} practical identifiability of the system,
  i.e. identifiability in a neighborhood of the maximum likelihood estimate;
  conversely, small eigenvalues are symptom  of practical non-identifiability,
  since the minimum of the NLL occurs in a shallow region.
  Numerically, the optimization procedure might even end up in a critical point where the Fisher Information Matrix is not even
  positive definite, e.g. it could be non-definite or it could have rank smaller than the number of parameters. In that case,
  the rank of Fisher Information Matrix gives an indication on the maximum number of parameters that can be simultaneously inferred, see e.g. \cite{eisenberg:FIM}.

\item[Correlation matrix $C$:] The inverse of the Fisher Information Matrix $H$ gives an approximation of the covariance matrix
  of the parameters, $\Sigma_G$, see Equation (\ref{eq:Sigma_MLE}). From $\Sigma_G$, it is possible to compute the \emph{correlation matrix}
  of the parameters, rescaling each entry as
  \[
    C_{ij} = \frac{\Sigma_{G,ij}}{\Sigma_{G,ii} \Sigma_{G,jj}}.
  \]
  If two parameters have correlation close to 1, they are linearly dependent, and cannot be estimated separately. Hence, they are practically non-identifiable.

\item[Optimization with multiple restart:]  As already mentioned in Section \ref{sect:posterior},
  one should repeat several times the optimization to determine the MLE for different initial guesses
  and check where the minimization ends.
  If the algorithm ends always at the same point, there is empirical evidence that the NLL has a unique minimum
  and the system is \emph{globally} practically identifiable (the minimum might be different from the nominal value of the parameter, i.e. the noise might introduce
  a bias in the estimate, see e.g. Example \ref{ex:inverseSIR}).
  If several local minima or a manifold of minima are detected we conclude that the system is not
  practically identifiable: in particular, in the case of a manifold of minima
  there are many sets of parameters that fit equally well the data, denoting a possible case of
  structural non-identifiability. 
  For instance, in Example \ref{ex:structural_identifiability_SIR} when only data of $I$ are considered,
  we expect that the region $K\beta=const$ will be a manifold of minimum points for the NLL.

\item[Profile log-likelihood for each parameter:] The previous discussion, pointing to the possibility that the NLL might have a
  manifold of minima, allows us to introduce the last tool to assess practical identifiability, i.e., the profile log-likelihood;
  this tool is actually ``in between'', and could also be used for structural identifiability, as it will be made clearer later on. 
  
  Whenever $\pparam$ is practically identifiable, the NLL should have a global minimum at $\pparam = \pparam_{true}$.
  Therefore, a visual inspection of the NLL can immediately tell whether the system is identifiable or not. Of course, this is not a viable solution
  for problems with more than two parameters. In this case, one can resort to the profile likelihood, which is a mono-dimensional slice of the log-likelihood
  function in the direction of the considered parameter $\vartheta_i$.
  The profile likelihood can be obtained by changing the parameter $\vartheta_i$ iteratively in a certain range of values 
  around its MLE value $\vartheta_{MLE,i}$, while reoptimizing all other parameters. Thus, the profile log-likelihood is defined as
  \[
    PL_i(\ttheta) = \min_{\{\mathbf{y} : y_i = \vartheta_{MLE,i}\}} NLL(\mathbf{y}).
  \]
  If the profile log-likelihood has a unique minimum, the parameter is practically identifiable, whereas
  more complex shapes, with shallow regions and multiple minima indicate that the parameter is practically non-identifiable. In particular, a flat profile means that the NLL has a manifold of minima that fit the data equally well and we can again conclude that the system is  
  structurally non-identifiable. For a schematic illustration we refer to Figure \ref{fig:profile-likelihood}.
  Crucially, if the NLL has a manifold of minima, the Fisher approximation of the posterior will be completely wrong,
  since it is based on the assumption that NLL has a unique minimum; the case of finitely many local minima could instead
  be fixed by acquiring more data, which should hopefully make the ``spurious peaks'' become smaller and smaller.
  
  In summary, the analysis of the profile log-likelihood is based on checking the ``flatness''/``shallowness'' of such function.
  To make this criterion more quantitative and discriminate between different levels of shallowness
  one can provide confidence thresholds of the profile log-likelihood $\text{CI}_{\text{PL}}$ using a $\chi_1^2$ test:
  \[
    \text{CI}_{\text{PL}}(y_i) = \{ \bf{y} | PL_i(\bf{y}) \leq PL(\vartheta_{MLE,i}) + \Delta_{\alpha}\chi_1^2\},
  \]
  where $\Delta_{\alpha}\chi_1^2$ is the $\alpha$ quantile of the $\chi_1^2$ distribution.
  If the likelihood profile of a parameter exceeds the confidence threshold $\Delta_{\alpha}\chi_1^2$ on both sides of $\pparam_{MLE,i}$ 
  the parameter is practically identifiable (the threshold is indicated by the black dotted lines in Figure \ref{fig:profile-likelihood}).
  This is an alternative approach to determine a posterior pdf for the MLE, consisting of a uniform pdf 
  instead of the Gaussian Fisher approximation discussed in Section \ref{sect:posterior}. Such uniform estimate might be more robust for limited datasets \cite{Raue1}.
  
  Finally, observe that if the data are generated in a noise-less way by sampling the trajectories of $\pparam=\pparam_{true}$,
  and they are sufficiently many, we end up in the scenario of structural identifiability.
  The profile log-likelihood can still be constructed and evaluated also in this case,
  as the NLL is just the sum of square misfits (without the $\frac{1}{\sigma^2}$ factor),
  and therefore it can be used as a computational tool to verify structural identifiability,
  i.e., for the existence of a manifold of minima.


\end{description} 

In case the system is structurally or practically non-identifiable, a possible workaround is to learn some of the parameters from independent studies,
thus reducing the number of parameters to be simultaneously identified. Sometimes even a hierarchical optimization approach might be effective:
in this approach, one does a first round of optimization to obtain the values of the entire set of parameters but retains the values obtained for the
identifiable parameters only. Upon fixing these parameters to the values just obtained, the optimization of the remaining parameters
can be repeated. An alternative is to reparametrize the model, replacing the original parameters with the combination that can be identified, see e.g. \cite{Tonsing}.
Finally, one could resort to so-called \emph{marginalization techniques}, see e.g. \cite{Iglesias,Kolehmainen,ruggeri2017}.

\begin{figure}[h]
	\centering
	\includegraphics[scale=0.6]{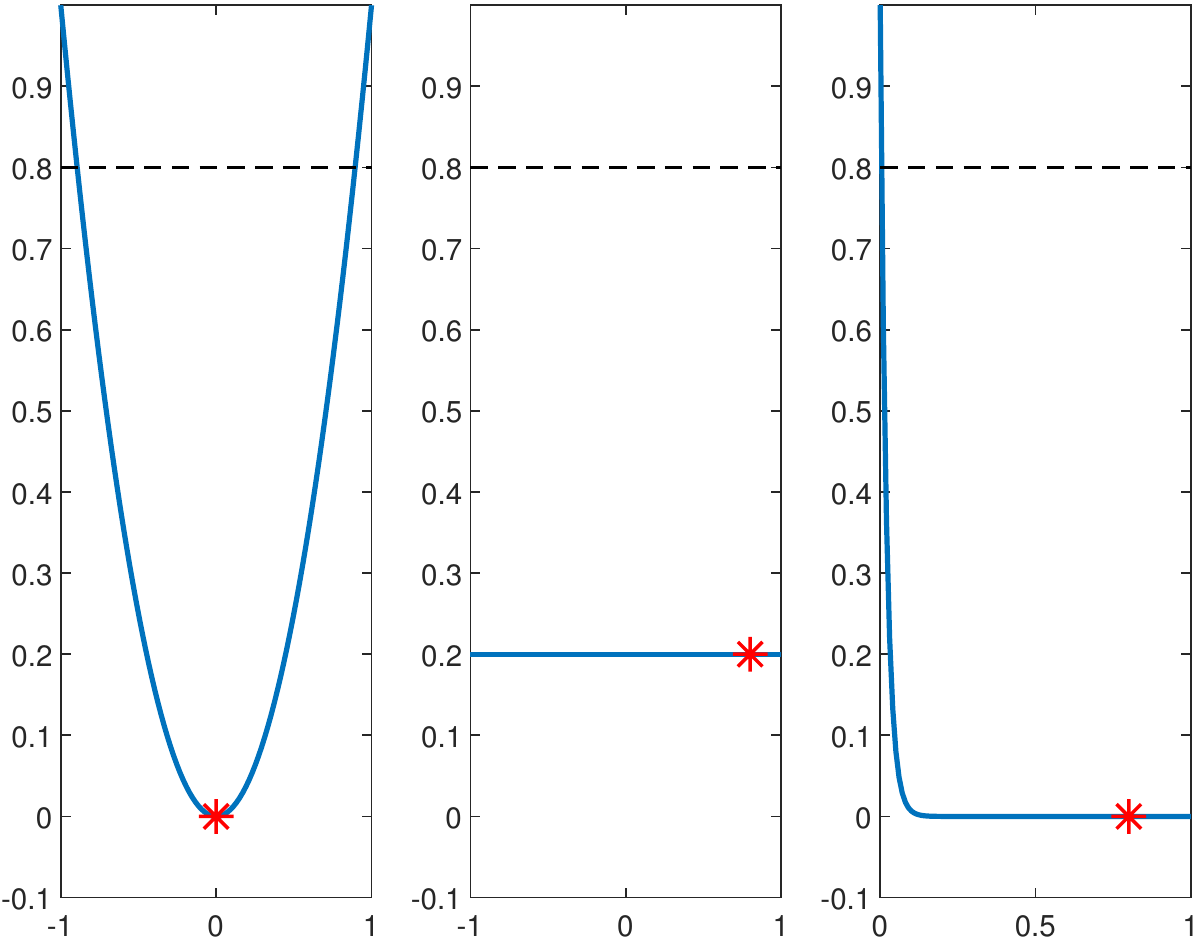}
	\caption{Three cases of profile likelihood (blue line), with nominal value of the parameter (red star), and confidence threshold
          $\Delta_{\alpha}\chi_1^2$ (dashed line ). Left: the parameter is practically identifiable;
          center: the parameter is not structurally identifiable; right: the parameter is practically non-identifiable. }
	\label{fig:profile-likelihood}
\end{figure}

\begin{example}[Practical non-identifiability of a SIR model with unknown under-reporting factor]\label{ex:SIR_identifiability}

  \begin{figure}[tbp]
    \centering
    \includegraphics[scale=0.6]{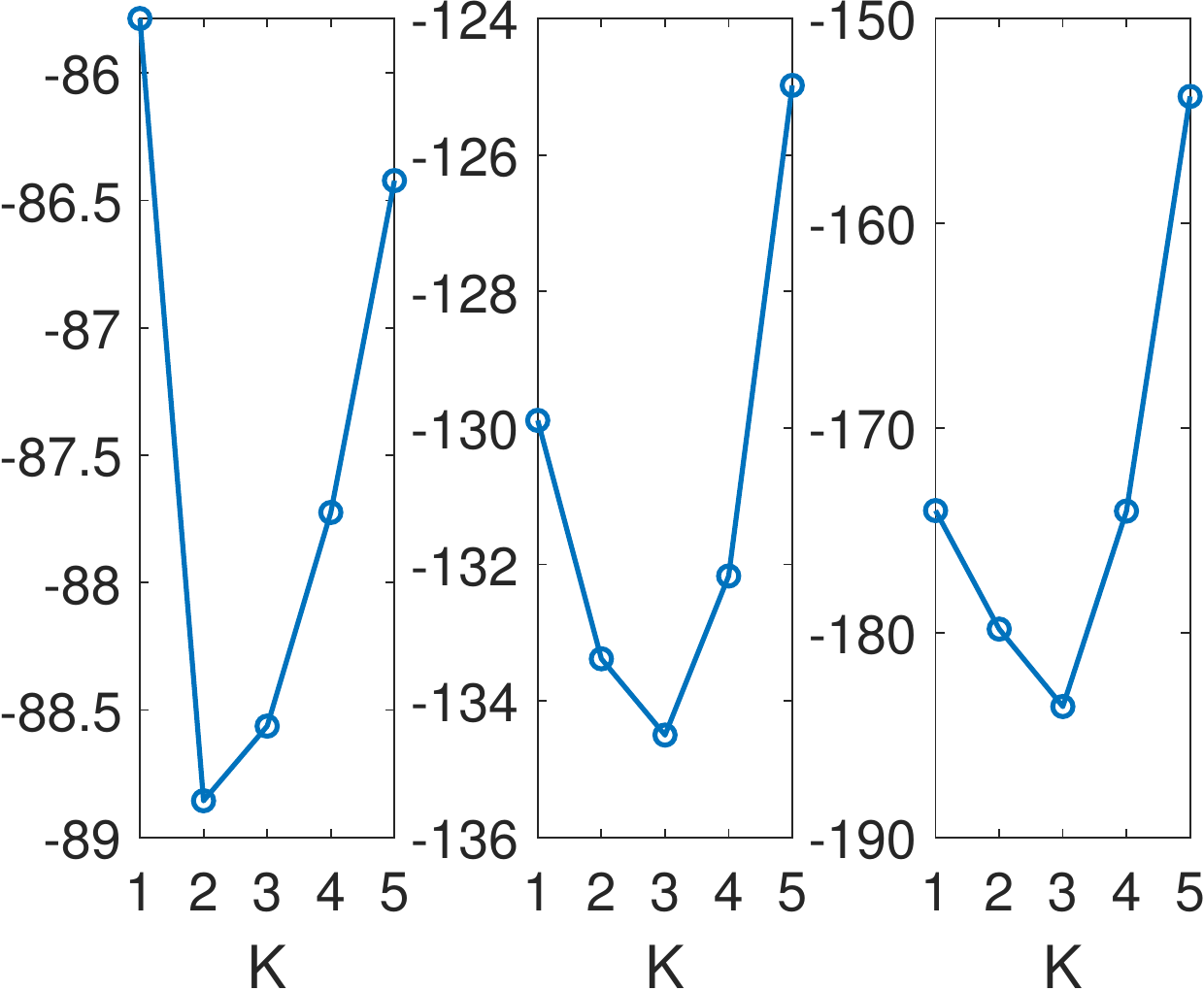}
    \caption{Profile Likelihood for $K$ for the three different scenarios, i.e. data collection ending:
      before peak of $I$ (left), around the peak of $I$ (center), after the peak of $I$ (right).}\label{fig:profile-likelihood-K}
  \end{figure}
  \begin{figure}[tbp]
    \centering
    \includegraphics[width=0.25\linewidth]{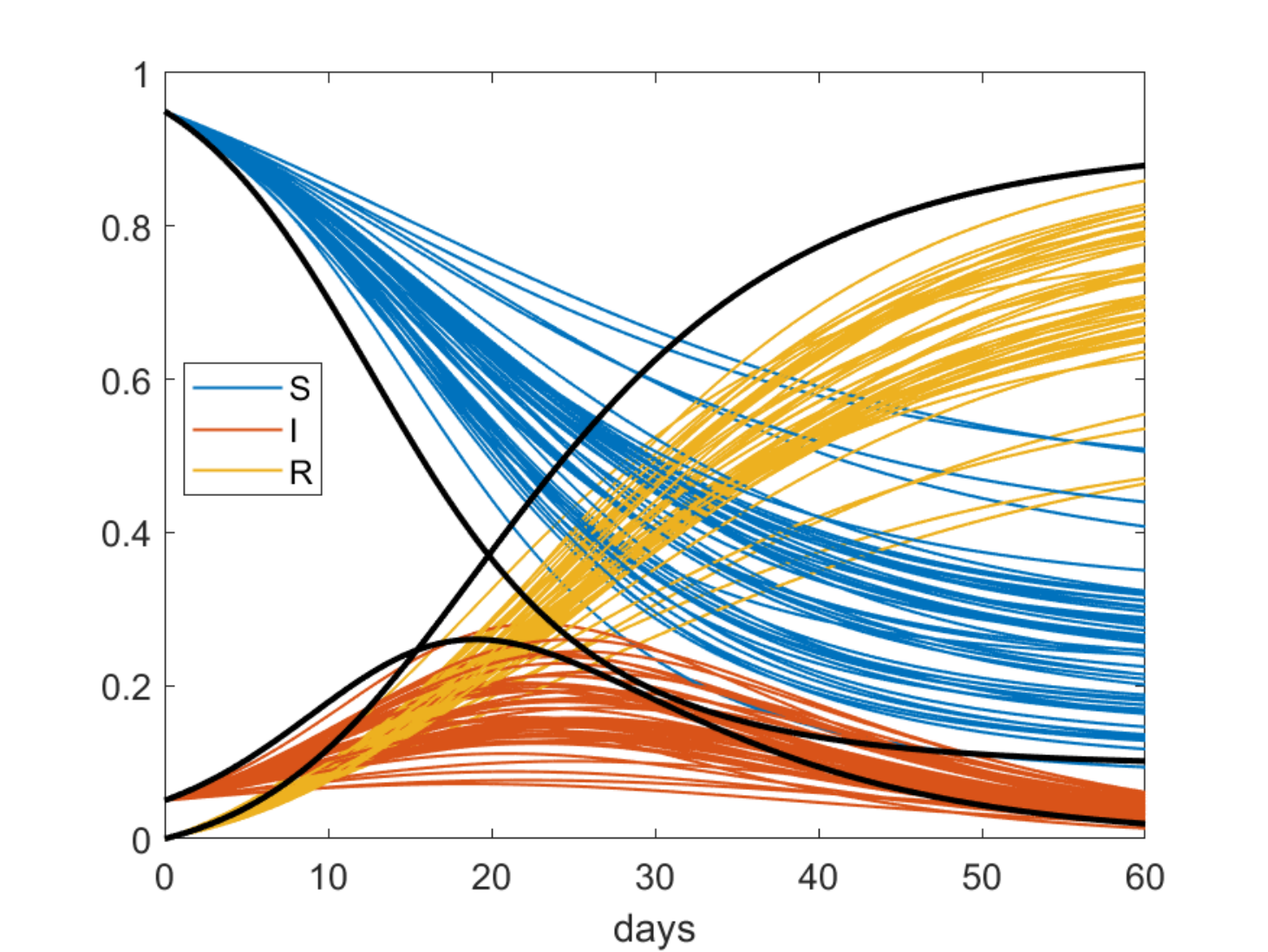}\hfill
    \includegraphics[width=0.22\linewidth]{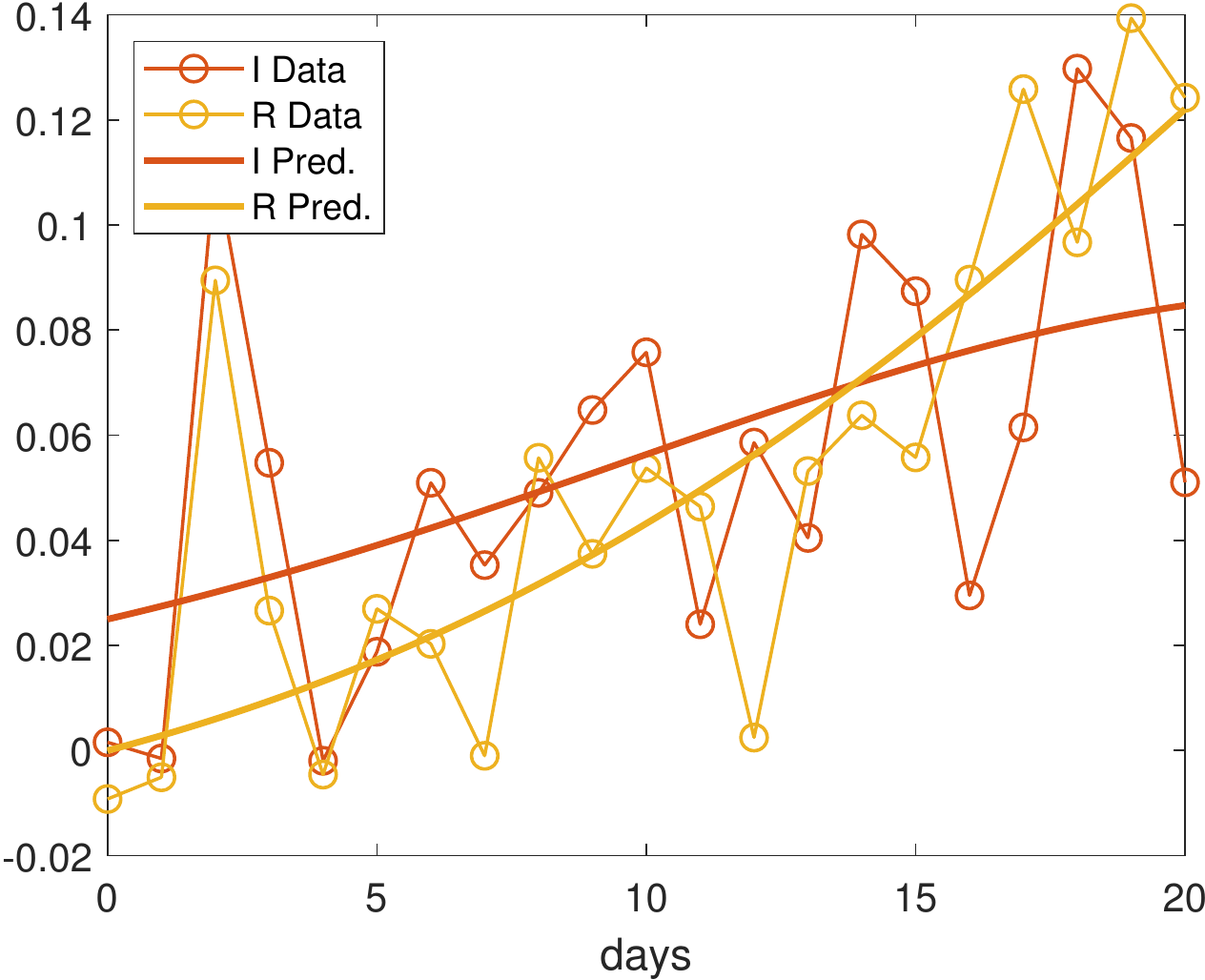}\hfill
    \includegraphics[width=0.22\linewidth]{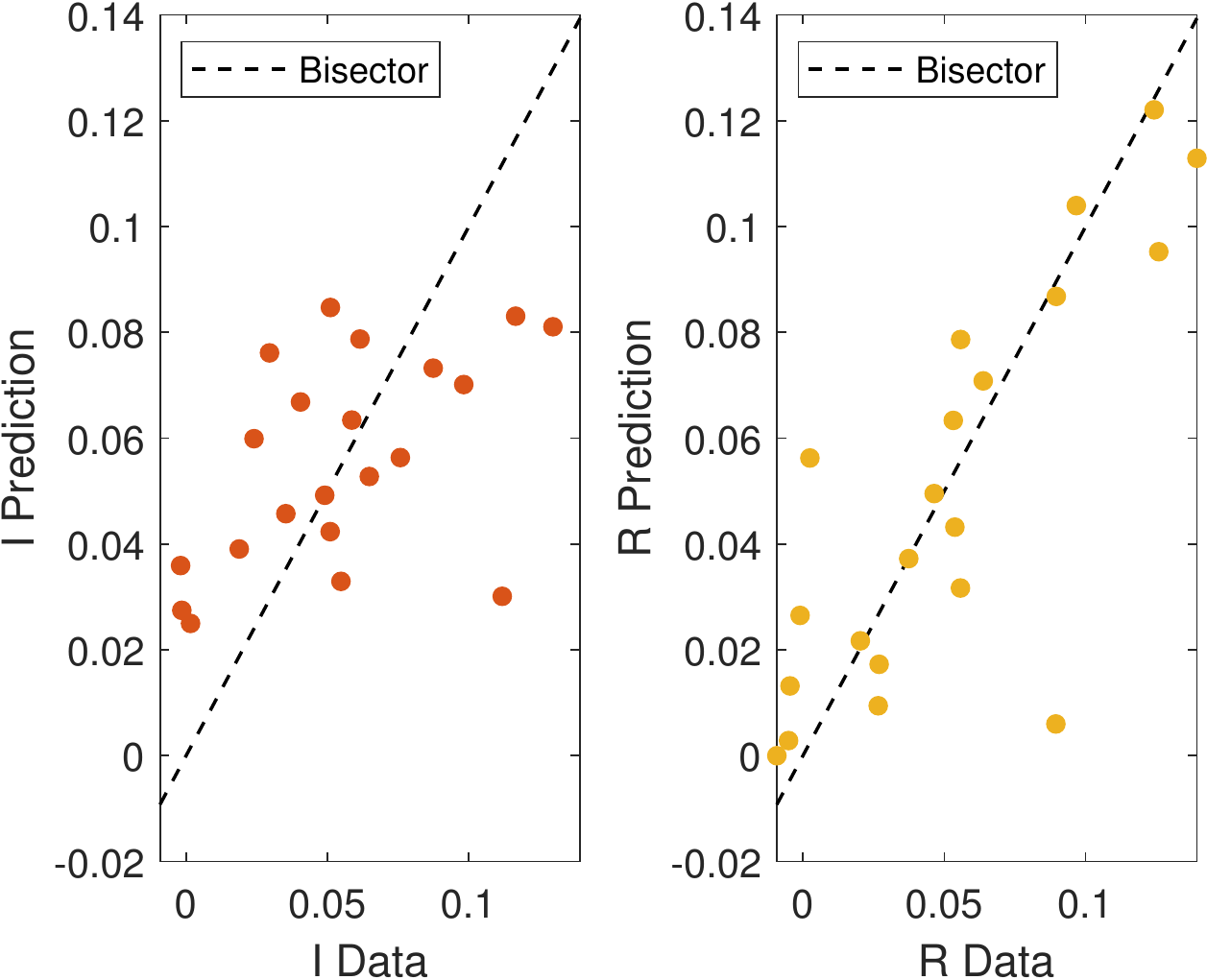}\hfill
    \includegraphics[width=0.22\linewidth]{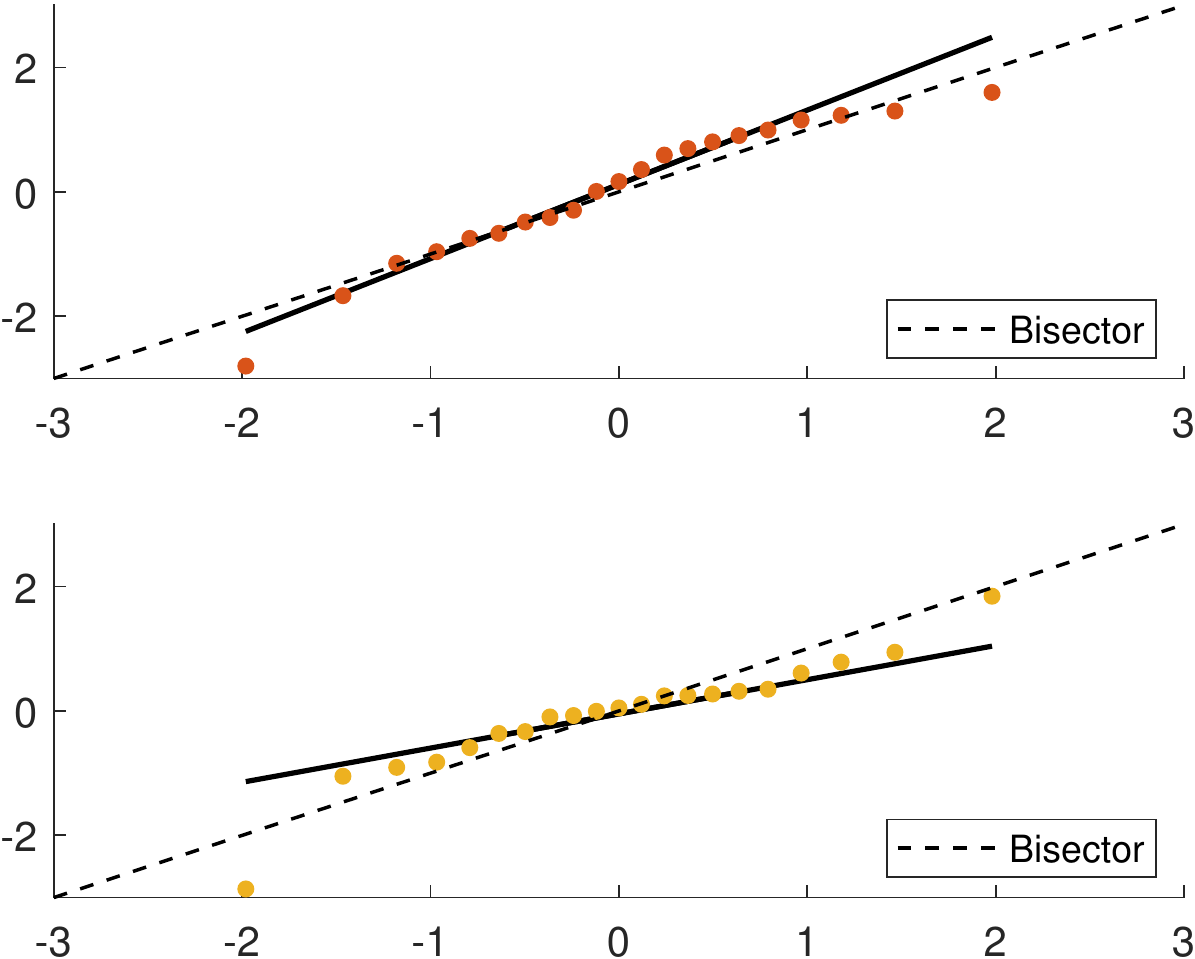} \\
    \includegraphics[width=0.25\linewidth]{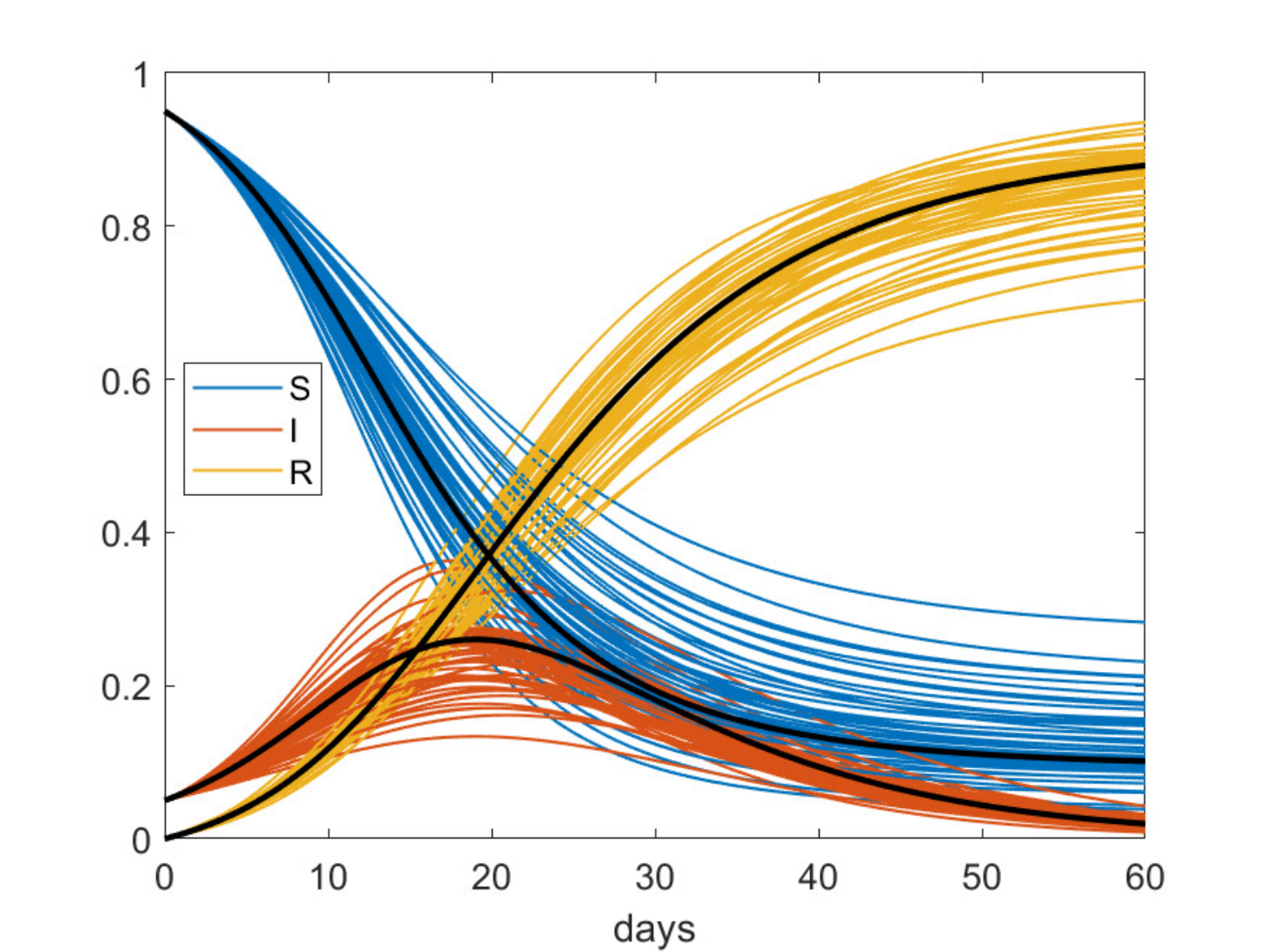}\hfill
    \includegraphics[width=0.22\linewidth]{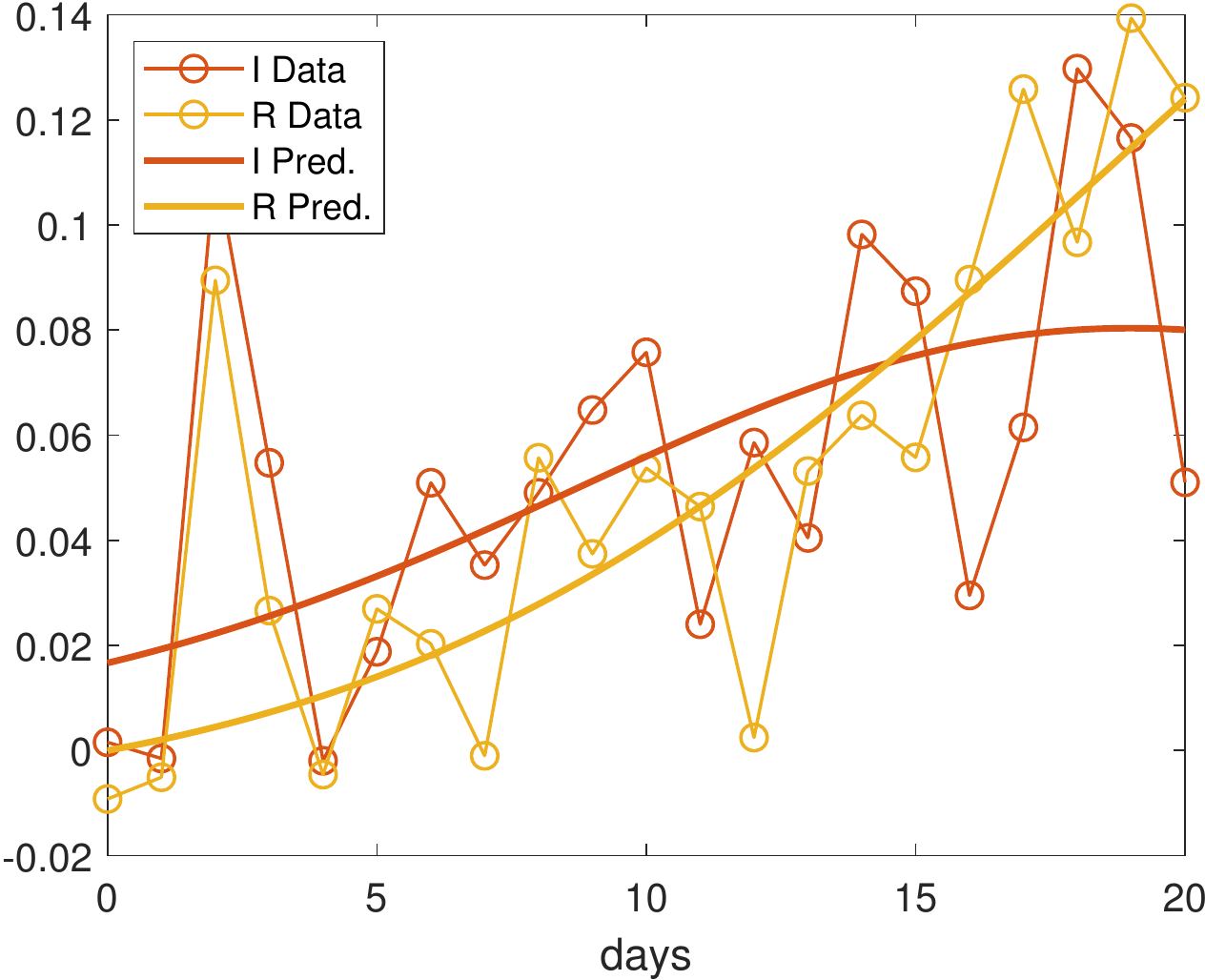}\hfill  
    \includegraphics[width=0.22\linewidth]{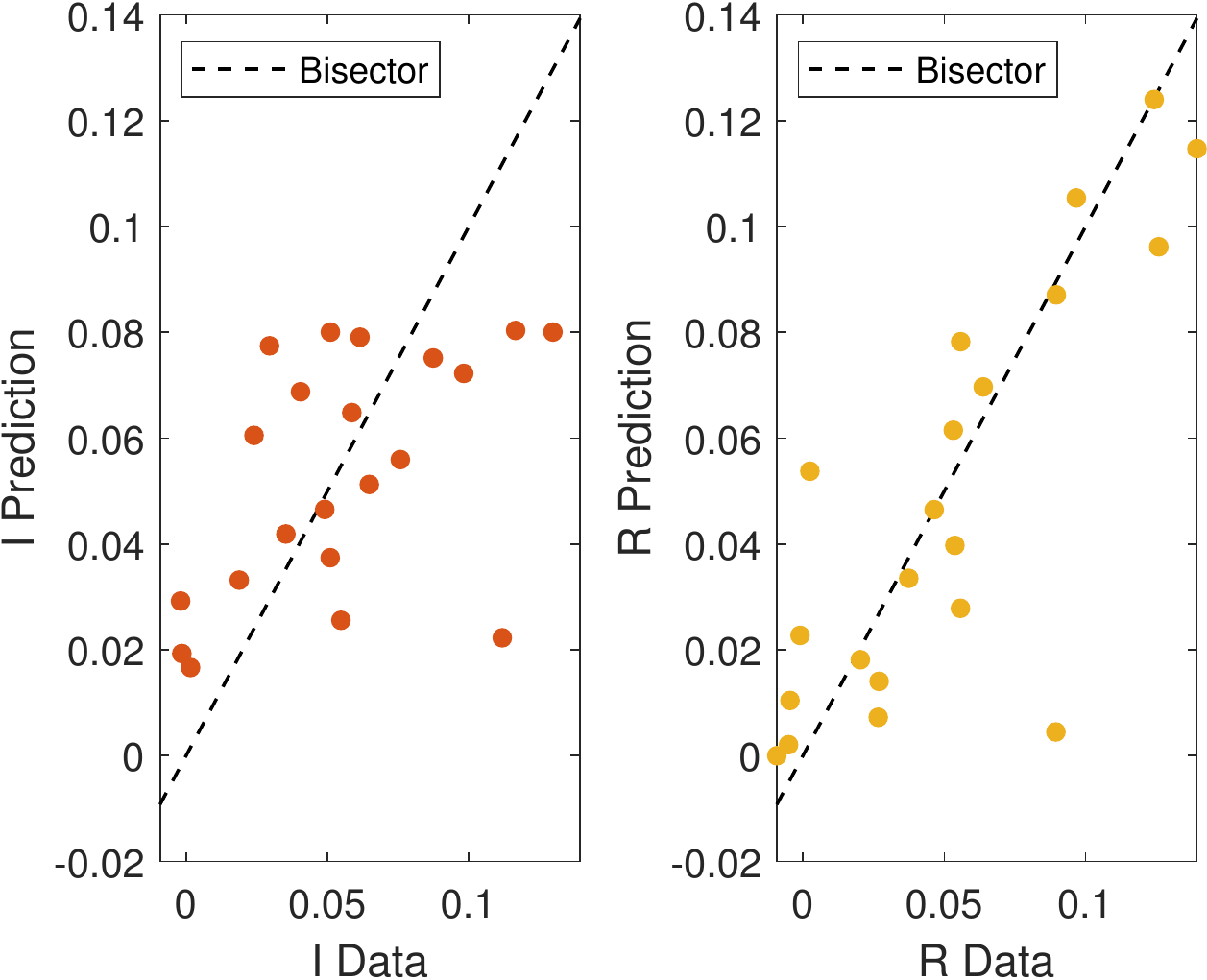}\hfill    
    \includegraphics[width=0.22\linewidth]{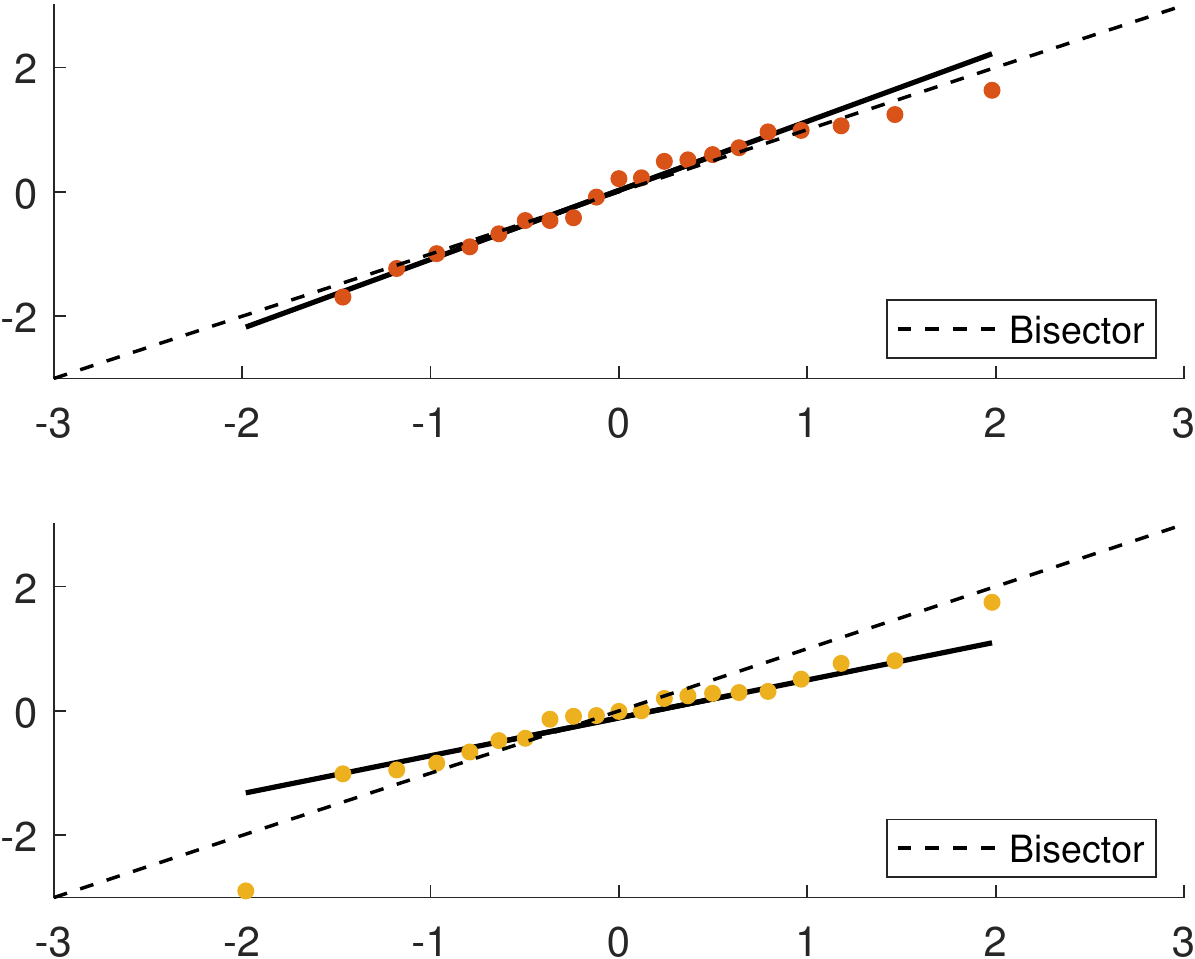}
    \caption{Comparison of quality of predictions and data fitting obtained for $K=2$ (top) and $K=3$ (bottom) in the scenario of data available only before the peak ($T=20$). From left to right: forward UQ based on the posterior pdf; the fitting of the data in the two cases; scatterplot of data vs predictions; qqplot of misfits. 
    }\label{fig:forward-UQ-K}
  \end{figure}
  \begin{figure}
    \centering
    \includegraphics[width=0.25\linewidth]{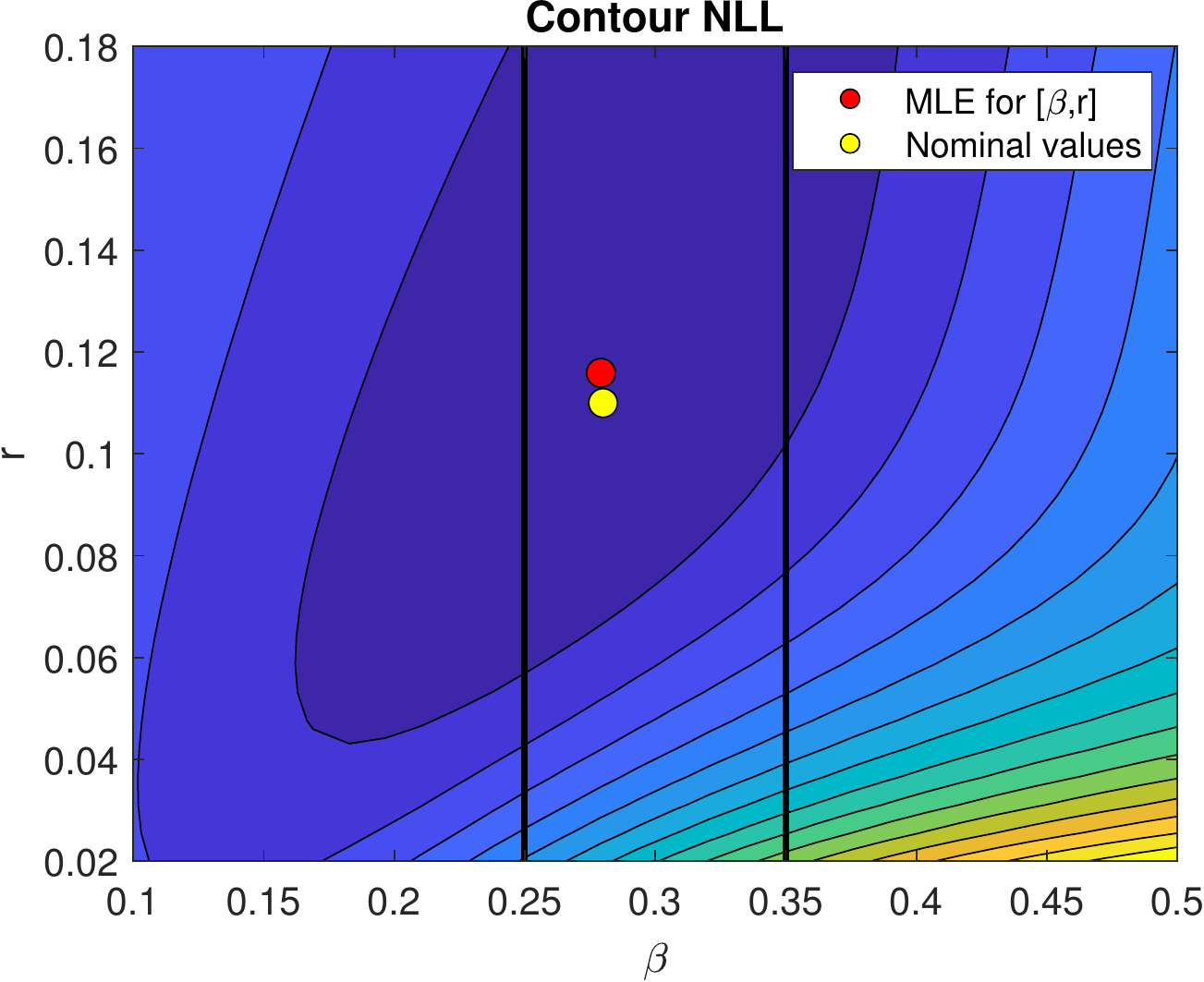}
    \includegraphics[width=0.25\linewidth]{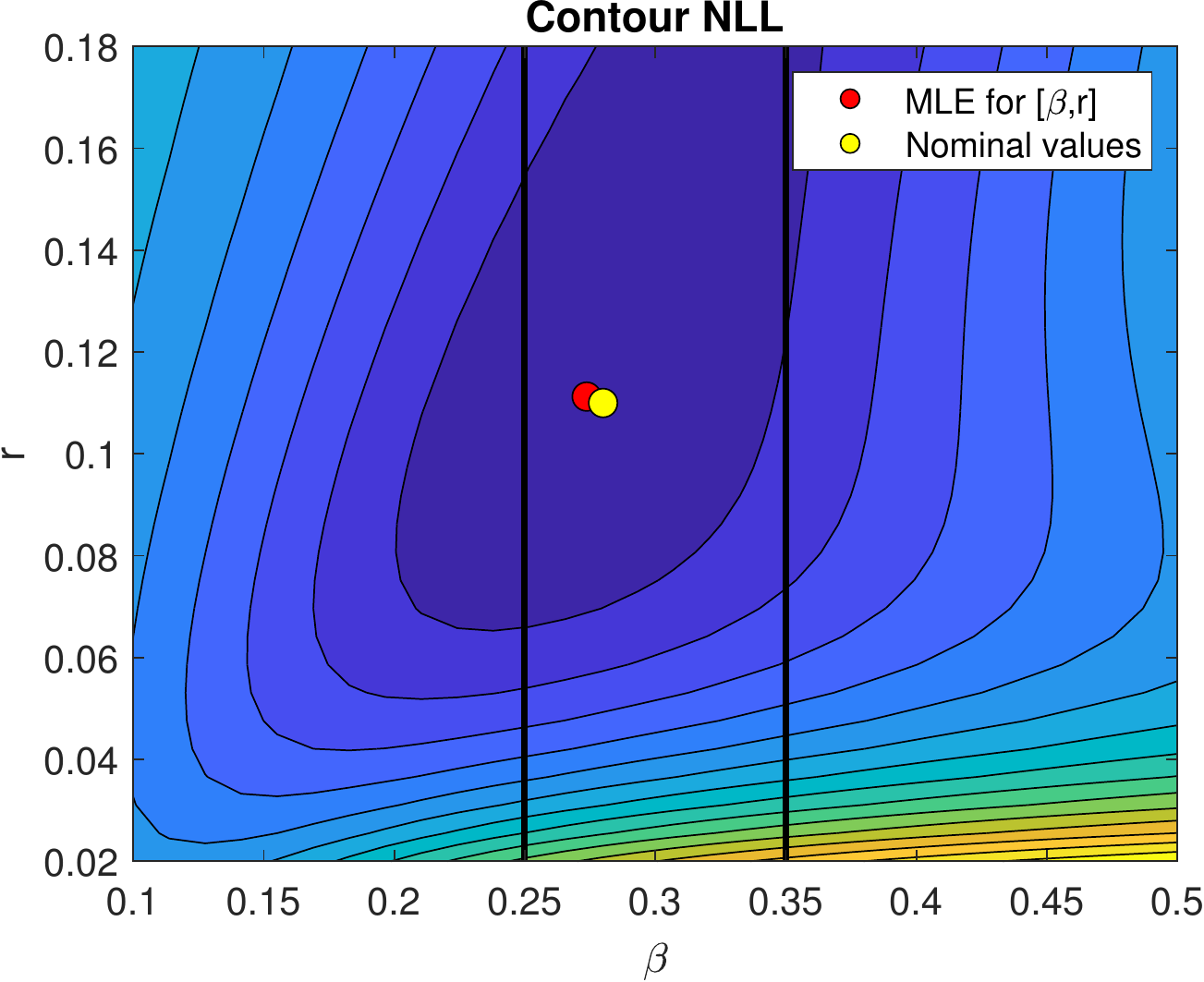}
    \includegraphics[width=0.25\linewidth]{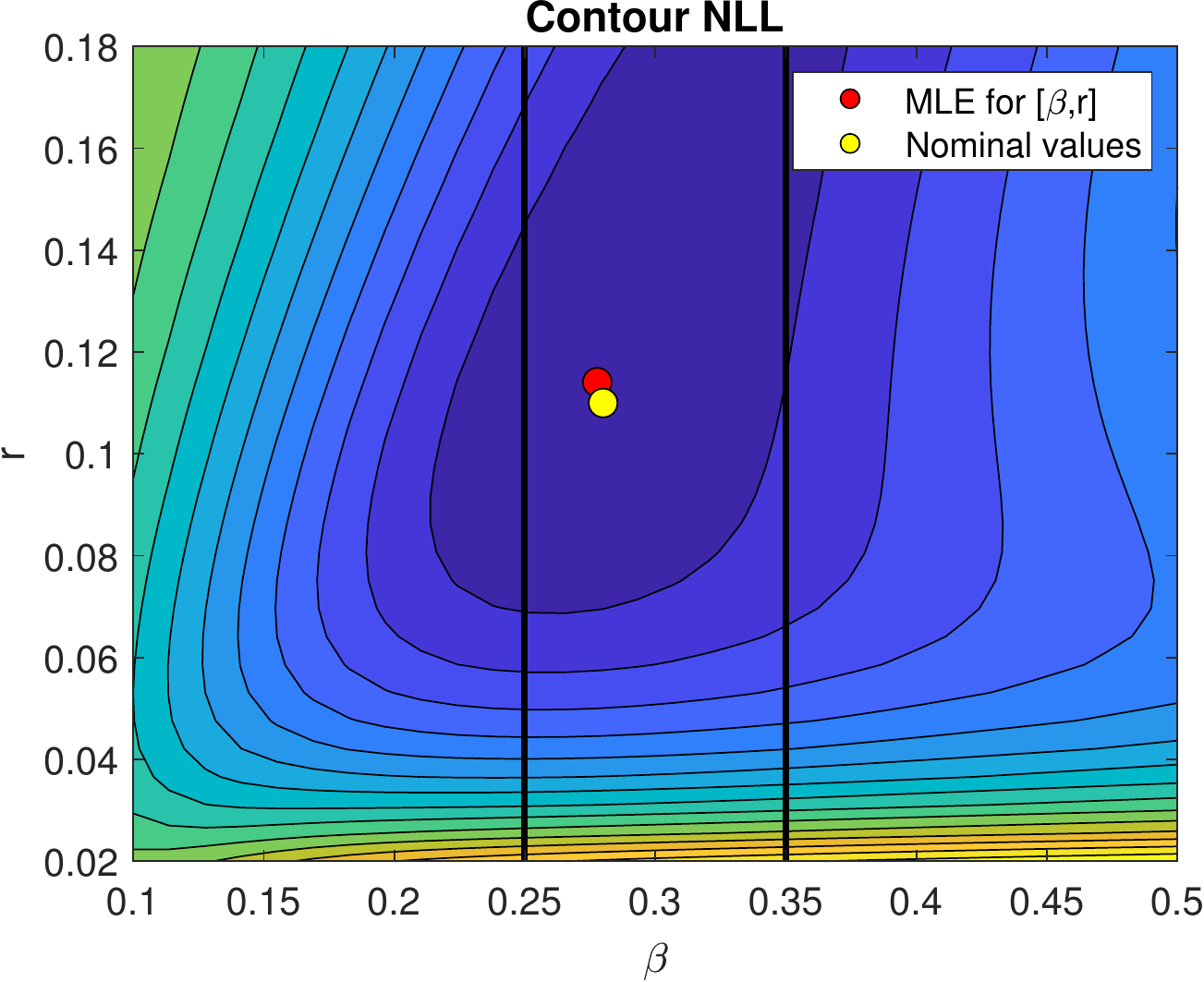}\\
    \includegraphics[width=0.32\linewidth]{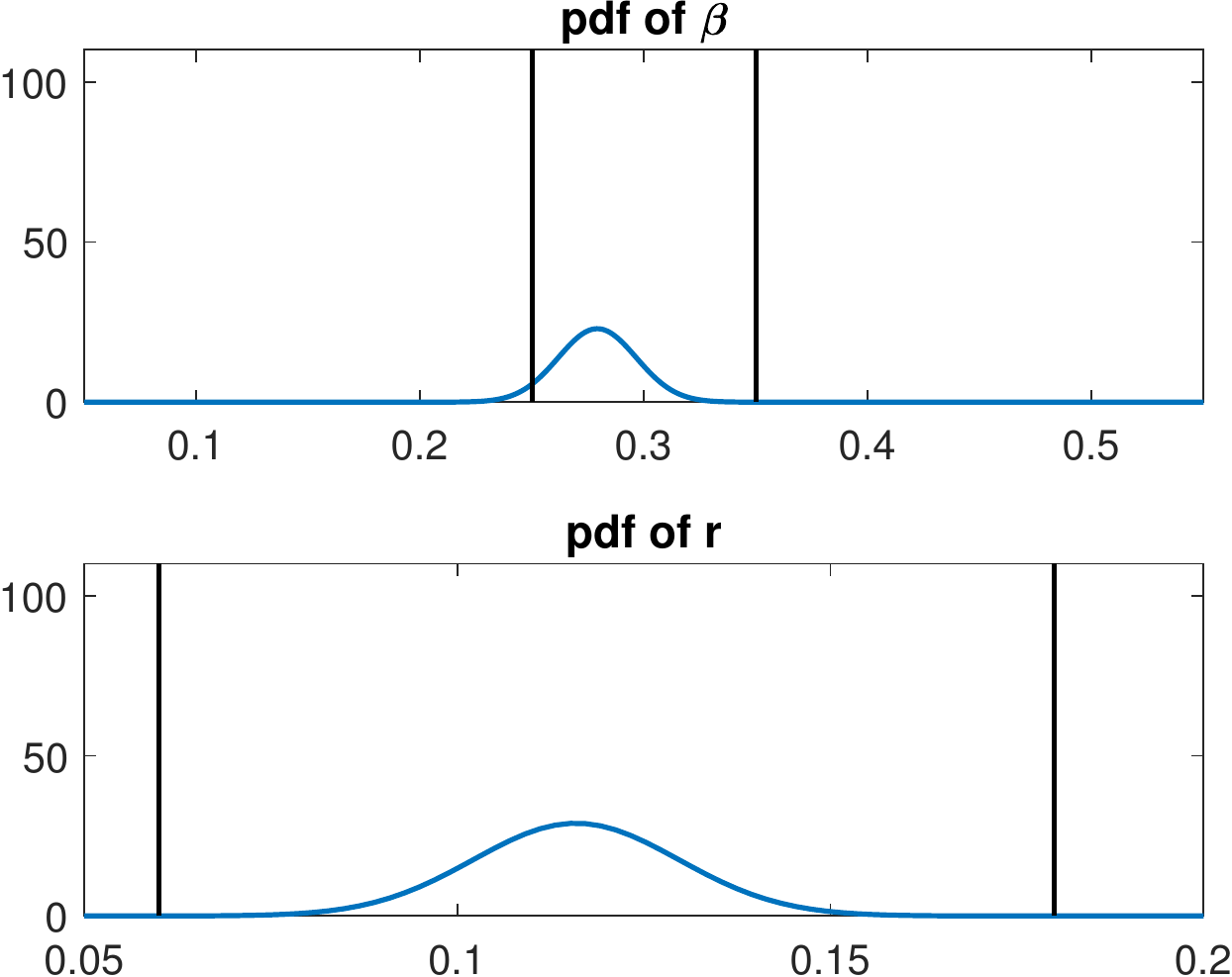}
    \includegraphics[width=0.32\linewidth]{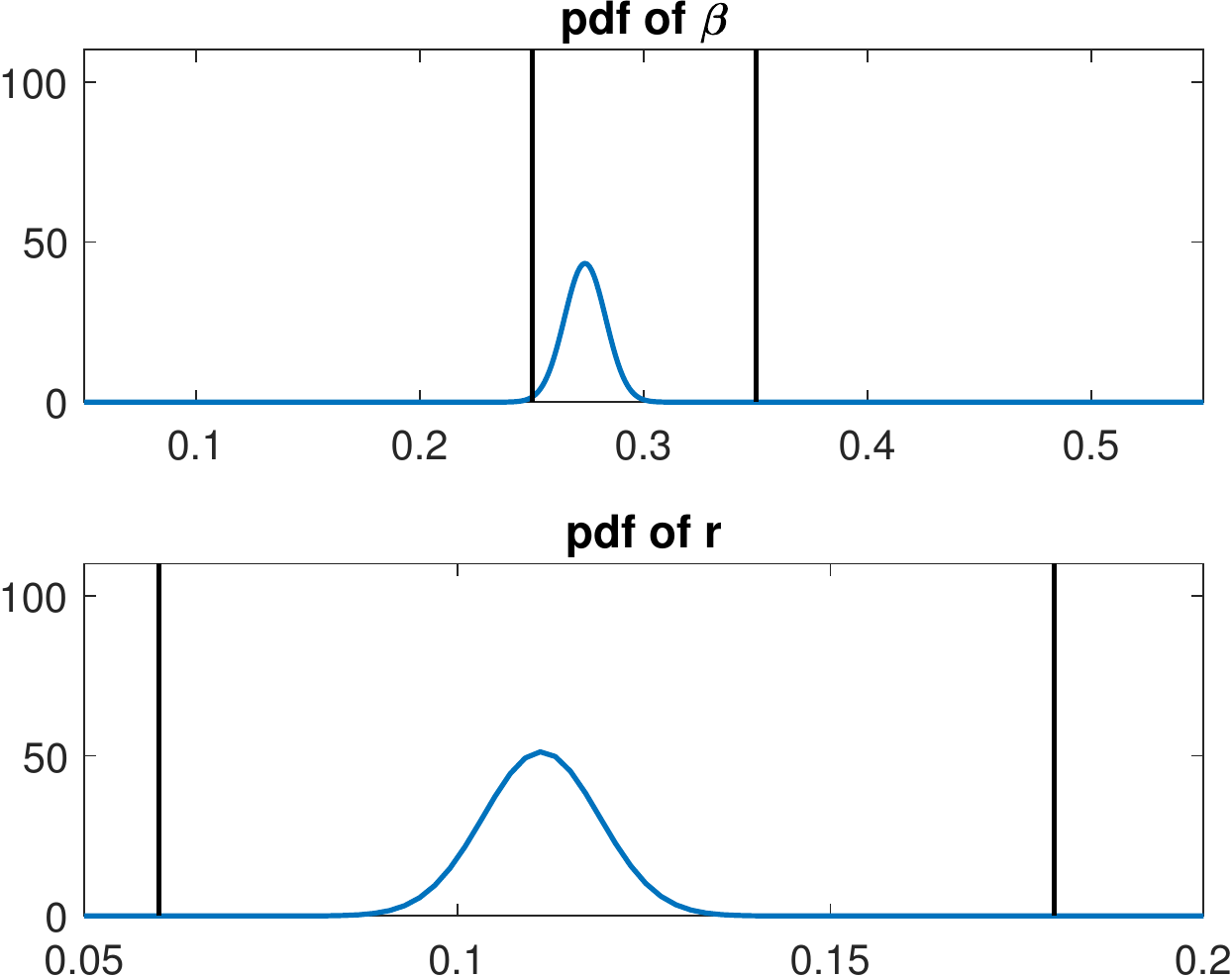}
    \includegraphics[width=0.32\linewidth]{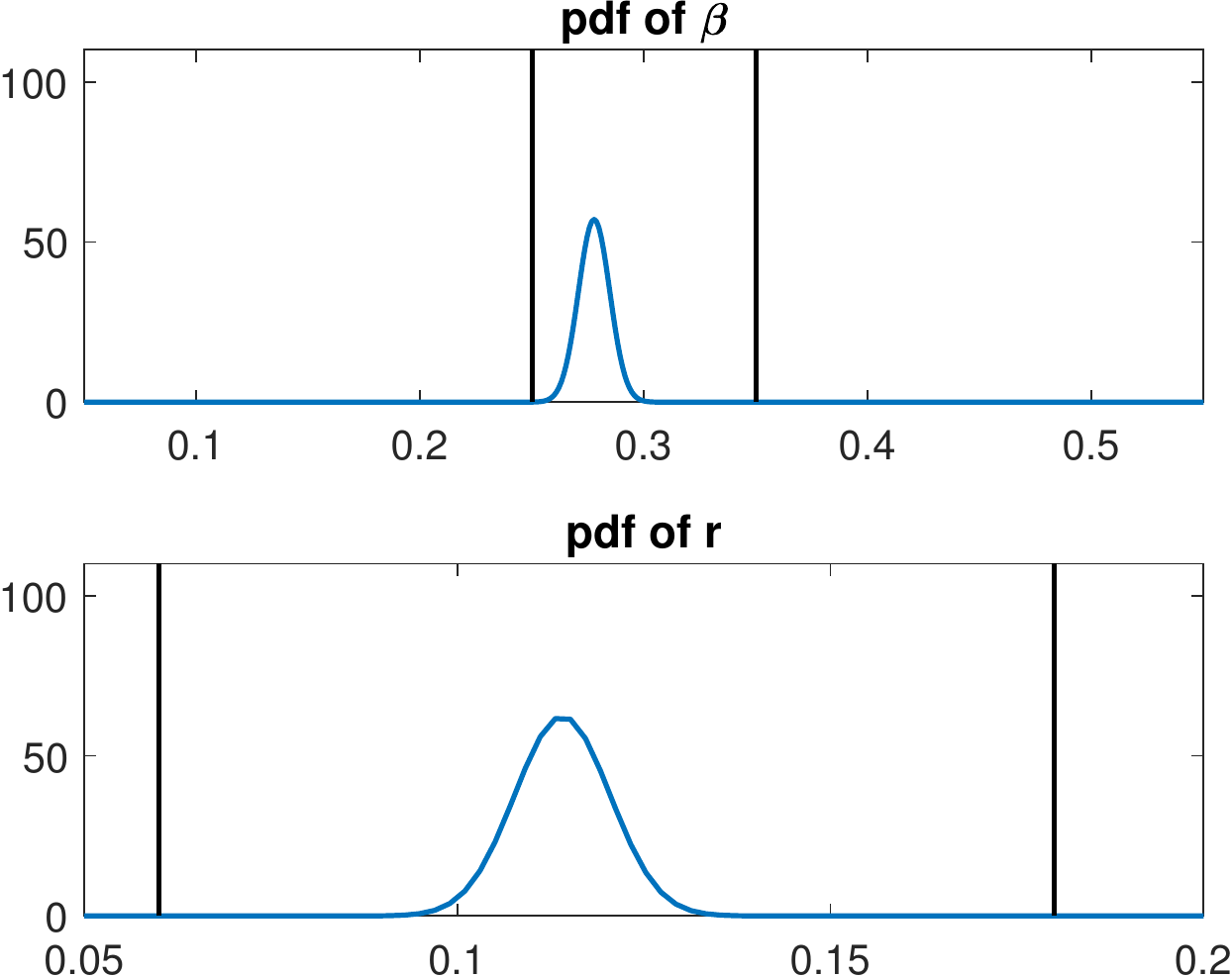}
    \caption{Results of inversion for the three cases $T=20,30,40$ for $K=3$.
      Top row: contours of the NLL in the three cases (left to right);
      the true value of the parameters is the yellow dot, the MLE is the red dot. The NLL contour lines suggest that the NLL has a unique minimum at the MLE,
      which is always close to the true value of the parameters. 
      Bottom rows: Gaussian approximation of the posterior pdfs of $\beta,r$ for the three cases $T=20,30,40$ for $K=3$.
    }\label{fig:identifiability_of_SIR_for_known_K}
  \end{figure}
  
  In the scenario of COVID-19, it has been often pointed out that data of infected and dead persons have been
  under-reported (even significantly), but the exact value of the under-reporting factor $K$ is not known.
  Some discussion on this aspect is provided in \cite{Russo,Gatto}.
  In this example, we consider a SIR model, generate synthetic data and investigate the practical identifiability
  of $\beta,r,K$ when measurements of $I$ and $R$ are considered, by computing the profile likelihood for $K$;
  we already know from Example \ref{ex:structural_identifiability_SIR}
  that in this scenario  \lorenzo{(somehow unrealistic, yet already rich enough for exposition purposes)}
  the parameters are structurally identifiable.
  We repeat practical identifiability analysis in three settings, that differ by the time-span covered by the data:
  until past the peak of $I$ ($T=20$), up to the peak of $I$ ($T=30$), and before the peak of $I$ ($T=40$).

  We fix $\pparam = [0.28,0.11]$, $\sigma=0.025$, and the discount factor to $K=3$. The results are reported in Figure \ref{fig:profile-likelihood-K}
  and suggest that for data before peak the profile likelihood of $K$ is shallow (look at the scale on the vertical axis)
  and even has a minimum at the wrong value $K=2$, denoting practical non-identifiability of $K$.
  For longer collection times instead, the profile likelihood
  shows an increasingly deep minimum around the correct value $K=3$. The fact that time might be important
  in determining whether a system is practically identifiable was already discussed in \cite{Capaldi,Chowell,Tuncer}.
  Fixing $K$ to the wrong value to perform the forward UQ analysis of course leads to predictions
  that are far from the true behavior of the system, see Figure \ref{fig:forward-UQ-K}-top-left. An important
  disclaimer to do here is that it would be hard to say that the set of parameters obtained for $K=2$ fits the
  data worse than those obtained for $K=3$, see Figure \ref{fig:forward-UQ-K}-top and bottom-second figure.
  Even using quantitative criteria to evaluate the goodness of fit of the two fittings, such as
  \begin{description}
    \item[Root Mean Square Error (RMSE)]: $\frac{1}{N_{meas}}\sqrt{\sum_{m=1}^{N_{meas}}\left( F_m(\ttheta_{MLE}) - \hat{F}_m \right)^2}$, for $F=I,R$
    \item[Mean Absolute Error (MAE)]: $ \frac{1}{N_{meas}}\sum_{m=1}^{N_{meas}}\lvert F_m(\ttheta_{MLE}) - \hat{F}_m \rvert$, for $F=I,R$
    \item[Mean Absolute Percentage Error (MAPE)]: $ \frac{1}{N_{meas}}\sum_{m=1}^{N_{meas}}\left( F_m(\ttheta_{MLE}) - \hat{F}_m \right)/\hat{F}_m$, for $F=I,R$.
  \end{description}
  would actually tell that the fitting of the case $K=2$ is slightly better than the case $K=3$ (numbers not reported for brevity).         
  The scatterplots of predictions vs data are qualitatively identical, and reasonably aligned with the bisector,
  see Figure \ref{fig:forward-UQ-K}-top and bottom third and fourth panel. The empirical distribution of the misfits in both cases are qualitatively identical and close to a Gaussian
  (see the quantile-quantile plots in Figure \ref{fig:forward-UQ-K}-top and bottom-right panels). In summary, all of these diagnostic
  tools give little-to-no evidence that $K=2$ is the wrong choice of the under-reporting parameter.
  We close this example with some remarks:
  \begin{itemize}
    \item the quality of the fitting of the cases $K=2,3$ cannot be assessed by the $R^2$ coefficient,
      which is not well-defined for non-linear least-squares problems \cite{spiess:againstR2}.
    \item if instead $K$ is known, the SIR system is practically identifiable regardless of time,
      since the NLL in the $\beta-r$ plan has always a unique minimum close to the true value, in all of the three scenarios,
      see Figure \ref{fig:identifiability_of_SIR_for_known_K}-top.
      Of course, the time span of data collection still has an impact on the quality of the results.
      Indeed, if time increases the minimum of the NLL is less and less shallow,
      which implies that the uncertainty on the parameters is smaller and smaller. This is visible in
      Figure \ref{fig:identifiability_of_SIR_for_known_K}-bottom, where we show the Gaussian approximation of the
      posterior pdfs of $\beta,r$. As the data time-span increases, the posterior pdfs are more and more concentrated
      and, equivalently, the system is more and more practically identifiable (cf. Fisher Information Matrix
      criterion for practical identifiability).
    \item \lorenzo{As already mentioned in Example \ref{ex:structural_identifiability_SIR},
        \cite{Tuncer} discusses whether the SIR model without under-reporting factor is identifiable from cumulative incidence data,
        rather than prevalence data; since cumulative incidence data are most typically reported in an outbreak, this question is very relevant
        for practical purposes. The finding is that while SIR is indeed structurally identifiable from cumulative incidence data,
        it is \emph{not} practically identifiable from this kind of data, and advocates for a broader diffusion of prevalence data,
        from which SIR is practically identifiable, as just discussed in this example.}
   \end{itemize}

\end{example}

\begin{example}[Structural and practical identifiability of a SEIRD model]\label{ex:SEIRDz}
  
  In this example, we discuss the identifiability of a slightly more complex model
  with incubation period (compartment ``E'', exposed), where we
  distinguish between recovered and dead persons (compartments ``R'' and ``D'', respectively). We also assume
  that at $T=T_{lock}$ the parameter $\beta$ changes, due to some restriction measure being enforced (lockdown).%
  \footnote{A change in $\beta$ of 90\% was observed on real-data e.g. in \cite{Crisanti}.}
  We call this model SEIRDz:
  \[
    \begin{cases}
      \displaystyle \dot{S} = - \frac{\beta(t)}{N_{pop}}IS\\[10pt]
      \displaystyle \dot{E} =  \frac{\beta(t)}{N_{pop}}IS - i E\\[6pt]
      \displaystyle \dot{I} =  i E - d I - r I\\[6pt]
      \displaystyle \dot{R} = r I \\[6pt]
      \displaystyle \dot{D} = d I
    \end{cases}
  \]
  where $\beta(t) = \beta_1$ for $t \leq T_{lock}$ and $\beta(t) = \beta_1 - z$ for $t > T_{lock}$.%
  \footnote{\lorenzo{Another option would be to consider $\beta(t)=\beta_1 (1-z)$ after $T_{lock}$,
      i.e., to use $z$ as the percentage decrease of $\beta_1$ instead of the absolute decrease.
      The treatment of the two models would be identical.}}
  We consider these ranges for the parameters (that we consider as uniform random variables):
  \[
    \beta_1 \in [0.25, 0.35] \quad
    r \in [0.06,  0.18] \quad
    d \in [0.01,  0.02] \quad
    i \in [0.14, 0.33] \quad
    z \in [0.1, 0.2]. 
  \]
  In this setting the rate $r$ is the recovery rate (inverse of the average days of sickness),
  the rate $d$ is the mortality rate and the rate $i$ is the inverse of the incubation time.
  With these intervals we are assuming that the average number of days of sickness is roughly between 5 and 16,
  while the average incubation time is roughly between 3 and 7 days.
  \lorenzo{Moreover, upon introducing the fatality ratio $d/(d+r)$, i.e. the proportion of infected that eventually die,
    this example considers a very lethal infection, with a fatality ratio roughly between 5\% and 25\%.}
  We consider $T_{lock} = 15$ and run the simulation until $T=100$.
  We set the initial conditions to $S(0) = 0.95, E(0) =0.04, I(0)=0.01, R(0)=0, D(0)=0$,
  \lorenzo{and we assume of having at disposal measurements of the compartments $I$, $R$, $D$.}
  
\begin{figure}
  \centering
  \includegraphics[width=0.23\linewidth]{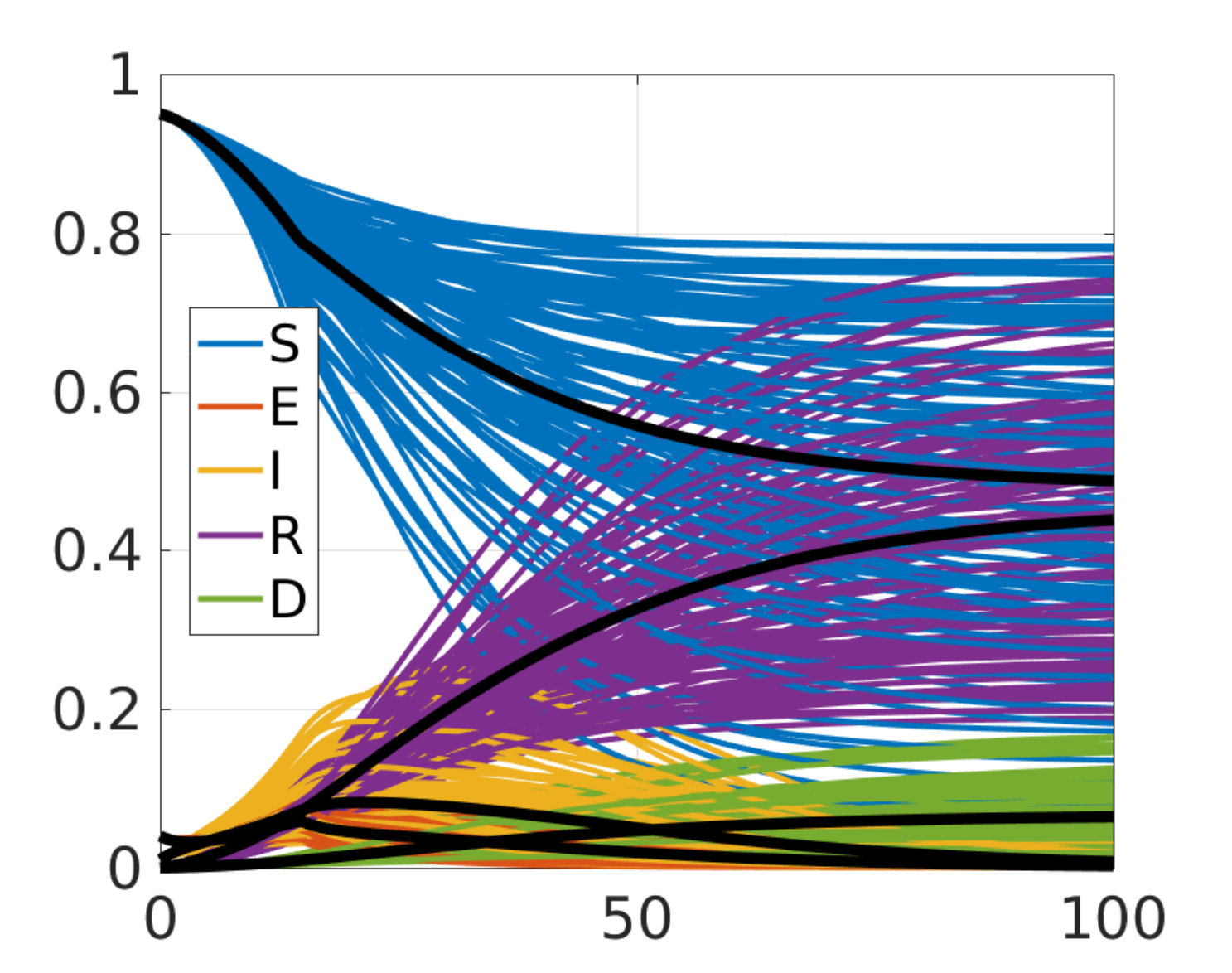} 
  \includegraphics[width=0.75\linewidth]{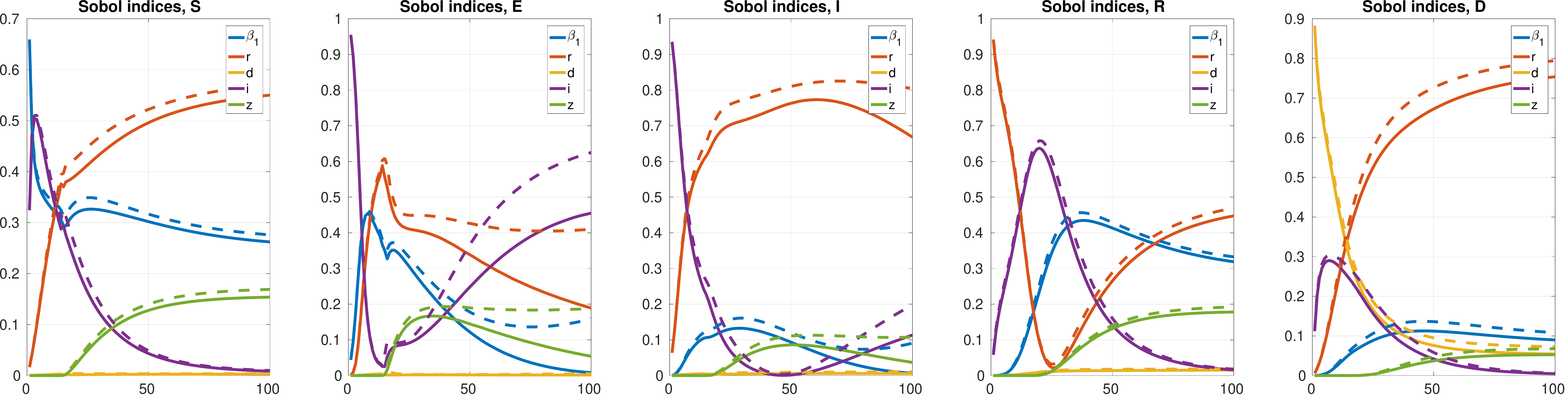} 
  \caption{Prior-based forward UQ for the SEIRDz model. Left-most panel: Monte Carlo realizations and expected value of the compartments.
    Remaining panels: time-evolution of the Sobol indices.}\label{fig:SEIRDz-prior-based-forward}
\end{figure}

If we assume that $\beta$ is constant in time, we can show that the system is structurally identifiable, by means of the
differential algebra approach. The computations are shown in details in  \ref{ex:structural_identifiability_SEIRD}.
Since we have assumed that we know the time $T_{lock}$ where the change in $\beta$ happens, we can apply
the structural analysis results to both the time intervals $t \leq T_{lock}$ and  $t > T_{lock}$ separately, and
conclude that the model  SEIRDz is structurally identifiable.

Figure \ref{fig:SEIRDz-prior-based-forward} shows some results for the preliminary prior-based forward UQ
analysis. The left-most panel shows 100 Monte Carlo trajectories, and the expected value of the compartments, computed with a sparse grid
(2433 model evaluations\footnote{the number of sparse grids points is larger than in Example \ref{ex:UQ_for_SIR}, where we considered
  a SIR model. This is because now we have to 
  sample a 5-dimensional parameters space, and sparse grids suffer to a certain degree the so-called ``curse of dimensionality'' \cite{wasi.wozniak:cost.bounds},
  i.e. loosely speaking, the number of sampling points grows rapidly (more than linearly) with the number of dimensions. A sampling method that fully
  suffers from this problem is cartesian sampling, where the number of points grows exponentially with the number of dimensions.}).
It is clearly visible that the trajectories are significantly scattered, and that the asymptotic values of the compartments vary considerably.
The remaining panels show the time-evolution of the Sobol indices, from which we can derive some information about the identifiability
of the model. We can see that not all parameters impact equally the variability
of the solution, and we expect in particular that it will be difficult to recover by the inversion procedure the value of
those parameters that have the smallest impact ($i,z,\beta_1$). Even more so given that we only measure the compartments $I$, $R$, $D$ (while $\beta_1$ would be best recovered from the $S$ compartment),
and that we will only measure them up to a certain time, and the Sobol indices are not constant
in time (for instance, $\beta_1$ has a significant impact on $R$ but only at late times, say $T \geq 30$, while we measure essentially early times).
Thus, while the system is structurally identifiable, it might be practically non-identifiable.
Next, we perform the inversion and check for practical identifiability. We fix the parameters as 
\[
\beta_{1,true} = 0.28, \quad
r_{true}  = 0.11, \quad
d_{true} = 0.018, \quad
i_{true} = 0.18, \quad
z_{true} = 0.18, \quad
K=3, \quad
\sigma = 0.01;
\]
and measure data until $T=40$ (after peak). We repeat the minimization procedure 20 times with different initial guesses for the parameters,
to investigate the presence of local minima of the NLL, and select as MLE the results that led to the smallest NLL.
The results of this procedure are shown in Figure \ref{fig:SEIRDz_inversion}-top, where we report the initial and final values of the
NLL, as well as the initial and final values of the parameters. We can conclude that while the final values of the NLL are all close (yet not identical),
the values of the parameters show a significant variability, denoting the fact that the NLL has a multiple minima
with similar NLL value. Moreover, most of the parameters are not correctly
identified. This further suggests that the model might be practically non-identifiable, at least for the value of $\sigma$ tested here. 
Therefore, the trajectories corresponding to the computed values of the parameters are quite far from the true one,
see Figure \ref{fig:SEIRDz_inversion}-bottom.

To confirm our diagnosis of non-identifiability, we compute the \emph{profile likelihood} of the problem.  
Results are shown in Figure \ref{fig:SEIRDz_pNLL}. The only parameter with a deep, narrow minimum is $r$ (compare the vertical scales)
while the other ones are in rather shallow regions. Moreover, the profile likelihoods for $i$, $\beta_1$ and $z$ are very noisy
(as expected, due to the fact that their Sobol indices are rather small).
This confirms that the only parameter that can be easily identified are $r$ and to a certain extent $d$ (given that
the shape of the profile likelihood is not noisy, although shallow), and overall the system is practically non-identifiable.

\begin{figure}
  \centering
  \includegraphics[width=0.16\linewidth]{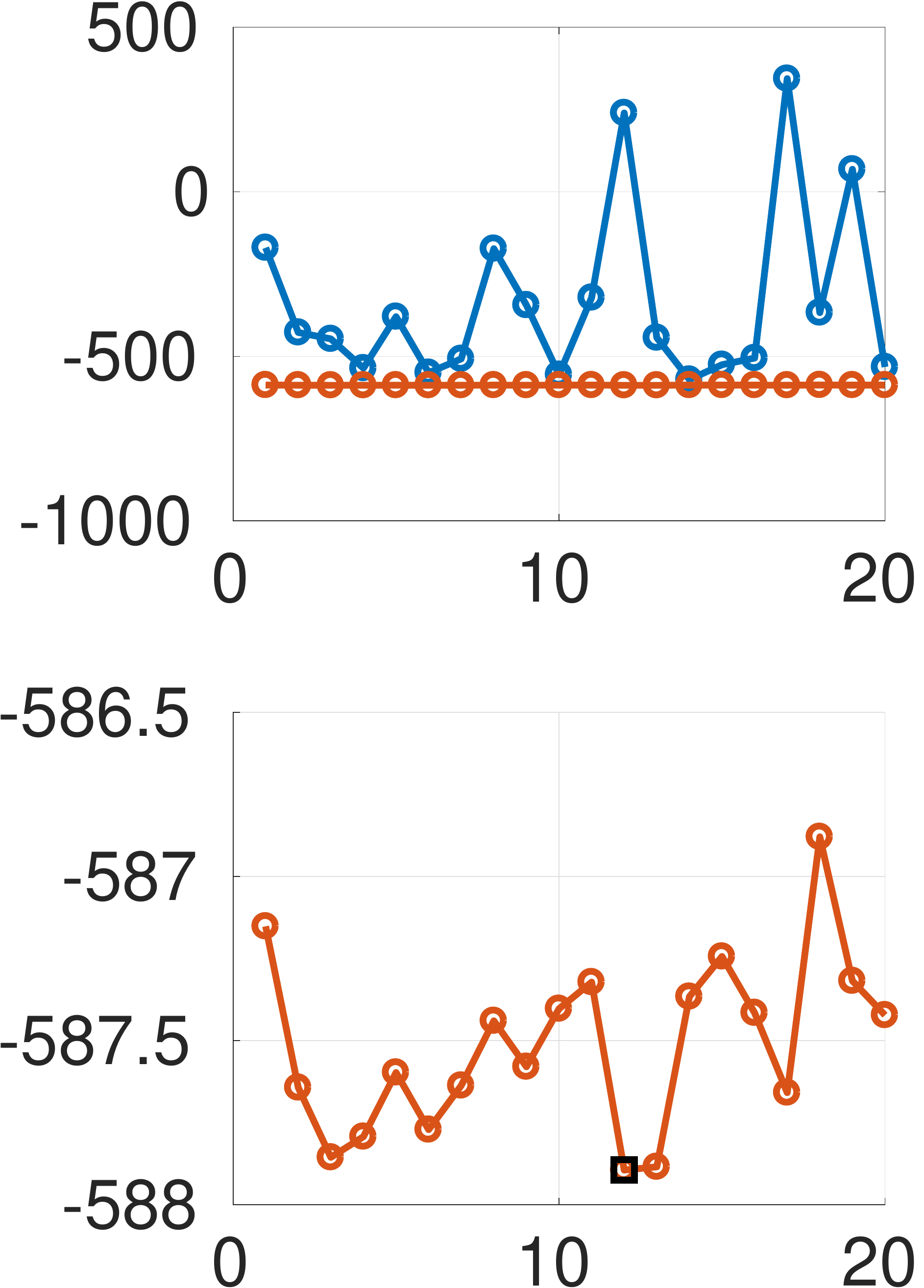}
  \includegraphics[width=0.80\linewidth]{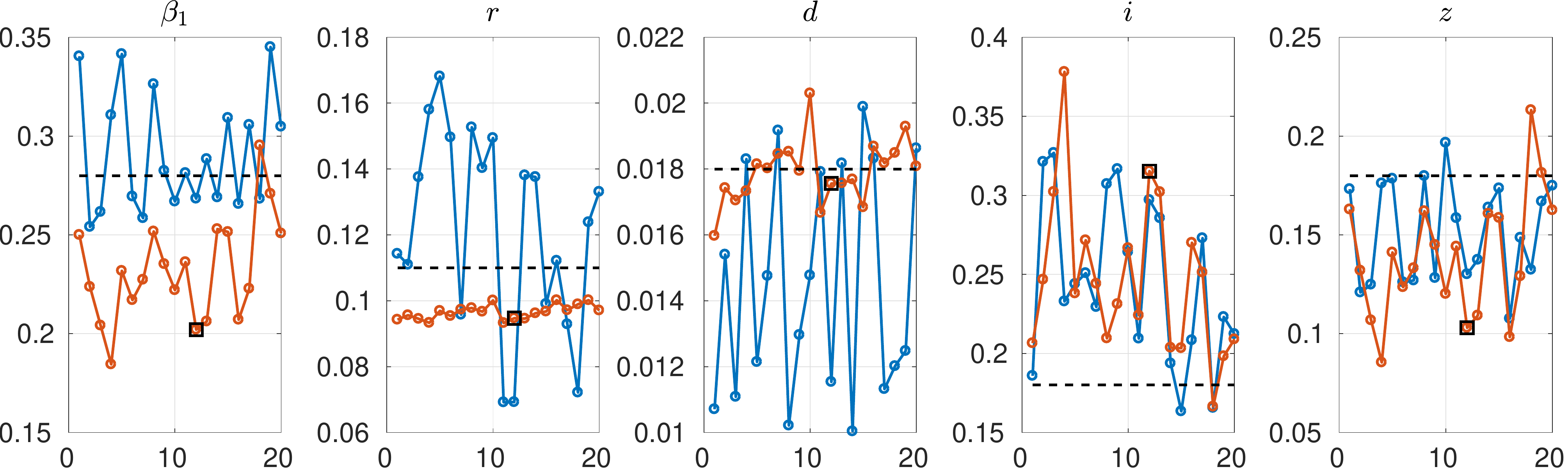}\\[6pt]
  \includegraphics[width=0.65\linewidth]{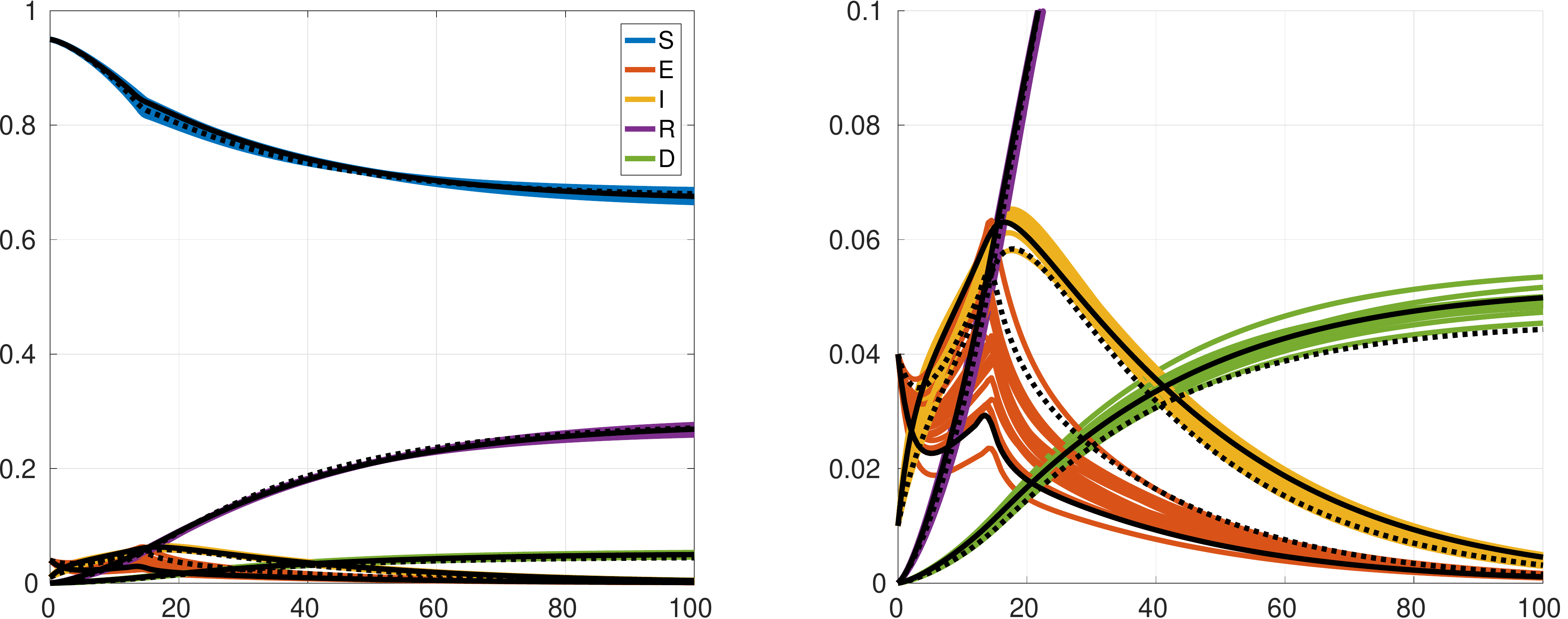}
  \caption{Results for inversion of SEIRDz. Top row, from left to right:
    the initial and final value of NLL for the 20 trials (blue and red line, respectively) and a zoom on the final values (the value with the
    smallest NLL is marked with a black square);
    the initial and final values of the parameters for the 20 trial (blue and red line, respectively), the correct values (black dash lines),
    and the values of the parameters that yield the smallest NLL (black square marker).
    Bottom row: the trajectories of the 20 MLEs and a zoom on the $E,I,D$ compartments.
    The colored trajectories are those obtained by the values of the parameters obtained by the 20 optimization trials.
    The MLE trajectories are reported in full black line, while the true ones in dashed black lines. 
	}  \label{fig:SEIRDz_inversion}
\end{figure}

\begin{figure}
  \includegraphics[width=0.95\linewidth]{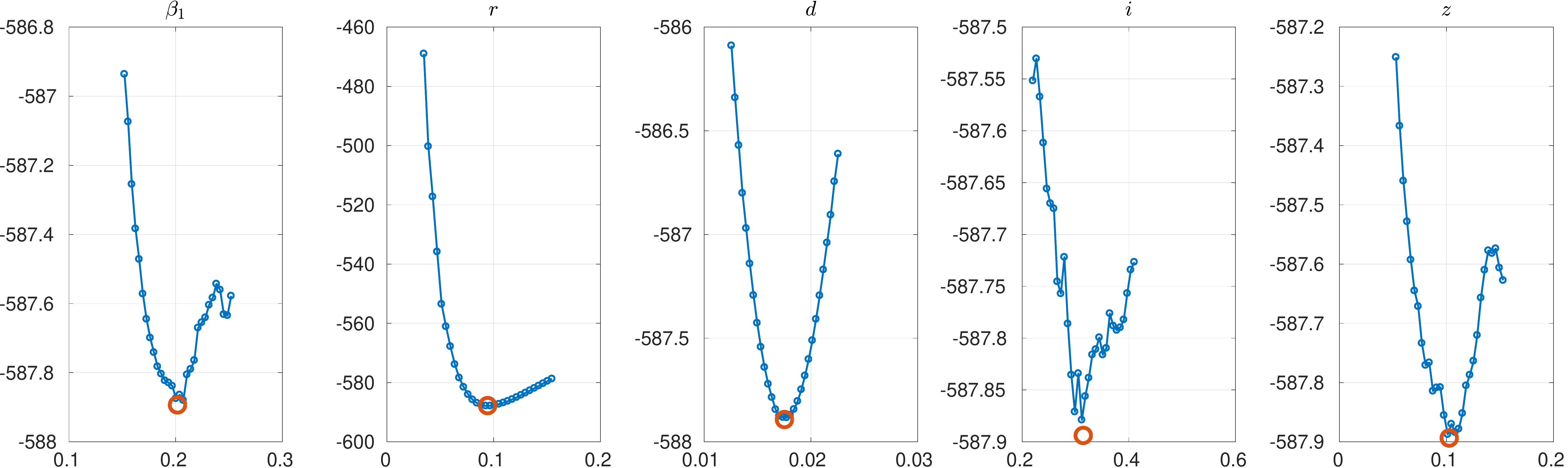}
  \caption{Profile likelihood for the SEIRDz identifiability problem, centered around the MLEs of the parameters
    (red circle markers). Observe that the red markers do not always coincide with the profile likelihood. This is because
    the optimization for the remaining parameters might land at different local optima (despite running the optimizer
    multiple times with random initial points).}\label{fig:SEIRDz_pNLL}
\end{figure}

As mentioned, a possible workaround which might help in this situation is to learn some of the parameters from independent studies,
and thus reduce the number of parameters to be simultaneously identified. For instance, medical studies
might give us estimates of the incubation time, recovery time and death rate, so that we are left to
identify only the contact probabilities $\beta_1$ and $z$. One has to pay attention to the fact that
setting the influential parameters to the wrong values can be however detrimental to the procedure.
Here, we fix the values of $i,r$, and $d$ to their exact values and repeat the identification procedure.
This is of course an over-optimistic scenario. Another possibility would be to use, for example, $r,d$ obtained
from the inversion procedure (whose profile likelihood is ``well-shaped'',  even though the one for $d$ is quite shallow;
we named this procedure as ``hierarchical optimization'' in the previous discussion).
The results obtained by fixing $i,r$, and $d$ to their exact values are reported in Figure \ref{fig:SEIRDz_inversion_reduced}, which shows the same information
of Figure \ref{fig:SEIRDz_inversion}, i.e. initial and final values of NLL and parameters, and trajectories
of the model. The presence of local minima in the NLL is greatly reduced, and the identification of $\beta_1$ and $z$
is more robust and closer to the true values (this might not always be the case though, depending on the noise level and
the quality of the data). As a result, the trajectories corresponding to this new
set of parameters are closer to the true ones than the previous results, as shown in the right-most panel.

\begin{figure}
  \centering
  \includegraphics[width=0.20\linewidth]{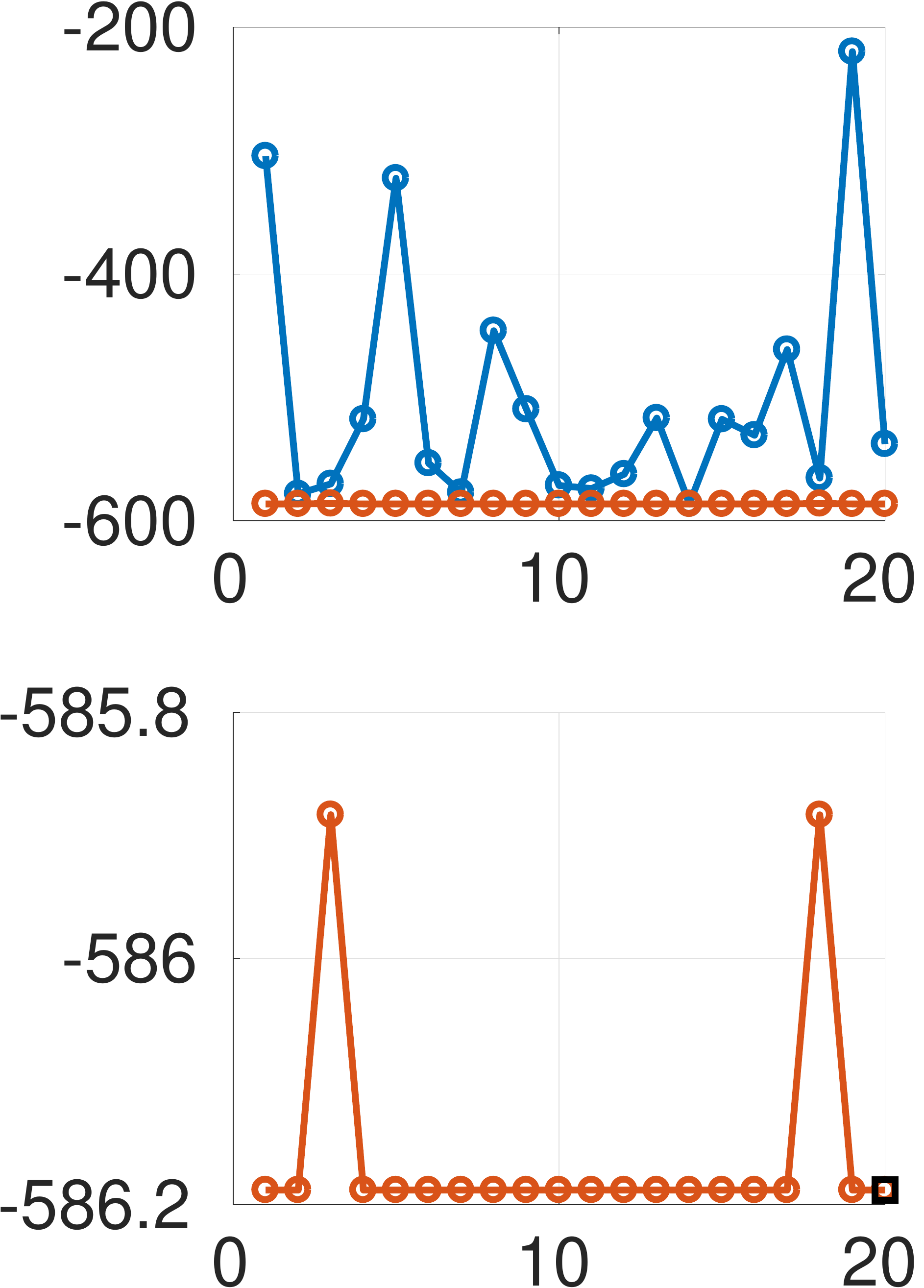}
  \includegraphics[width=0.41\linewidth]{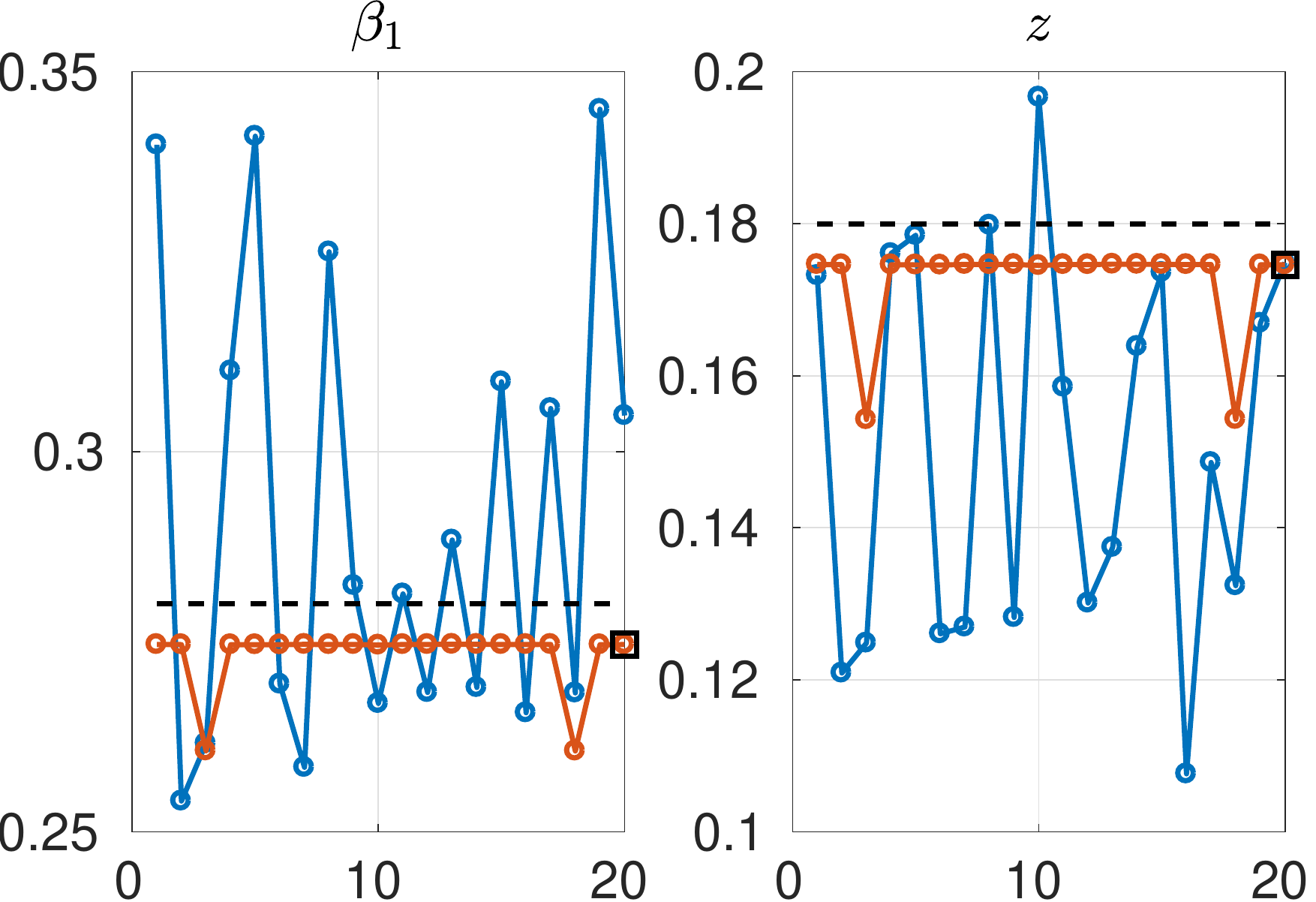}
  \includegraphics[width=0.35\linewidth]{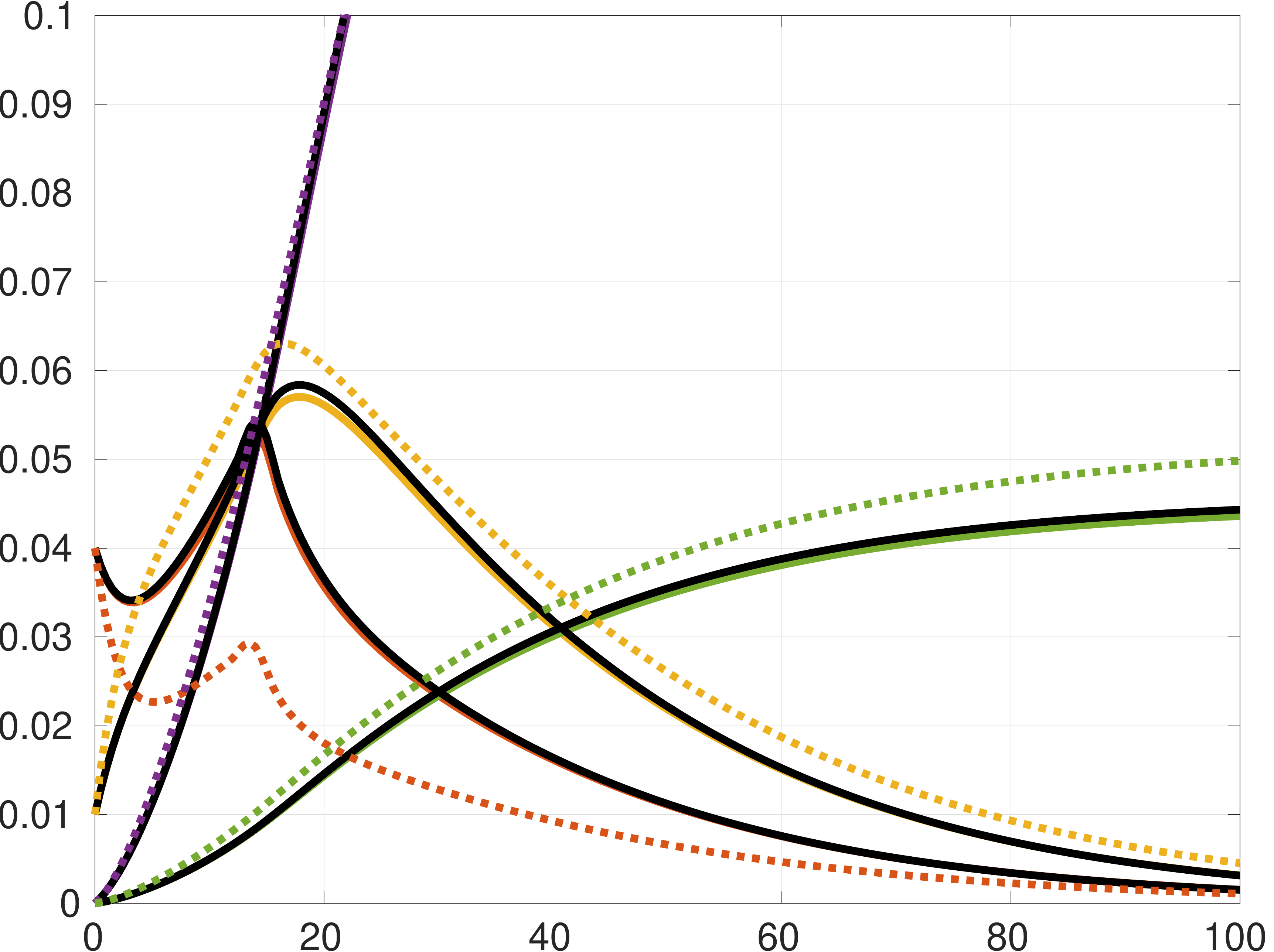}
  \caption{Results for inversion of SEIRDz upon blocking the parameters $r,d,i$. Left panel:
    the initial and final value of NLL for the 20 trials (blue and red line, respectively)
    and a zoom on the final values (the value with the smallest NLL is marked with a black square).
    Mid panel: the initial and final values of $\beta_1,z$ for the 20 trial (blue and red line, respectively),
    the correct values (black dash lines), and the values of the parameters that yield the smallest NLL (black square marker).
    Right panel: the true trajectories (black lines), the MLE when trying to identify all parameters (colored dotted lines)
    and the MLE when trying to identify $\beta_1,z$ only (full colored lines). 
	}  \label{fig:SEIRDz_inversion_reduced}
\end{figure}

\end{example}


\begin{biblio}
  \begin{itemize}
    \item For a general survey on practical identifiability,  we refer the reader to \cite{Miao,Tuncer,Roosa.Chowell}.
    \item Bootstrap approaches in the context of epidemiological models are considered e.g. in \cite{Tuncer,Roosa.Chowell}.
      The analysis in \cite{Tuncer} considers a criterion based on ARE to conclude on identifiability, whereas
      in \cite{Roosa.Chowell} a combination of the size of the confidence interval and mean squared error of the parameters
      is employed.
    \item An extensive explanation on the use of the likelihood profiles is given in \cite{Raue1} and further references are \cite{Eisenberg,Tonsing}.
    \item \cite{Chowell} provides a quite comprehensive step-by-step guide on fitting compartmental models for epidemiology, with an eye to identifiability.
    \item In the context of COVID-19, some discussion about identifiability is provided by \cite{Roda}.
    \item \cite{Eisenberg} provides an interesting example of the consequences of identifiability  
      in the context of a SEIR-based model for a Dengue outbreak: the system is practically non-identifiable,
      resulting in two sets of parameters that show an excellent fit of the data but provide dramatically
      different predictions when used to test a possible non-medical remediation strategy (removal of mosquitos).
    \item Optimal design of experiments can be used to improve the quality of the data to minimize the
      impact of practical identifiability issues, see e.g. \cite{long:expdesign}.
  \end{itemize}   
\end{biblio}

\section{Discussion and conclusions: a revisited UQ workflow}\label{section:disc}

In these notes we have reviewed some computational tools for prediction under uncertainty and parameter identification
for dynamical systems, which we referred to as forward and inverse UQ analyses, respectively.
Combined together, 
these tools provide a powerful framework for reliable predictions of the outputs of a dynamical system,
and complement the punctual predictions with confidence estimates with solid ground in probability/statistics theory.
However, investigators should always
carefully check whether their model is actually identifiable, both structurally and practically.
Dynamical systems might indeed be not fully identifiable, and blindly using the prediction tools
in this case can be harmful,
see e.g. Example \ref{ex:SIR_identifiability} or the above-mentioned case study on Dengue
reported in \cite{Eisenberg}.

If the system is not identifiable, the Fisher approach to the inversion problem is bound to fail,
because it intrinsically assumes identifiablity of the system, or in other words because it assumes
that the NLL has a unique minimum where the Gaussian approximation of the posterior should be centered,
whereas in case of structural non-identifiability the NLL has a manifold of minima. The case of NLL with
a finite number of local minima (symptom of practical non-identifiability,  see Section \ref{section:pract_id})
could instead hopefully be fixed by acquiring more data of the right kind, that should hopefully rule out the ``wrong minima''.
Conversely, MCMC methods make no assumptions on the shape of the NLL and therefore might be a partially safer technique,
However, MCMC algorithms come with a much larger computational cost and are not entirely safe either,
since they typically implement some sort of adaptive sampling, where most of the samples are collected in regions of large
likelihood, i.e., they cannot entirely escape the problem of computing the maxima of the likelihood. Therefore, unless they are properly designed and tuned,
they could fail to realize that there might be an entire manifold of minima. 
A compromise solution could be to use the Fisher approximation not plainly as the posterior pdf
of the parameters, but only in the context of an importance sampling strategy, where one
still generates samples from the Fisher approximation of the posterior but then rescales
them suitably to remove any bias \cite{beck:importance.sampling,beck:multi.level.importance}. 
In summary, the ideal UQ workflow sketched in Section \ref{section:UQ_workflow} (cf. Algorithm \ref{workflow:UQ})
can be adjusted as detailed in Algorithm \ref{workflow:UQ2}.
\begin{algorithm}
  \caption{Ideal UQ workflow}  \label{workflow:UQ2}
  Choose a model and the prior distributions for its parameter (literature, expert opinion)\;
  Determine whether the system is \textbf{structurally} identifiable (Sobol indices, profile likelihood, differential algebra, mapping approach, etc.)\;
  \eIf{the model is \textbf{structurally} identifiable}{
    \While{the model is not \textbf{practically} identifiable (bootstrap, profile likelihood, multiple restart, etc.)}{
        Acquire more data / get information on some parameters from independent studies / perform a hierarchical optimization\;
      }
     choose Fisher approximation as inversion method\;
  }{
    the likelihood has a manifold of minima: choose an appropriate MCMC algorithm as inversion method\;
  }
  perform the inverse UQ analysis\;%
  perform the forward UQ analysis based on the posterior distribution to obtain statistical information about the quantities of interest of the model
    (e.g. expected value, variance, full pdf of the outputs)\;
\end{algorithm}

\lorenzo{Of course, the underlying assumption here is that we know what is the exact model that generated the data,
  and we are dealing with parametric identification only. Discussing how model mis-specification affects identifiability
  further adds to the complexity of the problem and it is out of the scope of this work.}

Our theoretical discussion has been complemented with a number of small examples. None of these consider the
initial conditions of the system as unknown, but doing so would not pose any conceptual challenge from a numerical point of view.
As already hinted in Section \ref{sect:struct_identifiability}, the structural identifiability of parameters connected to the initial conditions can be investigated e.g. by the mapping approach.

An important point that we did not discuss is the issue of \emph{model-selection}. In the case when multiple
models are available (quite common in the case of epidemics modeling and, in particular, of COVID-19), 
is there any way to tell which one has the largest statistical evidence?
A large body of work is available on this topic in the statistical literature, where several criteria
have been developed to select the ``best model''. The underlying principle is that adding more parameters
might lead to a better fit of the data, but the more parameters, the larger the
chances that the model is overfitted, i.e., that it adjusts to the noise and gets limited predicting power.
Thus, one should restrain from blindly adding more parameters.\footnote{this is a mathematical formulation of the Occam's razor.}
Criteria that try to identify the optimal model among a pool of possible ones include, for example, the
Akaike Information Criterion (AIC), the Bayes Information Criterion (BIC), and the Kayshap Information Criterion (KIC). 
We refer the interested reader e.g. to \cite{schoniger.eal:model.selection,claeskens:model.selection,burnham:modelselection}
for a more thorough discussion, as well as to e.g. \cite{doi:10.1111/rssb.12187} for a discussion on a model selection strategy
in the case when some models are not identifiable.

\section*{Acknowledgments}

The authors acknowledge the many fruitful discussions with several colleagues,
and in particular the colleagues at CNR-IMATI that participated in the COVID-19 modeling study group. 


\section*{Declarations of interest}

None.

\section*{Funding}

Lorenzo Tamellini and Chiara Piazzola have been supported by the PRIN 2017 project 201752HKH8
``Numerical Analysis for Full and Reduced Order Methods for the efficient and accurate solution
of complex systems governed by Partial Differential Equations (NA-FROM-PDEs)''.
Lorenzo Tamellini also acknowledges the support of GNCS-INdAM (Gruppo Nazionale Calcolo Scientifico - Istituto Nazionale di Alta Matematica).
This work was supported by  the KAUST Office of Sponsored Research (OSR) under Award No. URF/1/2584-01-01
and the Alexander von Humboldt foundation. Ra\'ul Tempone is a member of the KAUST SRI Center for Uncertainty Quantification in Computational Science and Engineering.

\appendix
\section{Structural identifiability of a SEIRD model by differential algebra}\label{ex:structural_identifiability_SEIRD}
In this section we consider the SEIRD model, which is a simplified version of the SEIRDz model considered in Example \ref{ex:SEIRDz} with $\beta$ constant in time. We show by means of the differential algebra technique explained in Section \ref{sect:struct_identifiability} that it is structurally identifiable from prevalence data of $I$, $R$ and $D$. 

Let us consider the following system 
\[
\begin{cases}
\displaystyle \dot{S} = - \frac{\beta}{N_{pop}}IS\\[10pt]
\displaystyle \dot{I} =  i(N_{pop}-S-I-R-D) - d I - r I\\[6pt]
\displaystyle \dot{R} = r I \\[6pt]
\displaystyle \dot{D} = d I \\[6pt]
\displaystyle Y = \frac{1}{K} I \\[6pt]
\displaystyle Z = \frac{1}{K} R \\[6pt]
\displaystyle W = \frac{1}{K} D,
\end{cases}
\]
where we have removed the equation for the compartment $E$, as it holds that $N_{pop} = S+E+I+R+D$ (the same argument was used in Example \ref{ex:structural_identifiability_SIR},
when we removed the equation for $S$ while discussing identifiability of SIR with data of $I$ and $R$).
We then rewrite the differential equations in terms of the observed variables. In particular, we derive the following explicit expression for $S$ from the second equation: 
\[
S = -\frac{K}{i}\dot{Y}+N_{pop}-K\frac{i+d+r}{i}Y-KW-KZ.
\]
From this, we compute $\dot{S}$, insert both formulas in the first differential equation above, and obtain the first input-output equation:
\[
-\frac{K}{i}\ddot{Y}-K\frac{i+d+r}{i}\dot{Y}-K\dot{W}-K\dot{Z} - \frac{\beta K^2}{i N_{pop}} Y\dot{Y} + \beta K Y - \frac{\beta K^2}{i N_{pop}}(i+d+r)Y^2 = 0.
\]
The other two input-output equations follow from the third and fourth differential equation above and are: 
\[
K \dot{Z} - rKY = 0 \qquad \text{and} \qquad  K\dot{W} - dKY = 0.
\]
The set of these three input-output equations is not mutually reduced with respect to the ranking $Y<Z<W<\dot{Y}<\dot{Z}<\dot{W}<\ddot{Y}<\ddot{Z}<\ddot{W}$
(other ranking would lead to the same results).
The third equation is not reduced with respect to the first one, as its leader monomial $\dot{W}$ appears also in the first one. Similarly, the second equation is not reduced with respect to the first one. By doing some further substitutions to eliminate $\dot{Z}$ and $\dot{W}$ in the first equation we finally get a set of mutually reduced equations:
\[ 
\begin{cases}
\displaystyle -\frac{K}{i}\ddot{Y}-K\frac{i+d+r}{i}\dot{Y}-KdY-KrY - \frac{\beta K^2}{i N_{pop}} Y\dot{Y} + \beta K Y - \frac{\beta K^2}{i N_{pop}}(i+d+r)Y^2 = 0 \\[6pt]
K \dot{Z} - rKY = 0\\[6pt]
K\dot{W} - dKY = 0.
\end{cases}
\]
The last step is to make the polynomials monic with respect to their leaders, which are $\ddot{Y}$, $\dot{Z}$, and $\dot{W}$, respectively. It then follows that all the coefficients can be uniquely determined; hence, the SEIRD model is structurally identifiable from data of $I$, $R$, and $D$.

\bibliographystyle{elsarticle-num}
\bibliography{COVID_biblio,UQ_biblio}

\end{document}